\documentclass[pdflatex,sn-nature,icol]{sn-jnl}
\usepackage{graphicx}%
\usepackage{multirow}%
\usepackage{amsmath,amssymb,amsfonts}%
\usepackage{amsthm}%
\usepackage{mathrsfs}%
\usepackage[title]{appendix}%
\usepackage{xcolor}%
\usepackage{textcomp}%
\usepackage{manyfoot}%
\usepackage{booktabs}%
\usepackage{algorithm}%
\usepackage{algorithmicx}%
\usepackage{algpseudocode}%
\usepackage{listings}%
\usepackage{comment}
\usepackage{braket}
\usepackage{dblfloatfix} 
\makeatletter 
\renewcommand\@biblabel[1]{#1.} 
\makeatother

\begin{document}

\title[Title]{Quantum Dial}

\author*[1]{\fnm{Aashish} \sur{Sah}}\email{aashish.sah@aalto.fi}
\author[1]{\fnm{Suman} \sur{Kundu}}
\author[1,3]{\fnm{Priyank} \sur{Singh}}
\author[1]{\fnm{Eemeli} \sur{Forsbom}}
\author[1]{\fnm{Vasilii} \sur{Vadimov}}
\author[1,2]{\fnm{Mikko} \sur{M\"ott\"onen}}

\affil[1]{\orgdiv{QCD Labs, QTF Centre of Excellence, Department of Applied Physics}, \orgname{Aalto University}, \orgaddress{\street{Tietotie 3}, \city{Espoo}, \postcode{02150}, \country{Finland}}}

\affil[2]{\orgname{VTT Technical Research Centre of Finland Ltd}, \orgaddress{\street{Tietotie 3}, \city{Espoo}, \postcode{02150}, \country{Finland}}}
\affil[3]{\orgname{Center for Quantum Science and Technologies, Indian Institute of Technology}, \orgaddress{\street{Mandi}, \city{Himachal Pradesh}, \postcode{175075}, \country{India}}}

\abstract{
Accurate control and the resulting utilization of the vast quantum degrees of freedom appears promising for practical applications in sensing, communications, and computing~\cite{Degen2017QuantumSensing,Pirandola2020AdvancesCryptography,Preskill2018QuantumBeyond,Blais2021CircuitElectrodynamics}. Especially the realization of a practically beneficial quantum computer is one of the greatest on-going challenges of science and technology~\cite{Preskill2018QuantumBeyond}. 
The general criteria for constructing an operational quantum computer out of a register of qubits may seem contradicting: the qubits need to stay coherent for long time and hence be isolated from their environment, but at the same time, there needs to be an ability to control, readout, and reset the qubits, which calls for coupling to many auxiliary degrees of freedom~\cite{DivincenzoTheComputation,Krantz2019AQubits}. To date, this isolation-and-control (InC) challenge has been addressed individually by reset components, driving schemes, and Purcell filters, each with added complexity~\cite{Magnard2018FastQubit,Motzoi2009SimpleQubits,Chow2010OptimizedGates,Reed2010FastQubit,Sete2015QuantumReadout}. Unfortunately, decreasing the coupling strength of a qubit to its drive line for increased coherence leads to increased heating and cross coupling with other qubits~\cite{Krantz2019AQubits,Tripathi2022SuppressionDecoupling,Ding2020SystematicCompilation}. 
No single concept satisfactorily addressing the InC challenge has been demonstrated, till date. 
Here, we introduce and demonstrate the first generation of a quantum dial which on-demand mediates a desired strength of coupling of a qubit to the chosen auxiliary degrees of freedom. 
Our quantum dial consists of a band-stop filter between a high-coherence transmon qubit and a broadband transmission line, allowing to tune their coupling strength by several orders of magnitude in nanosecond timescales without any significant Stark shift on the qubit. By dialing the qubit into the reset configuration, we reduce its energy relaxation time $T_1$ from $>\!150~\mu\textbf{s}$ to roughly $200~\textbf{ns}$ and demonstrate qubit reset independent of its initial state, and control-configuration dial leads to $99.99\%$ idle fidelity and $99.9\%$ gate fidelities for $40~\textbf{ns}$ pulses of $\sim -110~\textbf{dBm}$. Importantly, the ability to quickly thermalize the qubit with its environment and to preserve long $T_1$ for a subsequent readout allows us to use the qubit as a thermometer of its environment with noise-equivalent temperature (NET) down to $0.6~\textbf{mK}/\sqrt{\textbf{Hz}}$ at $60~\textbf{mK}$ and essentially reaching the Cram\'er--Rao bound at higher temperatures.
Consequently, the quantum dial appears promising in providing on-demand isolation and strong coupling as desired in future quantum computers, potentially leading to reduced noise and errors in computations. 
}

\maketitle

\section*{Introduction}\label{Introduction}

Quantum systems coupled to auxiliary circuitry provide means for exploring fundamental physics and developing quantum technologies~\cite{Blais2021CircuitElectrodynamics, Devoret2013SuperconductingOutlook, Kjaergaard2020SuperconductingPlay}. The circuitry that enables coherent control and measurement also defines the environment, with which the system exchanges energy and phase information~\cite{Clerk2010IntroductionAmplification}. In superconducting quantum computers, each of the qubits is prototypically coupled to a readout resonator for state discrimination and to a $50~\Omega$ drive line for single-qubit gates~\cite{Blais2021CircuitElectrodynamics}. These couplings render the computation possible, but they also open dissipation and dephasing channels that bound the coherence times of the qubit~\cite{Martinis2005DecoherenceLoss, Ithier2005DecoherenceCircuit, Catelani2012DecoherenceTunneling}. Engineering the form and strength of this auxiliary circuitry to achieve long coherence times has therefore become central to both probing open-quantum-system dynamics and building large-scale quantum processors composed of arrays of qubits and supporting components~\cite{Reed2010FastQubit, Jeffrey2014FastQubits, Sete2015QuantumReadout, Bronn2015BroadbandElectrodynamics}.

Purcell decay of qubit energy to its readout and control environments is a principal limiting phenomenon in qubit design~\cite{Reed2010FastQubit}. For readout lines, numerous strategies have been devised to suppress this effect such as operating the readout resonator in the dispersive regime, which already reduces qubit loss through the resonator by orders-of-magnitude~\cite{Blais2021CircuitElectrodynamics}, and Purcell filters, which tailor the resonator environment and further minimize the real part of the admittance coupled to the qubit at its resonance frequency $f_\text{q}$~\cite{Jeffrey2014FastQubits, Sete2015QuantumReadout, Bronn2015BroadbandElectrodynamics}. Several flavors of Purcell filters have become widely used, ranging from notch and bandpass designs to stepped-impedance, multi-stage, and interferometric implementations, delivering substantial improvements in qubit lifetimes in state-of-the-art devices~\cite{Heinsoo2018RapidQubits, Sunada2022FastFilter, Yan2023BroadbandElectrodynamics, Zhou2024High-suppression-ratioReadout, Yen2025InterferometricQubit}. 

In contrast, qubit drive lines for implementing single-qubit gates are typically simple $50~\Omega$ transmission lines coupled capacitively~\cite{Koch2007Charge-insensitiveBox, Barends2014SuperconductingTolerance, Krantz2019AQubits, Blais2021CircuitElectrodynamics} or inductively~\cite{Manucharyan2009Fluxonium:Offsets, Nguyen2019High-CoherenceQubit, Moskalenko2022HighCoupler} to the qubit. Until recently, their coupling has been kept so weak that additional dissipation engineering seemed unnecessary. However, as fabrication advances have pushed intrinsic qubit coherence higher~\cite{Nguyen2019High-CoherenceQubit,Siddiqi2021EngineeringQubits,Premkumar2021MicroscopicQubits,Place2021NewMilliseconds,Wang2022TowardsMilliseconds,Bal2024SystematicEncapsulation,Tuokkola2025MethodsQubit}, the limitations owing to the Purcell loss through the drive line have become increasingly severe~\cite{Kono2020BreakingQubit}. Furthermore, keeping a low crosstalk throughout the chip while decreasing the coupling to the desired qubit poses a microwave engineering and fabrication challenge. 

Dissipation engineering of the drive line is distinctively different from that of the readout line since the protection and control both typically operate at the qubit frequency. Recent work aims to overcome this clash by separating the protection and control bands using subharmonic driving \cite{Xia2025FastDrives,Sah2024Decay-protectedFilters,Schirk2025SubharmonicLine}. In this approach, the control is applied at a fraction of the qubit frequency, $f_\text{q}/n$, while a low-pass filter in the cryogenic wiring chain or an on-chip stopband filter suppresses the environmental noise at $f_\text{q}$. Strong coupling at the subharmonic frequency supports fast gates, yet the filter protects relaxation evading the deterioration of qubit $T_1$. In Ref.~\cite{Sah2024Decay-protectedFilters}, we introduced quarter-wave and half-wave stub filters that realize a deep stopband at $f_\text{q}$ together with strong transmission at subharmonic frequencies. Alternatively, a power-selective solution may be considered, namely, the Josephson quantum filter (JQF) \cite{Kono2020BreakingQubit} which inserts an ancillary strongly coupled transmon directly in the control line a distance $d\approx\lambda_\text{q}/2$ from the qubit. In the small-signal regime the JQF acts as an atomic mirror, reflecting single photons emitted by the qubit and suppressing radiative decay; under strong resonant drive, the ancillary transmon saturates and becomes transparent, enabling fast control on the same line. However, it is not clear whether these static filters protect the qubit during the driving process. In addition, introducing several coupling elements to the qubit adds dissipation channels and compromises qubit coherence.

In this work, we propose the concept of a quantum dial which provides the qubit with only a single coupling element that can be temporally controlled to mediate the coupling according to the desired operation: reset, quantum logic, or readout. To this end, we introduce a tunable-frequency drive line filter that leverages the quarter-wave principle of Ref.~\cite{Sah2024Decay-protectedFilters} while adding \textit{in situ} tunability of the stopband via a series array of direct-current (dc) superconducting quantum interference devices (SQUIDs) integrated along the drive line. Tunability compensates for fabrication spread in $f_\text{q}$ and allows us to fully decouple the qubit from the drive line on demand, yielding almost three-orders-of-magnitude tuning of the measured Rabi frequency and $T_1$. With an on-chip fast-flux line for the tunable filter, we switch the dial in the nanosecond timescale from the idle configuration to (a) control configuration for fast gates and (b) qubit reset configuration for state initialization. We demonstrate fast control that bypasses the trade-off between gate speed, long coherence, and reduced cryogenic thermal load. We also realize sub-$\mu\text{s}$ reset from $|\text{e}\rangle$, $|\text{f}\rangle$, and $|\text{h}\rangle$ states, and carry out fast qubit thermometry up to hundreds of millikelvins with state-of-the-art sensitivity.

\section*{Results}\label{Results}

\begin{figure*}[ht]
\includegraphics[width=\linewidth]{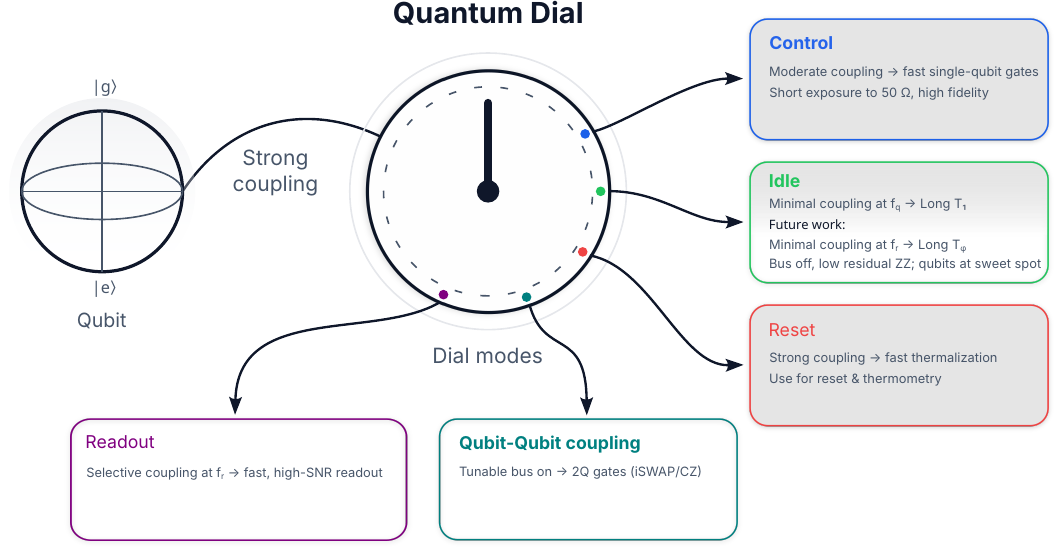}
\caption{\textbf{Quantum-dial concept.} The quantum dial (centre) chooses the environment coupled to a high-coherence qubit (left). In the first experiments reported here, we have access to three core dial configurations (shaded in gray): \textbf{Control} selects moderate coupling for fast, high-fidelity single-qubit gates, \textbf{Idle} isolates the qubit for state preservation, and \textbf{Reset} implements strong coupling to a cold $50~\Omega$ bath for fast initialization and thermometry. Two additional dial configurations completes the scheme (future work): \textbf{Readout} facilitates selective coupling at the resonator frequency $f_{\mathrm{r}}$ to open the measurement linewidth $\kappa$ only during readout while keeping the qubit protected at $f_\text{q}$, and \textbf{Qubit–qubit coupling} implements a tunable bus that is off at idle (low residual coupling) and turned on for two-qubit gate operations, for example, via parametric exchange (iSWAP) or a static $ZZ$ phase (CZ). This architecture enables duty-cycled protection of all elements, deterministic hardware reset, and two-qubit gates without moving qubits away from their flux sweet spots.}

\label{fig:quantum_dial}
\end{figure*}

\begin{figure*}[ht]
\includegraphics[width=\linewidth]{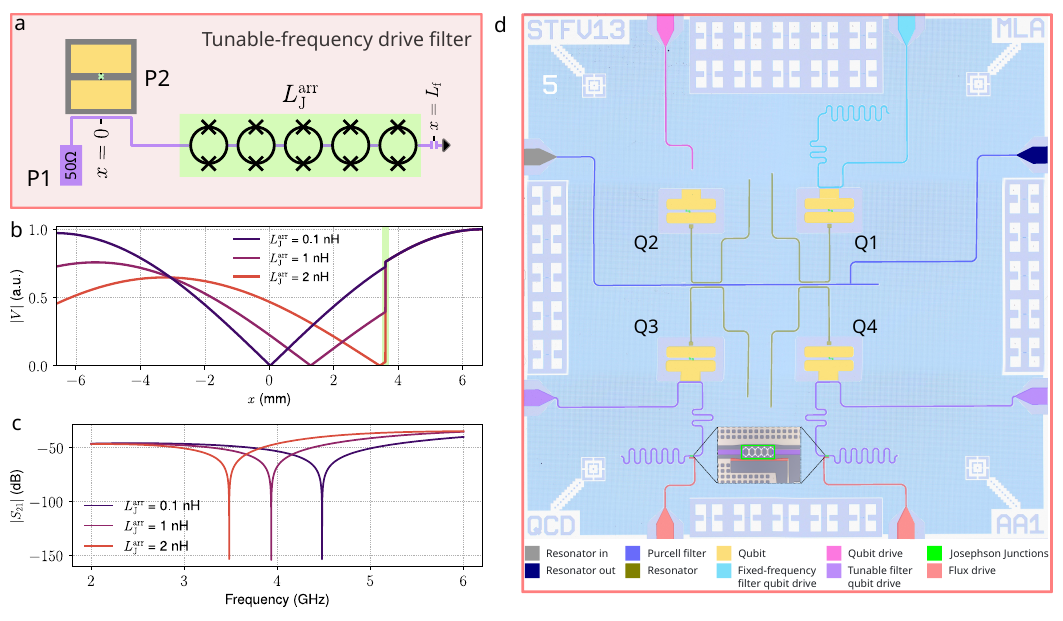}
\caption{\textbf{Device layout and operation principle.} \textbf{a}, Schematic diagram of a tunable-frequency drive-line filter implemented as a $50~\Omega$ transmission line that is strongly capacitively coupled to the qubit at the position $x=0$ (port P2). From the coupling point, the line extends a length $L_\mathrm{f}$ and is weakly terminated to ground through a small capacitance, forming a quarter-wave filter. A series array of direct-current (dc) superconducting quantum interference devices (SQUIDs) with an effective inductance $L_\mathrm{J}^{\mathrm{arr}}$ is integrated along the line to tune the filter. \textbf{b}, Voltage magnitude profile $|V|$ of the modes of the transmission line along the filter as a function of the position $x$ for several values of the flux-tunable $L_\mathrm{J}^{\mathrm{arr}}$. For a given $L_\mathrm{J}^{\mathrm{arr}}$, the filter decouples the qubit from the $50~\Omega$ line at the frequency that yields zero voltage at the qubit; here this occurs for a qubit designed near $4.5$~GHz. The SQUID placement and the number of elements are chosen to provide a large tunability while operating in the linear regime where the ac drive current is well below the critical current of the SQUIDs. \textbf{c}, Transmission magnitude $|S_{21}|$ from the source port P1 to the qubit port P2 as a function of drive frequency for different $L_\mathrm{J}^{\mathrm{arr}}$. Biasing the SQUID loops changes $L_\mathrm{J}^{\mathrm{arr}}$ and shifts the location of the minimum in the transmission. \textbf{d}, False-colored optical micrograph of the fabricated device with four double-pad transmon qubits (yellow). Qubit Q1 is flux-tunable, whereas Q2--Q4 are fixed-frequency. Each qubit couples to its own readout resonator (olive), which in turn couples to a common readout Purcell filter (blue). Q1 connects to a fixed-frequency drive filter (cyan), Q2 to a standard weakly-coupled drive line (magenta), and Q3--Q4 to tunable-frequency drive filters (purple). Each tunable-frequency drive line is inductively coupled to a fast-flux line (red) adjacent to the dc SQUID array.}
\label{fig:device_concept}
\end{figure*}

\subsection*{Concept and device}

The general concept of a quantum dial is presented in Fig.~\ref{fig:quantum_dial}, where a qubit is only coupled to the dial that mediates the coupling for the desired function one at a time. We implement a single-line version of the quantum dial for a transmon qubit by a tunable-frequency drive-line filter as shown in Fig.~\ref{fig:device_concept}. This filter generalizes the fixed-frequency quarter-wavelength filter of Ref.~\cite{Sah2024Decay-protectedFilters} by adding an in-line, flux-tunable inductive section. The key idea is to control the external admittance from the qubit to the line on demand. The filter is operated such that the qubit is effectively decoupled from the line during idle, and the filter is detuned from this idle configuration only during specific operations. This tunability addresses the fabrication spread without moving the qubit away from its flux-insensitive point where coherence properties are ideal and enables us to dial the qubit for (i) idle, where we have decay protection despite large coupling capacitor, (ii) resonant control, where we have fast low-power gates owing to strong coupling, and (iii) unconditional fast reset, where we have enhanced radiative decay and thus shortened $T_1$. In addition, we sequentially switch between configurations (iii) and (i) to implement fast and accurate qubit thermometry for probing the local microwave bath.

Figure~\ref{fig:device_concept}a schematically illustrates our experimental device consisting of a qubit coupled to drive line with a tunable-frequency filter. A $50~\Omega$ drive line (port P1) is capacitively coupled to the qubit (port P2) at $x=0$. Here, the line continues as a co-planar waveguide (CPW) of electrical length $L_\mathrm{f}$ and is weakly terminated to ground by a small capacitance, forming a quarter-wave filter element. A series array of dc SQUIDs is inserted along the CPW section with an effective inductance $L_\mathrm{J}^\mathrm{arr}$ that can be tuned by either a local fast-flux line or a global coil. Tuning $L_\mathrm{J}^\mathrm{arr}$ shifts the filter frequency and therefore the real part of the input admittance $Y_\mathrm{q}(\omega)$ presented to the qubit. If the quarter-wave frequency of the filter, \emph{the filter frequency}, is parked at the qubit frequency $f_\mathrm{q}$, the voltage of the drive line modes at the qubit location and frequency almost vanishes and hence the radiative coupling of the qubit to the $50~\Omega$ line is extinguished. On the other hand, if the filter is detuned away from this point, the coupling of the drive line to the qubit can be continuously controlled between essentially zero and a maximum value determined by the coupling capacitance of the qubit to the line. The number and placement of SQUID cells are chosen to provide a large tuning range while operating in the linear-inductance regime, with the peak alternating current (ac) per junction well below the critical current of the SQUIDs, see~\ref{supp_note:device_design}.

Figure~\ref{fig:device_concept}b illustrates the standing-wave voltage magnitude along the filter $|V(x)|$ as a function of position $x$ at a fixed probe frequency of $4.5~\mathrm{GHz}$ for several values of the array inductance $L_\mathrm{J}^\mathrm{arr}$. At the idle-configuration flux bias, $L_\mathrm{J}^\mathrm{arr}$ is vanishingly small and the qubit sits at a voltage node, i.e., $|V(0)|$ effectively vanishes. Thus, the qubit is effectively decoupled from the drive line. Increasing $L_\mathrm{J}^\mathrm{arr}$, increasing the voltage drop across the SQUID array and produces a finite voltage at the qubit, admitting strong coupling. Figure~\ref{fig:device_concept}c presents the corresponding transmission amplitude $|S_{21}|$ from port P1 to port P2 as a function of the probe frequency for fixed other parameters, highlighting the tunable frequency of the filter introduced by the dc SQUID array.

A false-color micrograph of the realized four-qubit device is shown in Fig.~\ref{fig:device_concept}d. For detailed fabrication steps, see~\ref{supp_note:device_fab}. Qubit~1 (Q1) is flux-tunable and couples to a fixed-frequency drive line filter. Qubit~2 (Q2) has a standard weakly-coupled drive line. Qubit~3 (Q3) and qubit~4 (Q4) are coupled to tunable-frequency drive line filters. Each qubit couples to an individual readout resonator that is connected to a common Purcell filter. Fast-flux lines are inductively coupled to the filter along the dc SQUID array to enable fast tunability. The experimental setup is described in~\ref{supp_note:meas_setup}, and the cryogenic wiring diagram is presented in Supplementary Fig.~\ref{fig:exp_setup}. The measured device parameters are summarized in Supplementary Table~\ref{tab:device_char}.

\subsection*{Tunable-frequency drive line filter}

\begin{figure*}[ht]
\includegraphics[width=\linewidth]{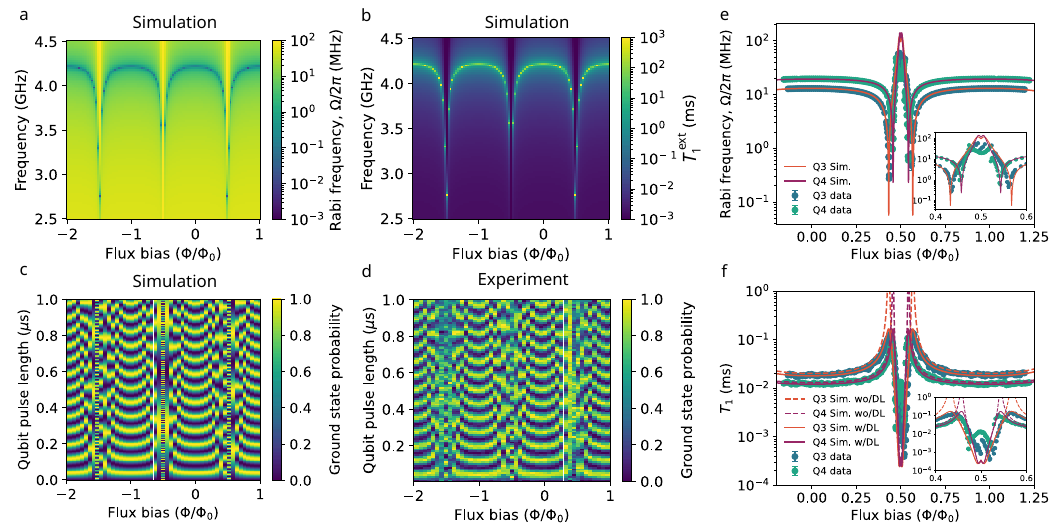}
\caption{\textbf{Simulation and experimental characterization of tunable-frequency drive line filters.} \textbf{a}, Simulated Rabi frequency $\Omega/2\pi$ as a function of the flux bias $\Phi$ and qubit frequency using a model (\ref{supp_note:device_design}), in which the dc SQUID array is represented by a tunable series inductance. The flux quantum is denoted by $\Phi_0$. The filter produces strong suppression of the Rabi frequency where the qubit and filter frequencies are equal. \textbf{b}, As \textbf{a}, but for the simulated external relaxation time $T_1^{\mathrm{ext}}$, showing an inverse trend to that of \textbf{a}, with large enhancements at the filter frequency. \textbf{c}, Simulated ground-state probability of the qubit as a function of qubit pulse length and flux bias, showing clear Rabi oscillations. \textbf{d}, As \textbf{c}, but for measurement data acquired from Q4, showing good agreement with simulated results in \textbf{c}. \textbf{e}, Measured Rabi frequency as a function of the flux bias for the fixed-frequency qubits Q3 and Q4. The markers denote experimental data obtained from a fit to Rabi oscillations similar to those in \textbf{d}, solid lines are simulations evaluated with the designed coupling and measured qubit frequency. The inset shows the behavior around the half flux quantum. \textbf{f}, Measured relaxation time $T_1$ as a function of flux bias for Q3 and Q4. The markers denote experimental data and the solid and dashed curves correspond to simulations with and without dielectric loss, respectively.  The inset shows the behavior around the half flux quantum. From the comparison we estimate the zero-bias filter frequency lies near $4.24$~GHz, corresponding to offsets of about $300$~MHz for Q3 and about $360$~MHz for Q4 from the design target. We observe almost three-orders-of-magnitude suppression of the Rabi frequency accompanied by a comparable enhancement of the relaxation times when the filter is decoupled from the qubit. The error bars represent $1\sigma$ fitting uncertainty.} 
\label{fig:tunable_drive_filter}
\end{figure*}

We model the filter behaviour with hybrid finite-element electromagnetic simulations and circuit-level modeling. In the simulations, we insert an ideal series inductor, the inductance of which is set to match the tunable Josephson inductance of the dc SQUID array, determined from the target single-junction critical current and the number of SQUIDs. From the resulting two-port transmission, we compute the real part of the admittance in series with the qubit port, $\mathrm{Re}\{Y_\text{q}(\omega,\Phi)\}$, and use it to estimate the Rabi angular frequency and the external relaxation time,
$\Omega \propto \mathrm{Re}\{Y_\text{q}\}^{1/2}$ and $T_1^\text{ext} \propto \mathrm{Re}\{Y_\text{q}\}^{-1}$, respectively. Figures~\ref{fig:tunable_drive_filter}a,b show these quantities as functions of the flux bias $\Phi$ and qubit frequency. Both quantities exhibit the expected periodic modulation with flux and orders-of-magnitude variation. At the point where the filter frequency matches the qubit frequency, $\Omega$ is strongly suppressed and $T_1^\text{ext}$ is enhanced. We also observe that the zero-bias filter frequency lies below our $4.5~\text{GHz}$ design target due to the finite array inductance at $\Phi=0$ which can be compensated by reducing the physical length of the filter, $L_\text{f}$.

Figures~\ref{fig:tunable_drive_filter}c,d show a comparison of the simulated and measured Rabi oscillations as functions of the flux bias and the length of the qubit drive pulse. The experimental data is obtained using a global coil mounted above the device package. The chevron patterns and their flux dependence agree well with the simulation.

Figures~\ref{fig:tunable_drive_filter}e,f present measurements of the Rabi frequency and relaxation time for Q3 and Q4 over a flux period, together with curves from the model, evaluated with the designed coupling, filter length, and the measured qubit frequency. We observe almost three orders-of-magnitude suppression of the Rabi frequency and a comparable enhancement of $T_1$ when the drive line is effectively decoupled from the qubit at the qubit frequency. Owing to the very good agreement with the experimental data, the simulation results yield zero-bias filter frequencies of approximately $4.24~\text{GHz}$, corresponding to offsets of about $300~\text{MHz}$ for Q3 and $360~\text{MHz}$ for Q4 from the designed value. These offsets are attributed to the geometric and zero-bias Josephson inductance of the dc SQUID array. The simulated trends are consistent with the closed-form expressions derived in~\ref{supp_note:device_design}.

We provide a comprehensive characterization of the qubits for each drive-line configuration in~\ref{supp_note:filter_char}. Across the fixed-frequency devices we observe a similar cap in the measured $T_1$, indicating that the tunable-frequency drive filter does not degrade relaxation in comparison to the standard drive line. As the filter is tuned from full coupling to somewhat decouple the qubit from the drive line, both the Rabi coherence time, $T_2^\text{R}$, and the echo coherence time, $T_2^\text{E}$, increase together with $T_1$, consistent with the radiative channel being dominant in this regime. The global coil and the on-chip flux biasing yield comparable Rabi frequencies and coherence times, see Supplementary Fig.~\ref{fig:filter_char_supp}, demonstrating that the local bias provides the expected performance without added dissipation or noticeable crosstalk. On the other hand, the flux-tunable qubit Q1 showed relatively lower $T_1$ compared to those of the fixed-frequency qubits.

We characterize the Lamb shift of the qubit frequency induced by its coupling to the tunable environment, with the flux bias dependence shown in Supplementary Fig.~\ref{fig:t1_ramsey_freq_supp}. The Lamb shift is $\lesssim 0.3~\mathrm{MHz}$ and is readily compensated with virtual-$Z$ and derivative removal by adiabatic gate (DRAG) phase corrections.

We also monitor qubit coherence over multiple days and cooldowns, see Supplementary Fig.~\ref{fig:long_coherence_supp}. The $T_1$ distributions for all qubits remain centered around stable mean values, pointing to reproducible biasing and small drift in the dc SQUID flux. In contrast, $T_2^\text{R}$ and $T_2^\text{E}$ for Q2 and Q3 are significantly shorter than $2T_1$, while Q4 performs better yet still below the $2T_1$ limit. To probe the dephasing spectrum we carry out dynamical-decoupling sequences: Carr-Purcell (CP), Carr-Purcell-Meiboom-Gill (CPMG), and Uhrig (UDD) and extract $T_2^\text{DD}$ as a function of pulse number as shown in Supplementary Fig.~\ref{fig:dynamical_decoup_supp}. All fixed-frequency qubits Q2--Q4 show an initial increase of $T_2^\text{DD}$ that saturates near $0.1~\text{ms}$, indicating effective cancellation of slow noise ($1/f$-like) with a residual, fast component that the sequences cannot refocus. Such a high-frequency floor is consistent with a combination of dielectric noise and residual photon noise in the measurement band.

\subsection*{Qubit control}

To implement fast and precise quantum gates we need to fulfill two competing requirements, the coupling needs to be strong enough for fast control yet sufficiently weak to protect coherence of the qubit. One way to resolve this dilemma is to use multi-photon (subharmonic) transitions, where the control pulses have a carrier frequency $f_\text{q}/n$ and qubit remains protected at $f_\text{q}$. This approach has been demonstrated for $n=3$ in Refs.~\cite{Xia2023FastDrives,Sah2024Decay-protectedFilters,Schirk2025SubharmonicLine}. In practice, however, the use of such subharmonic drive requires sizable calibration overhead because the ac-Stark shift can be comparable to the Rabi frequency, and hence the phase off-set of the gate accumulates strongly with amplitude and detuning. This requires frequent recalibration and careful cancellation of spurious transitions.

\begin{figure*}[ht]
\includegraphics[width=\linewidth]{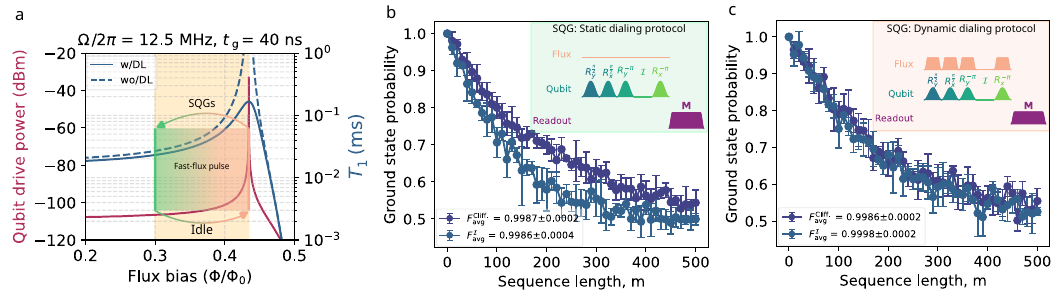}
\caption{\textbf{Resonant control and idle fidelity of Q4.} \textbf{a}, Required qubit-drive power at the sample (red line, left axis) to realize a target Rabi frequency $\Omega/2\pi=12.5$~MHz implementing gates of length $t_\text{g}=40$~ns as a function of the flux bias $\Phi$ of the tunable drive line filter. The qubit relaxation time $T_1$ (right axis) is also shown with (solid line) and without (dashed line) dielectric loss. The edges of the shaded regions indicate the positions of the dial: the idle configuration with the filter frequency parked at $f_\text{q}$ and the control configuration where the filter is detuned to provide a moderate coupling. The orange band denotes the fast-flux sweep used for dialing. \textbf{b}, Ground-state probability of the qubit as a function of the sequence length in randomized benchmarking for static dialing. Extracted average Clifford fidelity and interleaved idle fidelity are $\mathit{F}_\mathrm{avg}^\mathrm{Cliff.}=0.9987\pm0.0002$ and $\mathit{F}_\mathrm{avg}^\mathrm{I}=0.9986\pm0.0004$, respectively. The inset shows the pulse schedule (fixed filter bias, resonant single-qubit gates, and readout). \textbf{c}, As \textbf{b}, but for dynamic dialing, where a short flux pulse detunes the filter away from the idle dial configuration only during the gate pulse. The fits yield $\mathit{F}_\mathrm{avg}^\mathrm{Cliff.}=0.9986\pm0.0002$ and $\mathit{F}_\mathrm{avg}^\mathrm{I}=0.9998\pm0.0002$, reflecting the strongly suppressed idle error when the qubit remains at the idle position of the quantum dial. The insets depict the corresponding flux and single-qubit gate sequences together with the readout pulse. The error bars on the data represent $1\sigma$ standard deviation over a multiple repetitions.}
\label{fig:qubit_control}
\end{figure*}

In contrast, we drive resonant control enabled by the quantum dial of qubit Q4. Dialing the qubit to the idle configuration ($\Phi/\Phi_0 \approx 0.44$), the qubit is shunted by an extremely small drive admittance. Implementing $40$~ns gates in this configuration would require $60$~dBm more power at the sample compared to that in the control configuration ($\Phi/\Phi_0 \approx 0.3$) as shown in Fig.~\ref{fig:qubit_control}a. Such high power is impractical and thermally costly in a dilution refrigerator. The quantum dial in control configuration offers two practical protocols:

\textbf{Static dialing.} We fix the filter frequency away from the qubit frequency, to a flux bias that balances gate speed, thermal load, and qubit relaxation. With a target coupling that yields roughly $-110$~dBm at the sample for $40$~ns single-qubit gates, we measure $T_1=14~\mu\text{s}$. Standard randomized benchmarking (RB) yields an average fidelity per Clifford operation of $\mathit{F}_\mathrm{avg}^\mathrm{Cliff.}=0.9987\pm0.0002$ and interleaved idle fidelity of $\mathit{F}_\text{avg}^\text{I}=0.9986\pm0.0004$ as depicted in Fig.~\ref{fig:qubit_control}b. The pulse schedule for qubit drive and readout is shown alongside the data.

\textbf{Dynamic dialing.} We keep the qubit at the idle position except for the duration of the control pulses, for which we dial the qubit to the control position. Namely, a $40~\text{ns}$ flux pulse on the fast-flux line brings the filter into the control configuration only during the gate pulse. At the idle configuration set by the global coil, we measure $T_1=0.16$~ms. The control configuration is chosen to match the static-dialing gate speed to enable a direct comparison at the same power level of $-110$~dBm at the sample. The RB results in Fig.~\ref{fig:qubit_control}c show average per Clifford and interleaved idle fidelities of $\mathit{F}_\text{avg}^\text{Cliff.}=0.9986\pm0.0002$ and $\mathit{F}_\text{avg}^\text{I}=0.9998\pm0.0002$, respectively, about a $10\times$ improvement in the interleaved idle compared to the constant-bias case. The enhancement is consistent with strongly suppressed radiative decay during the many idle periods between the Clifford pulses. 

\subsection*{Qubit reset and single-shot readout}

\begin{figure*}[ht!]
\includegraphics[width=\linewidth]{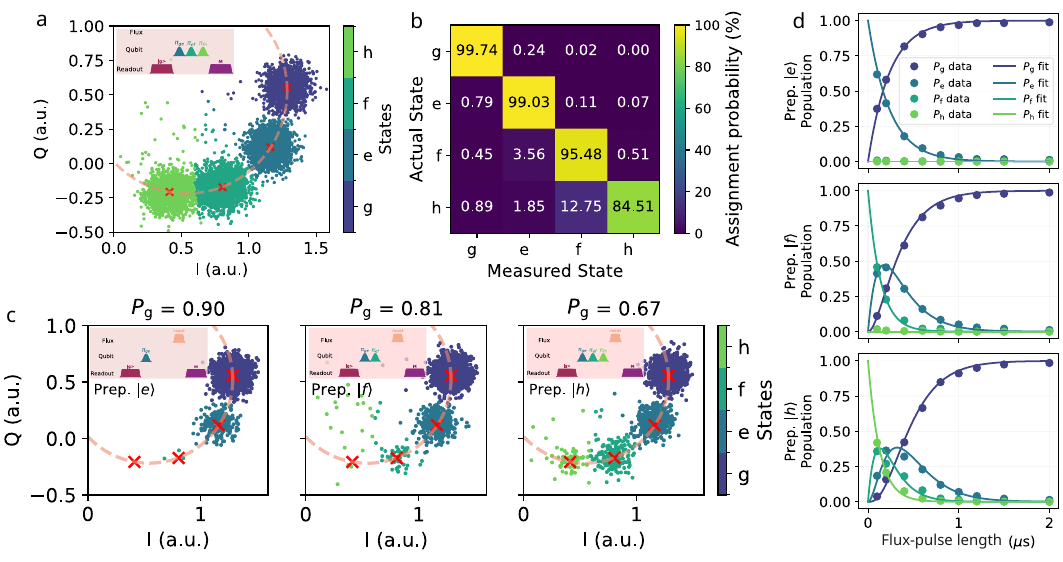}
\caption{\textbf{Single-shot measurements and reset of Q4}. \textbf{a}, Measured reference single-shot data in the in-phase--quadrature-phase (IQ) plane for the four lowest-energy states $\ket{\text{g}},\ket{\text{e}},\ket{\text{f}}$, and $\ket{\text{h}}$. This data is used to fit a four-component Gaussian mixture model (GMM). The dashed line is a circle fit to the means (red crosses) of each state returned by the GMM model fit. The inset shows the pulse sequence (weak reference readout, state preparation, final readout). \textbf{b}, Readout assignment matrix obtained by preparing each basis state and classifying the outcome. High fidelities are observed on the diagonal for $\ket{\text{g}}$ and $\ket{\text{e}}$, with slightly lower fidelities for $\ket{\text{f}}$ and $\ket{\text{h}}$ consistent with their shorter $T_1$ during readout. \textbf{c}, As \textbf{a}, but for preparations in $\ket{\text{e}}$, $\ket{\text{f}}$, and $\ket{\text{h}}$ after a fixed flux pulse of duration $t_\text{reset}=0.8~\mu\text{s}$; inferred ground-state populations $P_\mathrm{g}$ are indicated. \textbf{d}, Measured (markers) residual populations as functions of the flux-pulse length for the three state preparations in \textbf{c}. The solid curves represent a fit of a four-state Pauli master equation with only downward transitions allowed to the whole data set. The fit yields the following decay times $T_1^{mk}$ from the state $\ket{k}$ to the state $\ket{m}$: $T_1^\text{ge}=238.2\pm 3.3~\text{ns}$, $T_1^\text{ef}=136.8\pm 2.8~\text{ns}$, and $T_1^\text{fh}=128.8\pm 3.4~\text{ns}$. At $t_\text{reset}\approx 1.2~\mu\text{s}$ we obtain $P_\mathrm{g}\gtrsim 97\%$ from $\ket{\text{e}}$, $95\%$ from $\ket{\text{f}}$, and $91\%$ from $\ket{\text{h}}$, with saturation near $98.5\%$ set by a thermal floor corresponding to $T\approx 45~\text{mK}$.}
\label{fig:qubit_reset}
\end{figure*}

For the reset dial, the tunable-frequency drive-line filter can be used to realize a largely unconditional dissipative qubit reset without additional components. By moving the filter frequency with a brief flux pulse, the qubit--bath coupling is increased by orders-of-magnitude for the duration of the reset and is subsequently returned to the idle configuration. Because the reset couples the qubit directly to a broadband $50~\Omega$ environment through the same line used for control, it is essentially state-independent and adds no measurement overhead: the reset applies to $\ket{\text{f}}$, $\ket{\text{h}}$, etc., by way of sequential decay. For a qubit with energy relaxation time $T_1^\mathrm{low}$ during reset to a low-temperature environment, choosing $t_\mathrm{reset}=5T_1^\mathrm{low}$ leaves a residual excited-state probability $P_\mathrm{e}(t_\mathrm{reset})\approx \mathrm{e}^{-t_\mathrm{reset}/T_1^\mathrm{low}}P_\mathrm{e}(0)=\mathrm{e}^{-5}P_\mathrm{e}(0)<0.7\%$. While $t_\mathrm{reset}=10T_1^\mathrm{low}$ can reduce $P_\mathrm{e}(t_\mathrm{reset})$ below $10^{-4}$, it places more stringent requirements on the temperature of electromagnetic modes in the drive line; below we therefore use $t_\mathrm{reset}\approx 5T_1^\mathrm{low}$ instead. 

We begin the reset protocol with the qubit idling, apply a fixed-length flux pulse that brings the filter to the configuration providing the lowest measured $T_1$, and then return to the idle configuration. On qubit Q4, we measure $T_1^\mathrm{high}>0.15~\text{ms}$ at the idle configuration and $T_1^\mathrm{low}\approx 0.24~\mu\text{s}$ at the reset configuration. The corresponding wait times $t_\mathrm{reset}\approx 5T_1$ are $0.75~\text{ms}$ at idle and $1.2~\mu\text{s}$ at reset, representing an orders-of-magnitude suppression in reset time when utilizing the reset configuration. Because the coupling is only increased for the duration of the reset, dephasing and leakage during idle are not impacted.

We quantify reset using single-shot readout with model-based state classification. Figure~\ref{fig:qubit_reset}a shows representative data point clouds in the in-phase--quadrature-phase (IQ) plane for the four prepared states $\{\ket{\text{g}},\ket{\text{e}},\ket{\text{f}},\ket{\text{h}}\}$, and Fig.~\ref{fig:qubit_reset}b shows the assignment matrix used to benchmark the readout. With the filter parked to decouple at $f_\mathrm{ge}$, higher transitions have shorter lifetimes, and relaxation during the finite readout time therefore yields slightly lower assignment fidelity for $\ket{\text{f}}$ and $\ket{\text{h}}$ than for $\ket{\text{e}}$. Using this calibrated readout reference, we apply a fixed-length flux pulse after state preparation to implement reset and classify the post-reset populations for preparations in $\ket{\text{e}}$, $\ket{\text{f}}$, and $\ket{\text{h}}$. 

Figure~\ref{fig:qubit_reset}c shows the post-reset populations for $t_\mathrm{reset}=0.8~\mu\text{s}$ together with the residual ground-state population after reset, with the flux-pulse amplitude chosen to maximize coupling of the qubit to the tunable-frequency drive filter. Sweeping $t_\mathrm{reset}$ yields the reset dynamics in Fig.~\ref{fig:qubit_reset}d. A global fit to a four-state Pauli master equation allowing only sequential downward transitions yields $T_1^\mathrm{ge}=238.22\pm 3.28~\text{ns}$, $T_1^\mathrm{ef}=136.80\pm 2.80~\text{ns}$, and $T_1^\mathrm{fh}=128.84\pm 3.38~\text{ns}$, where the $1\sigma$ uncertainties are obtained from the Jacobian at the optimum. The hierarchy $T_1^\mathrm{ge}>T_1^\mathrm{ef}>T_1^\mathrm{fh}$ is consistent with the expected increase of radiative matrix elements for higher qubit transitions and the larger environmental spectral density at the corresponding frequencies. 

With a single flux pulse we demonstrate a state-independent reset: after $1.2~\mu\text{s}\approx 5T_1^\mathrm{ge}$, we reach ground-state populations of $97\%$ from $\ket{\text{e}}$, $95\%$ from $\ket{\text{f}}$, and $91\%$ from $\ket{\text{h}}$. For long pulses, the ground-state probability saturates near $98.5\%$ independent of the initial state, set by a thermal floor corresponding to $T\approx 45~\text{mK}$ at $3.9~\text{GHz}$ under a Boltzmann distribution.

\subsection*{Qubit thermometry}

\begin{figure*}[ht!]
\includegraphics[width=\linewidth]{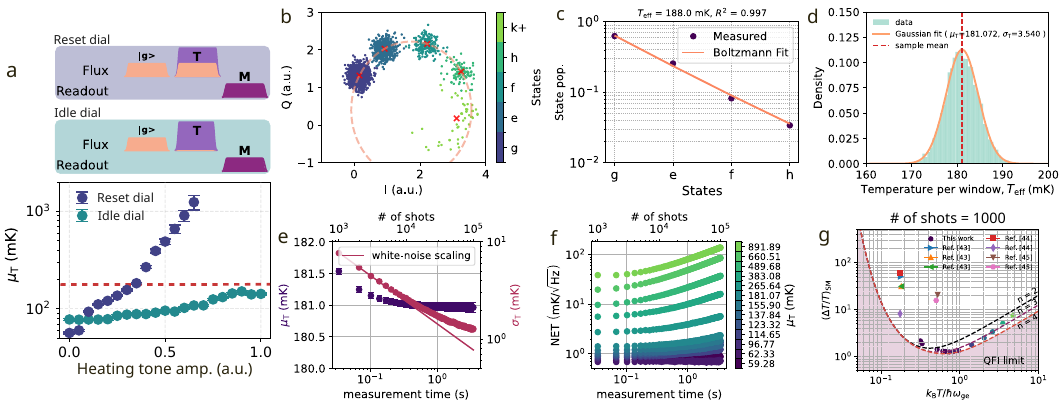}

\caption{\textbf{Near-optimal qubit thermometry utilizing the quantum dial for Q3.} \textbf{a}, Thermometry protocol and mean effective temperature $\mu_\mathrm{T}$ of the measured qubit environment as a function of the amplitude of a $5$~GHz heating tone applied on the fast-flux line, measured with the quantum dial in the reset configuration (strong coupling, $T_{1}\simeq200~\mathrm{ns}$) and in the idle configuration (weak coupling, $T_{1}\simeq200~\mu\mathrm{s}$). Each cycle begins with a $2~\mu\mathrm{s}$ reset pulse, followed by a heating pulse applied during the thermalization interval, and then a single-shot dispersive readout. Points show $\mu_\mathrm{T}$ and error bars denote the standard error of the mean $\sigma_\mu=\sigma_\mathrm{T}/\sqrt{N_\mathrm{win}}$, computed from $N_\mathrm{win}=5000$ non-overlapping windows of $5000$ shots. \textbf{b}, Example single-shot IQ distribution for one window ($5000$ shots) and Gaussian-mixture classification into $\ket{\text{g}},\ket{\text{e}},\ket{\text{f}},\ket{\text{h}}$, and an overflow cluster $\ket{\text{k}^+}$ aggregating higher levels. For temperature extraction, the $\ket{\text{g}}$--$\ket{\text{h}}$ populations are renormalized after excluding $\ket{\text{k}^+}$. \textbf{c}, Four-level Boltzmann fit for the window in \textbf{b} using the measured transition frequencies $f_\mathrm{ge},f_\mathrm{ef}$, and $f_\mathrm{fh}$; the fit returns $T_\mathrm{eff}$ and the fit-quality metric $R^2$. \textbf{d}, Probability density of the per-window effective temperature $T_\mathrm{eff}$ (teal) together with a Gaussian fit $\mathcal{N}(\mu_\mathrm{T},\sigma_\mathrm{T}^2)$ (orange), yielding $\mu_\mathrm{T}=181.072~\mathrm{mK}$ and $\sigma_\mathrm{T}=3.540~\mathrm{mK}$. The red dashed line marks the sample mean $\bar T$. \textbf{e}, Mean $\mu_\mathrm{T}$ (points with $\sigma_\mu$ bars) and standard deviation $\sigma_\mathrm{T}$ (from the Gaussian fit) as a function of measurement time per window $t_\mathrm{meas}=N_\mathrm{shot}t_\mathrm{shot}$ for window sizes $N_\mathrm{shot}=10^3$ to $10^5$. Here $t_\mathrm{shot}=34.2~\mu\mathrm{s}$ is the duration of one thermometry shot. \textbf{f}, Noise-equivalent temperature $\mathrm{NET}=\sigma_\mathrm{T}\sqrt{t_\mathrm{meas}}$ as a function of $t_\mathrm{meas}$ for several heating-tone amplitudes. \textbf{g}, Normalized single-measurement precision $(\Delta T/T)_\mathrm{SM}=(\sigma_\mathrm{T}/\mu_\mathrm{T})\sqrt{N_\mathrm{shot}}$ for windows of $N_\mathrm{shot}=1000$, shown for this work and for literature data as indicated in the legend (markers), compared with quantum Fisher information (QFI) bounds computed from $n=2,3,4$ qubit energy levels (solid curves).
}

\label{fig:qubit_thermometry}
\end{figure*}

The ability to tune the qubit-bath coupling on a timescale much shorter than $T_1$ renders our quantum dial a natural enabler of precise qubit thermometry. To this end, we operate the device as follows: the reset configuration to first let the qubit thermalize with the electromagnetic bath, the temperature of which we are measuring, followed by the idle configuration to allow accurate qubit readout. In the first step, we detune the filter frequency from $f_\text{ge}$ to open a radiative channel and hold it there for $10~\mu\text{s}$, long compared with $T_1\approx0.2~\mu\text{s}$. In this configuration, the populations approach their thermal steady state with finite occupation of qubit states. We then return the dial to the idle configuration, which suppresses further exchange with the bath and preserves the populations during readout. From single-shot IQ data we extract the state populations $\{P_\mathrm{g},P_\mathrm{e},P_\mathrm{f},P_\mathrm{h}\}$ and fit them to a Boltzmann distribution using the measured transition frequencies $\{f_\mathrm{ge},f_\mathrm{ef},f_\mathrm{fh}\}$ to obtain an effective qubit temperature $T_\mathrm{eff}$, see Methods. Repeating this protocol over many windows provides a well-defined statistic for the temperature readings along with its mean $\mu_\text{T}$ and variance $\sigma_\text{T}$, and other figures of merit such as the noise-equivalent temperature (NET). 

Figure~\ref{fig:qubit_thermometry}a shows the pulse sequence for qubit thermometry and reports the mean effective temperature $\mu_\mathrm{T}$ as a function of the amplitude of a heating pulse applied on the fast-flux line. To reduce initial thermal occupation of high-energy states, each thermometry cycle begins with a short pulse to the reset configuration. We then apply the heating pulse while operating either at reset-dial (strong coupling) or at idle-dial (weak coupling) during thermalization, and subsequently perform dispersive readout at the idle configuration. Under strong coupling at reset-dial we observe a substantially higher $\mu_\mathrm{T}$ than under weak coupling at idle-dial. In the latter case, the temperature saturates near $180~\mathrm{mK}$, indicating effective isolation from that environment.

Each point in Fig.~\ref{fig:qubit_thermometry}a is extracted by repeating the thermometry protocol over non-overlapping windows of $N_\mathrm{shot}$ single shots, yielding an array of effective temperatures $\{T_\mathrm{eff}^{(j)}\}$. Figures~\ref{fig:qubit_thermometry}b--d illustrate the analysis pipeline that converts single-shot populations to $T_\mathrm{eff}$ and then to the distribution of temperature outcomes. Figure~\ref{fig:qubit_thermometry}e shows that $\sigma_\mathrm{T}$ decreases with measurement time $t_\mathrm{meas}$ following the expected white-noise scaling $\sigma_\mathrm{T}\propto t_\mathrm{meas}^{-1/2}$ over more than a decade before approaching a floor set by slow drifts. 

Figure~\ref{fig:qubit_thermometry}f reports $\mathrm{NET}=\sigma_\mathrm{T}\sqrt{t_\mathrm{meas}}$, which is constant in the white-noise regime and reaches about $0.6~\text{mK}/\sqrt{\text{Hz}}$. Figure~\ref{fig:qubit_thermometry}g compares the normalized single-measurement precision $(\Delta T/T)_\mathrm{SM}$ to the quantum Fisher information bound computed from the measured energy levels, and benchmarks our performance against prior superconducting-qubit thermometry and effective-temperature measurements reported in Refs.~\cite{Lvov2025ThermometryQubit,Kulikov2020MeasuringCorrelations,Sultanov2021ProtocolCircuit}, shown as markers. The data closely follows the $n=4$ bound up to $400~\mathrm{mK}$ and remains within a small factor beyond, indicating that our readout protocol is near-optimal for temperature estimation.

The quantum dial, alternating between reset and idle configurations, provides fast thermalization together with high-fidelity readout, enabling fast and precise qubit thermometry.

\section*{Discussion}\label{Discussion}
Scaling superconducting quantum processors has largely followed a modular approach in which each new function, such as control, readout, reset, or filtering, is implemented by introducing an additional element with a fixed coupling to the qubit. While effective at small scale, every fixed coupling also opens an unwanted channel to the electromagnetic environment. This architecturally fixed coupling budget forces a static compromise between fast operation and protected idling. The quantum dial addresses this limitation by making the qubit–environment coupling programmable in time, so that the same physical interface can lead to weak coupling when coherence matters and strong coupling only when an operation demands it.

In this work we introduce and initially demonstrate the quantum dial as a system-level solution to the isolation-and-control challenge. By dynamically tuning the electromagnetic environment of the qubit rather than the qubit frequency itself, we dial a fixed-frequency transmon between three operational regimes using a single on-chip element. The idle configuration suppresses radiative decay while preserving operation at the flux-insensitive sweet spot. The control configuration enables resonant low-power gates using the existing drive line without requiring flux-tunable qubits away from their sweet spot. The reset configuration enables unconditional dissipative initialization by briefly opening a radiative channel to a $50~\Omega$ environment. Together, these regimes implement the core idea of the dial: temporally separating functions that otherwise compete when implemented with static coupling.

Interestingly, we observe that resonant single-qubit control achieved with the quantum dial operates near the fundamental energy--error limit set by the finite pulse energy. Using the average per Clifford fidelities
$F_\mathrm{avg}^\mathrm{Cliff}=0.9987$ for static dialing and $F_\mathrm{avg}^\mathrm{Cliff}=0.9986$ for dynamic dialing, together with a delivered control power of $-110~\mathrm{dBm}$ at the sample, the corresponding pulse energy implies a mean photon number of $n\approx155$. For coherent microwave pulses, this sets a maximum achievable fidelity per $\pi/2$-rotation of $F_\mathrm{max}^\mathrm{coh} = 0.9988$, estimated using disposable-pulse bound derived in Ref.~\cite{Ikonen2017Energy-efficientComputing}. Our results therefore indicate that single-qubit control is already operating near the quantum limit imposed by finite pulse energy, rather than being limited by classical control imperfections. Achieving higher fidelities would thus require either increased drive energy or non-disposable control strategies. In this context, the quantum-dial architecture enables high-fidelity operation at low drive power, mitigating ac Stark shifts appearing in strong and subharmonic driving schemes, while avoiding the trade-offs inherent to static coupling between control efficiency and idle-state coherence.

Beyond coherent control, the reset configuration shows that fast, state-independent initialization can be realized without measurement or feedback. By briefly dialing the qubit into a strongly coupled regime, we reset to ground-state populations above $97\%$ within microseconds, including from higher excited states via sequential decay. Unlike active reset protocols, this approach introduces no classical latency and requires no additional dissipative elements, highlighting the leverage provided by environmental engineering. The reset time and fidelity may be improved in the future by increasing the coupling capacitance and decreasing the temperature of the electromagnetic environment.

The quantum dial also enables precise qubit thermometry. By alternating between strong coupling, where the qubit rapidly thermalizes, and weak coupling, where populations are preserved during readout, the quantum dial implements a fast thermometer for the microwave environment. The temperature uncertainty exhibits the expected white-noise scaling over a wide range of integration times, and the single-measurement precision closely approaches the quantum Cramér--Rao bound for a many-level system. This establishes that many-level dispersive readout can be near-optimal for thermodynamic parameter estimation and positions the dial as a tool for continuous, in situ monitoring of cryogenic microwave environments.

Taken together, these demonstrations show that control, reset, and thermometry, functions often treated as separate subsystems, can be unified through dynamic control of the qubit--bath coupling using the existing drive line. The remaining functionalities suggested by this architecture follow naturally from the same principle: dialing the qubit--qubit interaction on demand for entangling gates, and dialing the measurement environment to achieve high-fidelity readout without compromising idle coherence. More broadly, the quantum dial offers a compact route to lowering idle error, accelerating error-correction cycles, and continuously benchmarking the electromagnetic environment in superconducting quantum processors.

\section*{Methods}\label{Methods}

\subsection*{Randomized benchmarking metrics and calibration procedures}
In Fig.~4 we report two RB-derived metrics. The average fidelity per Clifford operation $F_\mathrm{avg}^{\text{Cliff.}}$ is obtained from the reference RB decay parameter $p_\mathrm{ref}$ using
$F_\mathrm{avg}^{\text{Cliff.}}=1-(1-p_\mathrm{ref})/(2k)$ with $k=45/24$ the average number of primitive pulses per Clifford in our compilation~\cite{Barends2014SuperconductingTolerance}. The interleaved idle fidelity $F_\mathrm{avg}^{\mathrm{I}}$ is extracted from interleaved RB using the corresponding reference and interleaved decay parameters under the standard gate-independent-noise assumption~\cite{Magesan2011ScalableProcesses}. Occasional fidelity values slightly above unity are attributed to statistical fluctuations and slow drifts (e.g., $T_1$) combined with differing idle counts in the interleaved and reference sequences; see Supplementary Note~S5.

Since the qubit frequencies are $300$--$360~\text{MHz}$ away from the filter frequency at $\Phi=0$, we optimize both the global and local bias for maximal linear response of the dc SQUID array, avoiding excess ac current near $\Phi=\Phi_0/2$ where the SQUIDs are most sensitive. Before RB we follow standard calibrations: amplitude tuning of $\pi$- and $\pi/2$-pulses for a fixed gate length of $40~\text{ns}$, DRAG optimization to suppress leakage, and fine detuning to minimize residual phase errors; see \ref{supp_note:qubit_control}. The bias optimization and calibration data are summarized in Supplementary Fig.~\ref{fig:RB_Rabi_supp} and Supplementary Fig.~\ref{fig:RB_calib_supp}. We repeat RB overnight for both protocols to build statistics; the results are shown in Supplementary Fig.~\ref{fig:RB_results_supp}. As shown in Supplementary Fig.~\ref{fig:Q4_subharmonic_supp}, we also demonstrate multi-photon control with the tunable-frequency drive-line filter, including three-photon subharmonic driving.

\subsection*{Qubit reset: implementation, calibration, and analysis}
Reset is implemented by a fixed-length flux pulse that tunes the drive-line filter to the configuration that minimizes the measured qubit $T_1$, followed by a return to the idle point. In practice, we shape the flux-pulse edges to avoid non-adiabatic excitations and verify that repeated resets do not measurably heat the chip or the drive line. The reset dial allows trading reset time against residual excitation by selecting $T_1^\mathrm{low}$ through the filter bias.

Accurate preparation and classification of $\{\ket{\text{g}},\ket{\text{e}},\ket{\text{f}},\ket{\text{h}}\}$ require many-state calibration of transition frequencies, Rabi, Ramsey, and dispersive shifts, given in~\ref{supp_note:multilevel_char} and Supplementary Fig.~\ref{fig:Q4_multilevel_char_supp}. A summary of the extracted many-state qubit parameters for Q2--Q4 is presented in Supplementary Table~\ref{tab:multilevel_device_char}. The readout calibration pulse sequence consists of a weak readout tone, $1~\mu\text{s}$ resonator ring-down, state-preparation pulses, and a final readout. A four-component Gaussian-mixture model (GMM) is fitted to the calibration IQ distributions and then used to infer per-shot state probabilities for the reset data; averaging over shots yields post-reset populations.

We optimize the readout tone frequency for four-state high-fidelity single-shot readout in Supplementary Fig.~\ref{fig:Q4_singleshot_ro_freq_opt_supp}. To mitigate thermal excitations, we implement heralding and a ground-state preparation sequence via dissipative initialization using the quantum dial in the reset configuration, shown in Supplementary Fig.~\ref{fig:Q4_singleshot_supp}. The resulting multi-level classification performance is shown in Supplementary Fig.~\ref{fig:Q4_singleshot_classification_acc_supp}. An overall discussion of single-shot calibration and reset implementation is provided in~\ref{supp_note:qubit_reset}.

Supplementary Fig.~\ref{fig:Q4_t1_fpl_supp} shows the measured $T_1$ as a function of flux-pulse length for different pulse amplitudes, and extended data on post-reset populations across pulse lengths are given in Supplementary Fig.~\ref{fig:Q4_reset_IQ_fpl_supp}. Reset dynamics are modeled with a four-state Pauli master equation allowing only sequential downward transitions. Closed-form solutions for $\{P_\mathrm{g}(t),P_\mathrm{e}(t),P_\mathrm{f}(t),P_\mathrm{h}(t)\}$ for each initial state are used in a nonlinear least-squares fit that includes all populations and their covariances across the three initial states, and $1\sigma$ uncertainties are obtained from the Jacobian at the optimum. To remove heralding overheads, we repeat the reset experiment with purely dissipative initialization using the quantum dial in the reset configuration; the results are comparable, shown in Supplementary Fig.~\ref{fig:Q4_reset_decayfit_supp} and Supplementary Table~\ref{tab:qubit_reset_ps_fit}.

\subsection*{Qubit thermometry: calibration, protocol, and optimality benchmarks}
Single-shot thermometry uses the same model-based classification approach as in the reset experiment. Single-shot calibration data and representative IQ clouds for thermometry are shown in Supplementary Fig.~\ref{fig:Q3_thermometry_singleshot_supp}. From classified shots we extract $\{P_\mathrm{g},P_\mathrm{e},P_\mathrm{f},P_\mathrm{h}\}$ and fit to a Boltzmann distribution using the measured transition frequencies $\{f_\mathrm{ge},f_\mathrm{ef},f_\mathrm{fh}\}$ to obtain $T_\mathrm{eff}$.

During cooldown, we track the extracted temperature of the environment in both dial configurations together with summary statistics of the temperature estimates, shown in Supplementary Fig.~\ref{fig:Q3_temp_w_wo_therm}. We report the mean $\mu_\mathrm{T}$ and standard deviation $\sigma_\mathrm{T}$ over repeated, non-overlapping windows, and use these quantities to define figures of merit such as $\mathrm{NET}=\sigma_\mathrm{T}\sqrt{t_\mathrm{meas}}$.

Figure~\ref{fig:qubit_thermometry}a alternates a short initialization pulse at reset-dial, a thermalization interval at either reset configuration or an idle configuration, and dispersive readout at the idle configuration. Additional experimental details of the thermometry setup and the heating protocol are given in Supplementary Fig.~\ref{fig:Q3_thermometry_setup_supp}.

For each non-overlapping window of $N_\mathrm{shot}$ shots, we cluster IQ points with a Gaussian-mixture model into $\ket{\text{g}},\ket{\text{e}},\ket{\text{f}},\ket{\text{h}}$, and an overflow component $\ket{\text{k}^+}$ that aggregates higher levels. We exclude $\ket{\text{k}^+}$ and renormalize the $\ket{\text{g}}$--$\ket{\text{h}}$ populations prior to the Boltzmann fit, and we use the fit quality metric $R^2$ as reported in Fig.~\ref{fig:qubit_thermometry}c. We repeat the analysis over many windows to form $\{T_\mathrm{eff}^{(j)}\}$ and fit the resulting histogram to a Normal distribution $\mathcal{N}(\mu_\mathrm{T},\sigma_\mathrm{T})$ as in Fig.~\ref{fig:qubit_thermometry}d. Extended data for the full thermometry data set and additional scaling analysis are shown in Supplementary Fig.~\ref{fig:Q3_qt_full_supp}. The measurement time is $t_\mathrm{meas}=N_\mathrm{shot}t_\mathrm{shot}$, where $t_\mathrm{shot}$ includes state preparation, thermalization, readout, and inter-shot waiting.

Operating exclusively in the reset configuration degrades single-shot readout fidelity, shown in Supplementary Fig.~\ref{fig:Q3_singleshot_lowT1_supp}. For the comparison to the quantum Fisher information benchmark in Fig.~\ref{fig:qubit_thermometry}g, we use the normalized single-measurement precision $(\Delta T/T)_\mathrm{SM}=(\sigma_\mathrm{T}/\mu_\mathrm{T})\sqrt{N_\mathrm{shot}}$ and compute the bound $F_\mathrm{Q}(T)$ from the measured energy levels. Additional discussion of qubit thermometry and the alternating reset/idle protocol is provided in~\ref{supp_note:qubit_thermometry}.

\subsection*{Thermometry analysis: single-shot separation, Boltzmann fits, and QCRB benchmarks}
\subsubsection*{Single-shot IQ separation and required SNR.}

We use a Gaussian-mixture model for the classification in the IQ plane. Let $\boldsymbol{\mu}_i$ and $\Sigma_i$ be the mean and covariance of component $i\in\{\mathrm{g,e,f,h}\}$. For any pair $(i,j)$ we quantify the separation by the symmetric Mahalanobis distance
\begin{equation}
\delta_{ij}^{2}
= (\boldsymbol{\mu}_i-\boldsymbol{\mu}_j)^{\top}
\left( \frac{\Sigma_i+\Sigma_j}{2} \right)^{-1}
(\boldsymbol{\mu}_i-\boldsymbol{\mu}_j).
\label{eq:maha}
\end{equation}

For two Gaussians with equal priors and equal covariance, the Bayes-optimal misclassification probability is
\begin{equation}
\varepsilon_{ij}\;=\;\Phi\left(-\frac{\delta_{ij}}{2}\right) =\tfrac{1}{2}\,\mathrm{erfc}\left(\frac{\delta_{ij}}{2\sqrt{2}}\right),
\label{eq:bayes}
\end{equation}
where $\Phi$ is the standard normal CDF. In the multi-class case we bound the leakage out of class $i$ by $ \varepsilon_i \lesssim \sum_{j\neq i}\Phi(-\delta_{ij}/2)$; in practice the nearest neighbour dominates.

To ensure that assignment bias is negligible compared with the sampling noise in a window of $N_{\mathrm{shot}}$ shots, we require $\varepsilon_i \ll \sqrt{P_i(1-P_i)/N_{\mathrm{shot}}}$. For the typical range $P_i\sim 0.1\text{-}0.4$ and $N_{\mathrm{shot}}=5000$, the binomial standard deviations are $0.4\text{-}0.7\%$. Thus we target per-boundary misclassification $\varepsilon_{ij}\lesssim10^{-3}$, which corresponds (via Eq.~\eqref{eq:bayes}) to

\begin{equation}
\delta_{ij}\ \gtrsim\ 6.0
\quad\Longrightarrow\quad
\varepsilon_{ij}\ \sim\ 1\times 10^{-3}.
\label{eq:target}
\end{equation}
We therefore require a minimum pairwise Mahalanobis separation $\delta_{\min}=\min_{i\neq j}\delta_{ij}\gtrsim6.0$ among the four levels-manifold. In practice, this SNR target is met by adjusting readout integration time/bandwidth and amplifier gain so that the fitted GMM covariances yield $\delta_{\min}$ above the threshold. When $\delta_{\min}$ is smaller, we apply a calibrated confusion-matrix correction; however, our thermometry results use windows where $\delta_{\min}\ge6.0$, so assignment bias is subdominant to sampling noise.

\subsubsection*{Boltzmann distribution fit}
For each measurement window we acquire single-shot IQ points and classify them into the four
well-resolved levels $\{\text{g,e,f,h}\}$ using a Gaussian-mixture model. Shots assigned to higher or
ambiguous classes are discarded. Let
\[
\mathbf{P}^{\mathrm{meas}}
=\bigl(P_\text{g},P_\text{e},P_\text{f},P_\text{h}\bigr),\qquad
\sum_{i\in\{\text{g,e,f,h}\}} P_i = 1,
\]
denote the renormalized populations in that window.
The qubit energy levels are built from the measured transition frequencies
$\{f_\text{ge},f_\text{ef},f_\text{fh}\}$ as
\[
E_0 = 0,\qquad
E_1 = f_\text{ge},\qquad
E_2 = f_\text{ge}+f_\text{ef},\qquad
E_3 = f_\text{ge}+f_\text{ef}+f_\text{fh},
\]
expressed in GHz so that $k_\text{B}/h \equiv 20.84~\mathrm{GHz/K}$ converts between temperature and
dimensionless Boltzmann factors.

For a trial temperature $T$ the inverse “GHz-temperature” is
\[
\beta(T)=\frac{1}{(k_\text{B}/h)\,T},
\]
and the truncated four levels-manifold Boltzmann probabilities are
\begin{equation}
P_i^{\mathrm{th}}(T)
=\frac{\exp[-\beta(T)\,E_i]}{\sum\limits_{j\in\{\text{g,e,f,h}\}}\exp[-\beta(T)\,E_j]}\,,
\qquad i\in\{\text{g,e,f,h}\}.
\label{eq:pth}
\end{equation}

We determine the effective qubit temperature $T_\text{eff}$ by minimizing a $\chi^2$-like cost,
\begin{equation}
\chi^2(T)
=\sum_{i\in\{\text{g,e,f,h}\}}
\frac{\bigl(P_i^{\mathrm{meas}}-P_i^{\mathrm{th}}(T)\bigr)^2}{P_i^{\mathrm{th}}(T)}\,,
\label{eq:chi2}
\end{equation}
over a bounded interval $T\in[T_{\min},T_{\max}]$ (typically $1~\mathrm{mK}$-$20~\mathrm{K}$).
The estimate is
\begin{equation}
T_\mathrm{eff}=\arg\min_{T\in[T_{\min},T_{\max}]}\,\chi^2(T).
\end{equation}

For reporting, we compute a descriptive coefficient of determination,
\begin{equation}
R^2
=1-\frac{\sum_i\bigl(P_i^{\mathrm{meas}}-P_i^{\mathrm{th}}(T_\text{eff})\bigr)^2}
{\sum_i\bigl(P_i^{\mathrm{meas}}-\bar P^{\mathrm{meas}}\bigr)^2},
\qquad
\bar P^{\mathrm{meas}}=\tfrac14\sum_i P_i^{\mathrm{meas}},
\end{equation}
evaluated on the four levels.

Our fitting routine also quotes independent temperatures from each excited-to-ground ratio,
\begin{equation}
\frac{P_i^{\mathrm{meas}}}{P_\text{g}^{\mathrm{meas}}} =\exp\bigl[-\beta(T)\,(E_i-E_0)\bigr]
\;\Longrightarrow\;
T_i =\frac{-\,E_i}{(k_\text{B}/h)\,\ln\bigl(P_i^{\mathrm{meas}}/P_\text{g}^{\mathrm{meas}}\bigr)},
\qquad i\in\{\text{e,f,h}\},
\end{equation}
defined only when $P_i^{\mathrm{meas}},P_\text{g}^{\mathrm{meas}}>0$. Agreement of $T_i$ with $T_\text{eff}$ indicates a consistent Boltzmann slope across the manifold.

\subsubsection*{Quantum Cramér-Rao bound and Fisher Information}
The Quantum Cramér--Rao Bound (QCRB) establishes the fundamental lower limit on the variance of an estimator. For a parameter $T$ (temperature), it states that the variance $\mathrm{Var}(\hat{T})$ is inversely proportional to the Quantum Fisher Information (QFI), $F_{\mathrm{Q}}(T)$, which measures the sensitivity of a quantum state to changes in that parameter. They are related by:
\begin{equation}
\mathrm{Var}(\hat{T}) \ge \frac{1}{N F_{\mathrm{Q}}(T)}
\end{equation}
where $N$ is the number of independent measurements.

We benchmark the thermometry precision in Fig.~\ref{fig:qubit_thermometry}g against the quantum Cramér--Rao bound (QCRB) for an $n$-level probe in thermal equilibrium. The populations follow the Boltzmann distribution $P_i^{\mathrm{th}}(T) = e^{-E_i/(k_{\mathrm{B}}T)} / Z(T)$. For states diagonal in the energy basis, the QFI is determined by the energy variance $\mathrm{Var}_T(E) = \langle E^2 \rangle_T - \langle E \rangle_T^2$:
\begin{equation}
F_{\mathrm{Q}}(T) = \frac{\mathrm{Var}_T(E)}{k_{\mathrm{B}}^2 T^4}
\end{equation}
The normalized single-measurement precision, $(\Delta T/T)_{\mathrm{SM}} = (\sigma_T/T)\sqrt{N}$, must therefore satisfy:
\begin{equation}
(\Delta T/T)_{\mathrm{SM}} \ge \frac{k_{\mathrm{B}} T}{\sqrt{\mathrm{Var}_T(E)}}
\end{equation}

We define dimensionless energies $x_{\mathrm{ge}} = \frac{hf_{\mathrm{ge}}}{k_{\mathrm{B}}T}$, $x_{\mathrm{gf}} = \frac{h(f_{\mathrm{ge}}+f_{\mathrm{ef}})}{k_{\mathrm{B}}T}$, and $x_{\mathrm{gh}} = \frac{h(f_{\mathrm{ge}}+f_{\mathrm{ef}}+f_{\mathrm{fh}})}{k_{\mathrm{B}}T}$. The bounds for $n$ levels are:

\paragraph{Two-level ($n=2$, $\{\ket{\mathrm{g}},\ket{\mathrm{e}}\}$):}
\begin{equation}
(\Delta T/T)_{\mathrm{SM}}^2 \ge \frac{(1+e^{x_{\mathrm{ge}}})^2}{x_{\mathrm{ge}}^2 e^{x_{\mathrm{ge}}}}
\end{equation}

\paragraph{Three-level ($n=3$, $\{\ket{\mathrm{g}},\ket{\mathrm{e}},\ket{\mathrm{f}}\}$):}
\begin{equation}
(\Delta T/T)_{\mathrm{SM}}^2 \ge \frac{(e^{x_{\mathrm{ge}}+x_{\mathrm{gf}}}+e^{x_{\mathrm{ge}}}+e^{x_{\mathrm{gf}}})^2}{\Big\{x_{\mathrm{ge}}^2 e^{x_{\mathrm{gf}}}+x_{\mathrm{gf}}^2 e^{x_{\mathrm{ge}}}+(x_{\mathrm{ge}}-x_{\mathrm{gf}})^2\Big\} e^{x_{\mathrm{ge}}+x_{\mathrm{gf}}}}
\end{equation}

\paragraph{Four-level ($n=4$, $\{\ket{\mathrm{g}},\ket{\mathrm{e}},\ket{\mathrm{f}},\ket{\mathrm{h}}\}$):}
\begin{equation}
\begin{aligned}
(\Delta T/T)_{\mathrm{SM}}^2 \ge \frac{(e^{x_{\mathrm{ge}}+x_{\mathrm{gf}}+x_{\mathrm{gh}}}+e^{x_{\mathrm{ge}}+x_{\mathrm{gf}}}+e^{x_{\mathrm{ge}}+x_{\mathrm{gh}}}+e^{x_{\mathrm{gf}}+x_{\mathrm{gh}}})^2}{\mathcal{D}_4 e^{x_{\mathrm{ge}}+x_{\mathrm{gf}}+x_{\mathrm{gh}}}}
\end{aligned}
\end{equation}
where the denominator $\mathcal{D}_4$ is given by:
\begin{equation}
\begin{aligned}
\mathcal{D}_4 = \ &x_{\mathrm{ge}}^2 e^{x_{\mathrm{gf}}+x_{\mathrm{gh}}} + x_{\mathrm{gf}}^2 e^{x_{\mathrm{ge}}+x_{\mathrm{gh}}} + x_{\mathrm{gh}}^2 e^{x_{\mathrm{ge}}+x_{\mathrm{gf}}} \\
&+ (x_{\mathrm{ge}}-x_{\mathrm{gf}})^2 e^{x_{\mathrm{gh}}} + (x_{\mathrm{ge}}-x_{\mathrm{gh}})^2 e^{x_{\mathrm{gf}}} + (x_{\mathrm{gf}}-x_{\mathrm{gh}})^2 e^{x_{\mathrm{ge}}}
\end{aligned}
\end{equation}

\backmatter

\section*{Data Availability}

All relevant data and codes generating the figures in this article will be provided upon request.


\bibliography{main}

\section*{Acknowledgments}
We express our gratitude to the Jenny \& Antti Wihuri Foundation for their support through the Ph.D. grant awarded to the first author. We acknowledge the European Research Council under the Advanced Grant no. 101053801 (ConceptQ), Business Finland under the Quantum Technologies Industrial project (Grant no. 2118781) and Academy of Finland under its Centre of Excellence Quantum Technology Finland (Grant no. 352925) and through the Finnish Quantum Flagship. Special thanks are extended to Sergei Lemziakov, Dmitrii Lvov, and Joonas Peltonen for their invaluable assistance with the Otanano cryogenic facility. We thank Professor Jukka Pekola for insightful discussions on cryogenic thermometry. The authors acknowledge Jukka-Pekka Kaikkonen from VTT Technical Research for providing us with the unpatterned Niobium-on-Silicon wafer. We also acknowledge Visa Vesterinen from Arctic Instruments for fruitful discussions on the dc SQUID design and implementation.

\section*{Authors contributions}
A.S. and M.M. conceived the experiment. A.S. and S.K. designed and carried out the electromagnetic simulations of the device. A.S. and P.S. fabricated the device. E.F. performed initial measurements with the support of A.S. V.V. contributed to theoretical discussions on qubit thermometry. A.S. wrote the measurement codes, conducted the experiments, and analyzed the results with feedback from S.K. A.S.,~and M.M. wrote the manuscript with comments from all the authors. M.M. supervised the work.

\section*{Competing interests}
M.M. declares that he is a Co-Founder and Shareholder of IQM Finland Oy and QMill Oy. All other authors declare no competing interests.

\clearpage 
\appendix 
\setcounter{page}{1} 

\section*{Supplementary Information for "Quantum Dial"}
\renewcommand{\thefigure}{\textbf{S\arabic{figure}}}
\pretocmd\theHfigure{S}{}{}
\renewcommand{\thetable}{\textbf{S\arabic{table}}}
\pretocmd\theHtable{S}{}{}
\renewcommand{\figurename}{\textbf{Supplementary Fig.}}
\renewcommand{\tablename}{\textbf{Supplementary Table}}
\renewcommand{\thesection}{\textbf{Supplementary Note S\arabic{section}}}
\setcounter{section}{0}
\setcounter{figure}{0} 
\setcounter{table}{0}

\section{\hspace{-9pt}: Device design and simulations}\label{supp_note:device_design}

We designed a four-transmon device with double-pad capacitors to compare three drive-line implementations: a standard weakly coupled line, a fixed-frequency drive-line filter, and two tunable-frequency drive-line filters. In the main text, \emph{quantum dial} refers to the broader architecture in which a qubit couples strongly to a tunable intermediary that can be temporally connected to auxiliary functions. In this device, the realized dial element is a tunable-frequency drive-line filter that enables three regimes using the same physical line: idle, control, and reset.

For the filtered lines, we target the same qubit--drive coupling. The standard weakly coupled line is instead designed for an external relaxation time $T_1^{\mathrm{ext}} \sim 5~\mathrm{ms}$ near $4.5~\mathrm{GHz}$. This corresponds to $C_{\mathrm{qd}} \approx 0.03~\mathrm{fF}$ for the weakly coupled line and $C_{\mathrm{qd}} \approx 4.4~\mathrm{fF}$ for the strongly coupled filtered lines. Because the tunable-frequency drive-line filters provide frequency agility without tuning the qubit, all qubits are fixed-frequency except the qubit paired with the fixed-frequency drive filter, which is realized as a tunable-frequency qubit for matching to the fixed stopband. Target frequency bands are $7.0$--$7.7~\mathrm{GHz}$ for the readout resonators and $4.0$--$4.5~\mathrm{GHz}$ for the qubits.

The layout is prepared in KQCircuits within KLayout and simulated with a Sonnet--AWR workflow. Chip substructures are simulated in Sonnet, exported as Touchstone S-parameters, and assembled in AWR Microwave Office for system-level sweeps. Josephson junctions and dc SQUID arrays are modeled as lumped, flux-dependent inductors. This workflow predicts resonator frequencies and linewidths, qubit--resonator coupling, external quality factors, and the frequency and tunability of the drive-line filters.

We optimize the qubit capacitance to set $E_C$ (and hence the anharmonicity), choose the qubit--resonator coupling to reach the target dispersive shift for readout, and set the qubit--drive coupling to achieve the required Rabi rates while satisfying Purcell constraints. We then simulate the full chip, including Purcell filters, readout resonators, qubit pads, and the drive lines and filters, to verify the combined response against the target bands.

\subsection*{Theory and Simulations of tunable-frequency drive filter}

We model the drive network as a stub attached to the qubit node as depicted in Supplementary Fig.~\ref{fig:tdf_sim}a and from here on, refer to it as a quarter--wavelength filter. The qubit node is at $x=0$ and is driven from a $50~\Omega$ drive port P1. The qubit port P2 is connected to this drive port at the qubit node by a small series coupling capacitor $C_\mathrm{d}$. From the node, the filter consists of a first uniform transmission line (characteristic impedance $Z_0$, phase constant $\beta=\omega/v_\text{p}$) of length $L_\ell = x_\mathrm{s}$, followed by a series inductor $L_\mathrm{s}$ located at $x=x_\mathrm{s}$, and then a second uniform transmission line of length $L_\mathrm{r} = L_\mathrm{f}-x_\mathrm{s}$ that ends at $x=L_\mathrm{f}$. The physical open end at $x=L_\mathrm{f}$ can be modeled with a small shunt capacitance $C_\mathrm{g}$ to ground. To first order, this acts as a small effective length extension of the filter. At the operating point where $L_\mathrm{s}=0$ and $C_\mathrm{g}$ is small, the filter length $L_\mathrm{f}\approx \lambda/4$ at the filter frequency $f_\mathrm{f}$, producing a current antinode at the qubit node and a current node at the far--end.

\subsubsection*{Filter frequency with a fixed series inductance}
In this section, we derive the filter frequency of the quarter--wavelength filter with an added inductor. This quarter--wavelength filter of length $L_\mathrm{f}$ has input impedance
\begin{equation}
Z_\text{f}(\omega)= \text{i} Z_0 \tan(\beta L_\mathrm{f}).
\end{equation}

It is an open circuit when $\beta L_\mathrm{f}=\pi/2$, i.e. at the filter frequency
\begin{equation}
\omega_\mathrm{0}=\frac{\pi v_\text{p}}{2L_\mathrm{f}},\qquad f_\mathrm{0}=\frac{v_\text{p}}{4L_\mathrm{f}}.
\end{equation}

By inserting a small series inductance $L_\mathrm{s}$ in the filter path, one adds a small series reactance that is equivalent, for phase purposes, to a short additional piece $\delta l$ of line. A short length $\delta l$ of line adds the series impedance $Z_\text{s}=\mathrm{i} Z_0 \tan(\beta \,\delta l)\approx \mathrm{i} Z_0\,\beta\,\delta l$ (for $\beta\,\delta l \ll 1$). Equating $Z_\text{s}$ to the inductor impedance $j\omega L_\mathrm{s}$ gives an equivalent added phase
\begin{equation}
\delta\theta \equiv \beta\,\delta l \;\approx\; \frac{\omega L_\mathrm{s}}{Z_0}.
\end{equation}

The open-circuit condition is that the total phase through the filter equals $\pi/2$ giving
\begin{equation}
\beta L_\mathrm{f} + \delta\theta = \frac{\pi}{2}.
\end{equation}

Solving for the new filter frequency $\omega_\mathrm{f}$,
\begin{equation}
\beta=\frac{1}{v_\text{p}}\omega_\mathrm{f} = \frac{1}{L_\mathrm{f}}\Big(\frac{\pi}{2}-\delta\theta\Big) \;\Rightarrow\;\omega_\mathrm{f} = \frac{v_\text{p}}{L_\mathrm{f}}\Big(\frac{\pi}{2}-\delta\theta\Big).
\end{equation}

Divide by $\omega_0=\tfrac{v_\text{p}}{L_\mathrm{f}}\tfrac{\pi}{2}$ and use $\delta\theta\approx \omega_\mathrm{f} L_\mathrm{s}/Z_0$; keeping terms to first order in the small parameter $\omega_\mathrm{f} L_\mathrm{s}/Z_0$ yields
\begin{equation}
\frac{\omega_\mathrm{f}}{\omega_0} = 1 - \frac{2}{\pi}\,\delta\theta \approx 1 - \frac{2}{\pi}\,\frac{\omega_0 L_\mathrm{s}}{Z_0}.
\end{equation}

Equivalently, written in the convenient linear frequency form,
\begin{equation}\label{eq:filter_freq}
f_\mathrm{f} \approx \frac{f_0}{1+\dfrac{4 f_0 L_\mathrm{s}}{Z_0}}.
\end{equation}

Adding an inductor in the filter transmission line effectively pulls the filter frequency down by a factor of $\frac{1}{1+\frac{4f_0L_\text{s}}{Z_0}}$.

\subsubsection*{Series inductance from a dc SQUID array}

We introduce the tunability to the filter's filter frequency by integrating a series of dc SQUIDs (two identical Josephson junctions in parallel) in the filter transmission line. A single junction in a dc SQUID follows the Josephson relation:
\begin{align}
    I &= I_\mathrm{c} \sin\delta,\\
    V &=\frac{\Phi_0}{2\pi}\,\dot{\delta},\\
    \Phi&=\frac{\Phi_0}{2\pi}\,\delta,
\end{align}
with critical current $I_\mathrm{c}$, superconducting phase $\delta$, and flux quantum $\Phi_0$. Using these relations, we obtain the flux-current relation:
\begin{equation}
\Phi(I) = \frac{\Phi_0}{2\pi}\arcsin\Big(\frac{I}{I_\mathrm{c}}\Big).
\end{equation}

We expand the flux-current relation for $|I|<I_\mathrm{c}$ up to the fifth-order term,
\begin{equation}
\Phi(I) = \underbrace{\frac{\Phi_0}{2\pi I_\mathrm{c}}}_{L_\mathrm{J}^{(0)}} I +\underbrace{\frac{\Phi_0}{12\pi I_\mathrm{c}^{3}}}_{\alpha_\mathrm{J}} I^{3} + O(I^{5}).
\end{equation}

The Josephson inductance of a junction is the differential inductance seen by the filter circuit and is given by,
\begin{equation}\label{eq:Lj_snj}
L_\mathrm{J}(I) \equiv \frac{\mathrm{d}\Phi}{\mathrm{d}I} = L_\mathrm{J}^{(0)} + 3\alpha_\mathrm{J} I^{2} + O(I^{4}) = L_\mathrm{J}^{(0)}\left[1 + \tfrac{1}{2}\Big(\tfrac{I}{I_\mathrm{c}}\Big)^2\right] + O(I^{4}).
\end{equation}

A symmetric dc SQUID (two identical junctions in parallel, negligible loop inductance $\beta_L = 2L_\text{Loop}I_\mathrm{c}/\Phi_0\ll 1$) behaves like a single effective junction with flux-dependent critical current
\begin{equation}
I_\mathrm{c}^{\mathrm{SQ}}(\Phi_\mathrm{ext})=2I_c\big|\cos(\pi\Phi_\mathrm{ext}/\Phi_0)\big|.
\end{equation}

Plugging this in, the Josephson inductance of a single junction Eq.~\eqref{eq:Lj_snj} yields the Josephson inductance of a single dc SQUID
\begin{equation}
L_\mathrm{J}^\mathrm{SQ}(I,\Phi_{\mathrm{ext}})=
\underbrace{L_{\mathrm{fixed}}}_{\text{fixed}}
+
\underbrace{\frac{\Phi_0}{2\pi\,I_\mathrm{c}^{\mathrm{SQ}}(\Phi_{\mathrm{ext}})}}_{\text{linear Josephson}}
+
\underbrace{\frac{\Phi_0}{4\pi}\,\frac{I^{2}}{\big[I_\mathrm{c}^{\mathrm{SQ}}(\Phi_{\mathrm{ext}})\big]^3}}_{\text{nonlinear (Kerr)}}
+
O(I^{4}).
\end{equation}
where $L_\mathrm{fixed}$ is the flux-independent series inductance arising from the geometric and kinetic contributions of the leads of the dc SQUID array. In the limit $|I|\ll I_\mathrm{c}$, the nonlinear contribution can be neglected yielding
\begin{equation}
L_\mathrm{J}^\mathrm{SQ}(\Phi_{\mathrm{ext}})=L_{\mathrm{fixed}}+\frac{\Phi_0}{2\pi\,I_\mathrm{c}^{\mathrm{SQ}}(\Phi_{\mathrm{ext}})}.
\end{equation}

Finally, in a series array of $N$ identical dc SQUIDs, the same current flows through each dc SQUID and therefore the flux drop across the node adds up, yielding the Josephson inductance of a series array,
\begin{equation}\label{eq:ind_sq_arr}
L_\mathrm{J}^\mathrm{arr}(\Phi_{\mathrm{ext}})= N L_{\mathrm{fixed}}+N\frac{\Phi_0}{2\pi\,I_\mathrm{c}^{\mathrm{SQ}}(\Phi_{\mathrm{ext}})}.
\end{equation}

Replacing the series inductance introduced in the derivation of the filter frequency in Eq.~\eqref{eq:filter_freq}, we obtain the flux-dependent filter frequency
\begin{equation}\label{eq:tunable_filter_freq}
f_\mathrm{f}(\Phi_\mathrm{ext}) \approx \frac{f_0}{1+\dfrac{4f_0 L_\mathrm{J}^\mathrm{arr}(\Phi_\mathrm{ext})}{Z_0}}.
\end{equation}

Thus the filter frequency decreases as $L_\mathrm{J}^\mathrm{arr}$ increases with applied external flux $\Phi_\mathrm{ext}$. This first-order result is very accurate for $\omega_0 L_\mathrm{J}^\mathrm{arr}/Z_0 \lesssim 1$ and when the series of dc SQUIDs is placed near a current antinode of the quarter--wavelength filter mode. For large inductances where $L_\mathrm{J}^\text{arr} > 1~\text{nH}$, we later derive an accurate model for transcendental equation that finds the filter frequency as a function of dc SQUIDs position $x_\text{s}$. 

Junction asymmetry and finite loop inductance modify $I_\mathrm{c}^{\mathrm{SQ}}(\Phi_{\mathrm{ext}})$ (removing the divergence at half flux quantum $\Phi_\mathrm{ext} = \Phi_0/2$). In practice, keeping $I\lesssim 0.1\,I_\mathrm{c}$ limits the fractional inductance change to $\lesssim0.5\%$, ensuring operation in the linear regime.

\subsection*{Current profile}

The quarter--wavelength filter sets up a standing wave between the node and the open end. The open boundary enforces $I(L_\mathrm{f})=0$, consequently the current in the second section ($x\in[x_\mathrm{s},L_\mathrm{f}]$) takes the standing--wave form
\begin{align}\label{eq:curr_prf_f}
I(x)&=I_\mathrm{s}\,\frac{\sin\!\big[\beta(L_\mathrm{f}-x)\big]}{\sin\!\big[\beta(L_\mathrm{f}-x_\mathrm{s})\big]},
\end{align}
where $I_\mathrm{s}$ is the drive current at the inductor and is given by
\begin{equation}
I_\mathrm{s}=\frac{I(0)}{\displaystyle \cos(\beta x_\mathrm{s})
+\Big\{\cot\!\big[\beta(L_\mathrm{f}-x_\mathrm{s})\big]-\frac{\omega L_\mathrm{s}}{Z_0}\Big\}\sin(\beta x_\mathrm{s})}.
\end{equation}
This exhibits a current node at the open end and, near the $\lambda/4$ condition, a current anti-node at the qubit location. Across the series inductor the current is continuous while the voltage drops by $j\omega L_\mathrm{s}\,I_\mathrm{s}$. The inductor current $I_\mathrm{s}$ propagating back through the first section ($x\in[0,x_\mathrm{s}]$) gives
\begin{equation}
I(x) = I_\mathrm{s}\!\left[\cos\!\big(\beta(x_\mathrm{s}-x)\big)
+\Big(\cot\!\big[\beta(L_\mathrm{f}-x_\mathrm{s})\big]-\frac{\omega L_\mathrm{s}}{Z_0}\Big)\sin\!\big(\beta(x_\mathrm{s}-x)\big)\right].
\end{equation}
At exact quarter--wavelength with $L_\mathrm{s}=0$, this reduces to the simple profile
\begin{equation}\label{eq:curr_prf_l}
I(x)=I(0)\,\frac{\sin\!\big[\beta(L_\mathrm{f}-x)\big]}{\sin(\beta L_\mathrm{f})},\qquad
\beta L_\mathrm{f}=\frac{\pi}{2},
\end{equation}
so that the current is maximal at the node and vanishes at the open end as presented in Supplementary Fig.~\ref{fig:tdf_sim}a (top panel). A small series inductance perturbs this distribution by locally increasing the electrical length in proportion to the local current, which is why placing the inductance near regions of large current produces a larger filter frequency pull than placing it near a current node as shown in Supplementary Fig.~\ref{fig:tdf_sim}a (bottom panel).

\subsubsection*{Position dependence of the filter frequency shift}

The effective frequency shift is proportional to the inductive participation of the dc SQUID array in the mode, so the contribution of the tunable element is weighted by the local current amplitude $|I(x)|$. Placing the dc SQUIDs exactly at a current anti-node (voltage node) maximizes participation and tunability but also maximizes RF current through the junctions. Moving the array away from the anti-node reduces $|I|$ and therefore reduces the energy participation of $L_\mathrm{J}$ in the mode; tunability decreases because a smaller fraction of the total inductive energy resides in the adjustable element. Here, we determine the filter frequency as a function of the position $x_{\mathrm{s}}$ of a lumped series inductance $L_{\mathrm{s}}$ along a lossless quarter-wavelength filter. 

The right section referenced at $x=x_{\mathrm{s}}$ has input impedance
\begin{align}
Z_2(\omega) &= Z_0\frac{Z_{\mathrm{end}}+\mathrm{i} Z_0\tan(\beta L_\mathrm{r})}{Z_0+\mathrm{i}Z_{\mathrm{end}}\tan(\beta L_\mathrm{r})},
\end{align}
where the impedance due to the shunt capacitance at the open end is $Z_{\mathrm{end}} =\frac{1}{\mathrm{i}\omega C_{\mathrm{g}}}$.

Adding the series inductor gives the net series impedance at the location of the inductor
\begin{equation}
Z_1(\omega)=\mathrm{i}\omega L_{\mathrm{s}}+Z_2(\omega),
\end{equation}
which is purely imaginary for a lossless network. Transforming $Z_1$ through the left section to the drive node yields the input impedance
\begin{equation}
Z_{\mathrm{in}}(\omega)=Z_0\,\frac{Z_1(\omega)+\mathrm{i}Z_0\tan(\beta x_{\mathrm{s}})}{Z_0+\mathrm{i}Z_1(\omega)\tan(\beta x_{\mathrm{s}})}.
\end{equation}
The filter frequency is defined when the node is effectively short, i.e. when $|Z_{\mathrm{in}}| = 0$, that is, when the numerator vanishes:
\begin{equation}
Z_1(\omega)+\mathrm{i}Z_0\tan(\beta x_{\mathrm{s}})=0.
\end{equation}
Writing $Z_2(\omega)=\mathrm{i}X_2(\omega)$ this becomes the scalar transcendental equation
\begin{equation}
\omega L_\mathrm{s}+\big[X_2(\omega)+Z_0\tan(\beta x_\mathrm{s})\big]=0,
\label{eq:Fzero}
\end{equation}
whose solution $\omega=\omega_{\mathrm{f}}(x_{\mathrm{s}},L_{\mathrm{s}})$ gives the filter frequency. For the ideal-open case (no additional shunt capacitance $C_{\mathrm{g}}=0$), $X_2(\omega)=-Z_0\cot(\beta L_\mathrm{r})$ and Eq.~\eqref{eq:Fzero} reduces to
\begin{equation}
\frac{\omega L_\mathrm{s}}{Z_0} = \cot(\beta L_\mathrm{r}) - \tan(\beta x_\mathrm{s})
\end{equation}

The root near the fundamental $\omega_0=\tfrac{\pi v_\text{p}}{2L_{\mathrm{f}}}$ is obtained by a one-dimensional numerical solve (bisection or Newton) of the transcendental function 
\begin{align}
F(\omega) &= \tan(\beta x_{\mathrm{s}})\left[\cot(\beta\ell_2)-\frac{\omega L_{\mathrm{s}}}{Z_0}\right]+1.
\end{align}
Expanding $F(\omega)$ to first-order about $\omega_0$ recovers the compact perturbative pull
\begin{equation}
\frac{f_{\mathrm{f}}(x_{\mathrm{s}})-f_0}{f_0}\approx
-\frac{2}{\pi}\,\frac{L_{\mathrm{s}}}{Z_0}\,\cos^2\!\left(\frac{\pi x_{\mathrm{s}}}{2L_{\mathrm{f}}}\right),
\end{equation}
which is accurate when $L_{\mathrm{s}}$ is small but, for quantitative agreement across all $(x_{\mathrm{s}},L_{\mathrm{s}})$, the transcendental condition above is used. The first-order approximation matches well with the version we derived in Eq.~\eqref{eq:tunable_filter_freq}. For series inductance in range $L_\text{J}^\text{arr} \in [0,10]~\text{nH}$, we plot the shift in the filter frequency $\Delta f_\text{f} = f_\mathrm{f}-f_0$ by solving Eq.~\eqref{eq:Fzero} numerically and present the theoretical (solid curves) together with the simulation (markers) in Supplementary Fig.~\ref{fig:tdf_sim}a (bottom panel). As expected, we see lower tunability away from the current anti-node due to lower inductive participation of the series inductor to the overall electrical length of the filter. 

\subsubsection*{Admittance seen by the qubit through the coupling capacitor}
Next, we derive the admittance $Y_\mathrm{q}$ seen by the qubit through the coupling capacitor $\mathrm{C_\mathrm{d}}$. The qubit--drive filter coupling can be obtained from the admittance using the expression $\gamma_\mathrm{qf} = \text{Re}(Y_\mathrm{q})/C_\mathrm{q}$ where $C_\mathrm{q}$ is the total qubit capacitance. From the qubit--drive coupling, we can estimate the Rabi frequency $\Omega/2\pi \propto \sqrt{\gamma_\mathrm{qf}}$ and the extrinsic relaxation rate $T_1^\mathrm{ext} = \gamma_\mathrm{qf}^{-1}$. The analytical expression gives us a tool to fit the measured Rabi frequency and the relaxation rate.

When looking from the node toward the left, back into the matched drive port P1, the impedance to ground is simply the port/reference impedance $Z_s$ (typically $Z_s=Z_0=50~\Omega$). Since the node is the junction of the left and right paths, the net node impedance is the parallel combination $\big(Z_s \parallel Z_{\mathrm{in}}(\omega)\big)$ yielding
\begin{equation}
Z_\mathrm{node}(\omega) = \frac{Z_s\,Z_{\mathrm{in}}(\omega)}{Z_s + Z_{\mathrm{in}}(\omega)} .
\end{equation}

From the qubit port, the environment is the series combination of $C_\mathrm{d}$ and the node network. The admittance seen by the qubit is therefore
\begin{equation}
Y_\mathrm{q}(\omega) = \frac{1}{\,Z_{C_\mathrm{d}} + Z_\mathrm{node}(\omega)\,} =
\frac{\mathrm{i}\omega C_\mathrm{d}}{\,1 + \mathrm{i}\omega C_\mathrm{d}\,Z_\mathrm{node}(\omega)\,},
\end{equation}
where the impedance due to the coupling capacitance is $Z_{C_\mathrm{d}}=\frac{1}{\mathrm{i}\omega C_\mathrm{d}}$.

At the filter frequency, $Z_{\mathrm{in}}\to0$, hence $Z_{\mathrm{node}}\to0$ and $Y_{\mathrm{q}}\to \mathrm{i}\omega C_{\mathrm{d}}$ (purely reactive, $\mathrm{Re}\{Y_{\mathrm{q}}\}=0$), which corresponds to a complete cancellation of the qubit-drive coupling. A slight detuning of the notch or any finite loss makes $Z_{\mathrm{node}}$ acquire a small real part, so that $\mathrm{Re}\{Y_{\mathrm{q}}\}>0$ and a finite radiative decay channel opens for the qubit. 

\subsubsection*{Design considerations}

\begin{figure*}[ht]
  \centering
  \includegraphics[width=\linewidth]{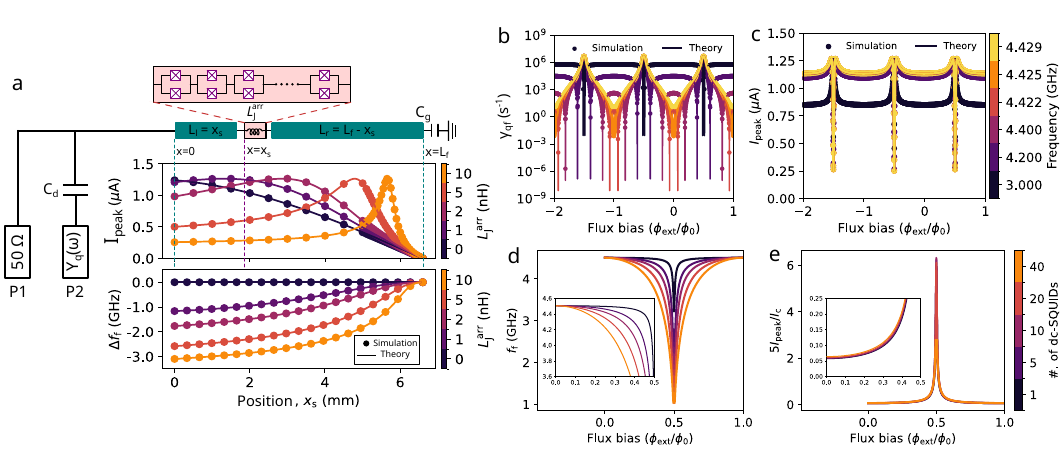}
  \caption{
  \textbf{Tunable-frequency drive filter with a series dc SQUID array.} \textbf{a}, Schematic of the drive port P1 (50~$\Omega$) shunt-coupled to the qubit port P2 through a small capacitor $C_\mathrm{d}$. The line continues as an open-ended $\lambda/4$ stub of electrical length $L_\mathrm{f}$; a series dc SQUID array with effective inductance $L_\mathrm{J}^{\mathrm{arr}}$ is inserted at position $x_\mathrm{s}$, splitting the section into $L_\mathrm{\ell}=x_\mathrm{s}$ and $L_\mathrm{r}=L_\mathrm{f}-x_\mathrm{s}$. A small end capacitor $C_\mathrm{g}$ models the open-end capacitance to the ground. Top: peak ac-current at the inductor, $I_{\mathrm{peak}}$, versus $x_\mathrm{s}$ for several $L_\mathrm{J}^{\mathrm{arr}}$ (color scale). Bottom: corresponding filter frequency shift $\Delta f_\mathrm{f}=f_\mathrm{f}-f_0$ with $f_0=v_\text{p}/(4L_\mathrm{f})$. Markers are AWR simulations; solid curves are the analytic solution. \textbf{b}, Qubit--drive filter coupling $\gamma_\mathrm{qf}(\Phi_{\mathrm{ext}})=\text{Re}(Y_\mathrm{q})(\omega)/C_\mathrm{q}$ versus reduced flux $\Phi_{\mathrm{ext}}/\Phi_0$ at several drive frequencies (color bar) from simulation (markers) and theory (solid curves) at the fixed inductor position. 
  \textbf{c}, Peak ac current $I_{\mathrm{peak}}(\Phi_{\mathrm{ext}})$ for the same frequencies, showing flux-dependence of the drive current through the inductor at the fixed position. Due to increasing inductance towards the half flux quantum $\Phi_0/2$, we see a decrease in the current through the inductor. \textbf{d}, Filter frequency $f_\mathrm{f}(\Phi_{\mathrm{ext}})$ as the inductance of the dc SQUID array is tuned; inset zooms the small $C_\mathrm{g}$-induced shift near zero bias. \textbf{e}, Nonlinearity margin: $5I_{\mathrm{peak}}/I_\mathrm{c}$ versus flux for different numbers of series dc SQUIDs (color bar); inset shows the low-flux regime. Across panels, markers denote simulations and solid curves the closed-form/transcendental model evaluated under the shunt-short condition.}
  \label{fig:tdf_sim}
\end{figure*}

The preceding sections establish how the filter frequency $f_\text{f}$ follows from the shunt-short condition in Eq.~\eqref{eq:Fzero}, how the series inductance from a dc SQUID array enters $L_{\rm J}^\text{arr}(\Phi_{\rm ext})$ according to Eq.~\eqref{eq:ind_sq_arr}, how the line current profile is obtained from Eqs.~\eqref{eq:curr_prf_f}-\eqref{eq:curr_prf_l}, and how the environment admittance $Y_{\rm q}(\omega)$ dictates the drive strength and the extrinsic relaxation time. Here we translate those results into the practical implementation of the tunable-frequency drive filter.

We set the required tunability to cover qubit frequency spread, here $0.2$-$0.5~\mathrm{GHz}$ around the target qubit frequency. Given the available drive at the sample (Gaussian pulses, $-80~\mathrm{dBm}$), we compute $I_{\rm peak}(x_{\rm s})$ from the current profile and require a linear-response margin $5 I_{\rm peak}/I_{\rm c} < 0.25$ over the entire tuning range. This bound fixes both the allowed inductance change and the placement of the array away from the current antinode.

We pick $L_\text{f} \approx 6.5$~mm so that the bare filter frequency $f_0=v_\text{p}/(4L_\text{f})$ sits slightly above the target qubit frequency around $4.5$~GHz, $f_\text{f}$ is then pulled down across the required band by $L_{\rm s}(\Phi_{\rm ext})$ according to Eq.~\eqref{eq:Fzero}. The position $x_\text{s} \approx 2$~mm trades tunability for linearity via the leverage factor implicit in the current profile (maximal near $x=0$, negligible near the open end). For this choice of $x_\text{s}$, we plot the qubit-drive coupling $\gamma_{\mathrm{qf}}$ as a function of reduced flux $\Phi_{\mathrm{ext}}/\Phi_{0}$ for several drive frequencies, and compare the theoretical curves (solid lines) with AWR simulations (markers) in Supplementary Fig.~\ref{fig:tdf_sim}b. The coupling is periodically modulated with reduced flux ($\Phi_{\mathrm{ext}}/\Phi_{0}$), when the drive frequency is detuned from the filter frequency at zero bias, deep minima in $\gamma_{\mathrm{qf}}$ occur at biases where the tuned filter frequency crosses the drive frequency, yielding extremely low coupling rates. In Supplementary Fig.~\ref{fig:tdf_sim}c, we also observe a periodic change in the peak ac current at that location, indicating flux-dependent drive current in the dc SQUIDs. 

For our choice of tunability and linear-response margin on the drive current as shown in Supplementary Fig.~\ref{fig:tdf_sim}d-e, we target junctions with per junction critical current $I_{c}=10~\mu\mathrm{A}$, corresponding to Josephson inductance $L_\text{J}\approx0.03~\mathrm{nH}$. A dc SQUID contributes $\approx0.015~\mathrm{nH}$ at zero flux, an array of $N=5$ dc SQUIDs therefore yields $L_{\rm s}\approx0.075~\mathrm{nH}$, when placed at the optimized $x_\text{s}$ provides $\sim 0.5~\mathrm{GHz}$ of tuning while satisfying $5I_{\rm peak}/I_\text{c}<0.25$. Larger ranges are achievable by increasing $N$ subject to the same margin.

Finally, a fast-flux line is integrated along the dc SQUID array to tune the filter frequency on a few-ns time scales. This enables gate protocols in which the filter frequency is briefly detuned from qubit frequency so that the qubit-drive coupling is stronger, allowing single-qubit gates at low drive power. After the gate, the flux returns the filter to the filter frequency, restoring a long $T_1$. The same rapid control of the external relaxation rate supports fast qubit reset and efficient qubit thermometry experiments. The fast-flux line is designed with sufficient inductive coupling to the dc SQUIDs to thread the required magnetic flux with modest ac/dc currents, while maintaining low crosstalk and finite bandwidth supporting flux pulses of tens-of-nanoseconds.

\section{\hspace{-9pt}: Device fabrication}\label{supp_note:device_fab}

A $200$-nm-thick Nb film on a $6$-inch, $(100)$-oriented, high-resistivity ($>10~\mathrm{k}\Omega\mathrm{cm}$) intrinsic-silicon wafer was procured from VTT Technical Research Centre of Finland. The wafer was spin-coated with AZ-5214E (few-micron protective layer) prior to dicing, then cut into $33\times33$~mm dies using a Disco DAD3220 saw. Post-dicing, the dies were soaked overnight in acetone (ACE), sonicated in ACE for $10$~min, rinsed in isopropanol (IPA), and dried with a nitrogen gun.

For patterning Nb capacitors, coplanar transmission lines, resonators, qubit pads, and test-junction pads, we used optical lithography on a Maskless Aligner MLA150 (Heidelberg). Each die was HMDS-primed (YES-3), then spin-coated with AZ-5214E at $4000$~rpm for $40$~s (LabSpin6), yielding a resist thickness of $\sim1.4~\mu\mathrm{m}$, followed by a $90^{\circ}$C soft bake for $1.5$~min. A CAD layout comprising a $3\times3$ array of $10\times10$~mm device fields was converted to the MLA150 job format and exposed with the $375$-nm laser at a dose of $120~\mathrm{mJ/cm^2}$. Development was performed for $50$~s in a $1{:}5$ AZ-351B:DI-$\mathrm{H_2O}$ solution.

Nb etching was carried out on an Oxford Plasmalab 80 Plus using a three-step reactive-ion process in a single pump-down, with $60$-s cool-downs between steps:
(1) O$_2$ ash: $100$~mTorr, $40$~SCCM, $150$~W, $10$~s;
(2) CF$_4$ etch: $50$~mTorr, $20$~SCCM, $30$~W, $\sim11$~min;
(3) O$_2$ ash: same as step (1), $\sim2$~min.
Before processing, the recipe was run without a sample to clean the chamber and check etch stability, then repeated with the patterned die to transfer features (resist protects Nb; exposed regions are removed).

After etch, the resist was stripped in a heated N-methyl-2-pyrrolidone (NMP) remover at $80^{\circ}$~C for $\sim8$~h, followed by $3$~min high-power sonication in the same bath. The dies were then cleaned by sequential $3$~min sonications in ACE and IPA and dried with nitrogen.

Immediately after Nb pattern transfer, the sample was subjected to a buffered-oxide-etch step to remove native oxides from Nb and Si: it was immersed in an ammonium-fluoride solution ($90{:}10$) for $5.5$~min, then rinsed twice in DI water ($2\times5$~min). To minimize re-oxidation, the die was transferred in a DI-water-filled beaker to the spin coater, spin-dried at $4000$~rpm for $60$~s, and immediately coated with MMA 8.5 E11 at $1550$~rpm for $45$~s, followed by a $160^{\circ}$C bake for $60$~s.

Electron-beam lithography (Raith) was used to define Manhattan-style Josephson junctions and galvanic-contact patches. Exposures were performed at $100$~kV with a $300~\mu\mathrm{m}$ aperture, $1$~nA beam current, and a dose of $1400~\mu\mathrm{C/cm^2}$. Owing to the larger junction area (about $10\times$) in the tunable-frequency drive filter relative to the qubit, we patterned the filter and the qubit junctions in a first EBL step and used a second, similar step for the galvanic-contact patches.

The resist development was carried out in MIBK:IPA ($1{:}3$) for $27$~s, followed by a $5$~s IPA dip. The formation of the junction was performed in a Plassys MEB700S2-III UHV system. The sample was loaded into the loadlock and pumped overnight ($12$-$15$~h) to below $1\times10^{-7}$~mbar. Residual resist organics were removed by ozone ashing ($10$~mbar, $1$~min), after which the chamber was re-pumped to high vacuum. A Ti gettering pre-evaporation ($0.1$~nm/s, $2$~min) was used to reduce the base pressure to below $5\times10^{-8}$~mbar before the first Al deposition. The first Al electrode was then deposited at $45^{\circ}$ and $-135^{\circ}$ planetary rotation with $45^{\circ}$ tilt, $25$~nm per rotation at $0.2$~nm/s, followed by a controlled oxidation at $1.2$~mbar for $10$~min to form the tunnel barrier. A second Ti gettering step preceded the second electrode deposition at $-45^{\circ}$ and $135^{\circ}$ both at $45^{\circ}$ tilt, with $30$~nm deposited at each angle at the same rate. Finally, a passivation oxidation was applied at $20$~mbar for $10$~min.

For the galvanic-contact process, we followed the same sequence for electron-beam writing and deposition, except that during deposition after ozone ashing we removed native oxides from the exposed junction electrodes and Nb surface by Ar ion milling (beam voltage $400$~V, beam current $120$~mA, $2$~min, $0^{\circ}$ planetary rotation, $0^{\circ}$ tilt) immediately prior to Al deposition. Al electrodes were deposited at $0^{\circ}$ and $180^{\circ}$ both at $45^{\circ}$ tilt, with $100$~nm deposited at each angle at a rate of $0.2$~nm/s.

The lift-off process was performed in heated Remover PG on a hot plate at $80^{\circ}$C for $3$-$4$~h, followed by $3$~min high-power sonication in the same bath. The sample was then cleaned by $3$~min sonications in ACE and in IPA, and dried under nitrogen flow. For dicing protection, the die was coated with AZ-5214E and cut on a Disco DAD3220; post-dicing, it was cleaned again in ACE and IPA under sonication before proceeding to packaging.

\section{\hspace{-9pt}: Experimental setup}\label{supp_note:meas_setup}

\begin{figure*}[ht]
\includegraphics[width=0.65\linewidth]{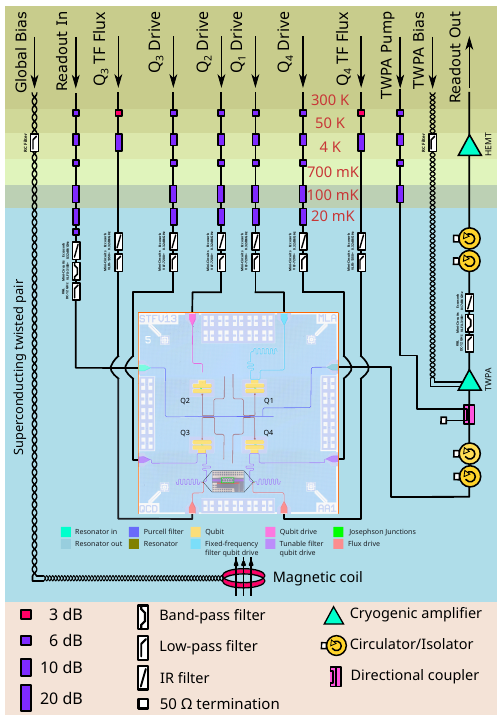}
\centering
\caption{\textbf{Cryogenic wiring and measurement setup.} Schematic of the dilution-refrigerator wiring for the qubit drive, readout, and fast-flux lines of the four-qubit device (Q1--Q4), including the distribution of cryogenic attenuation and filtering. The readout output chain includes circulators/isolators, a traveling-wave parametric amplifier (TWPA) at $20$~mK, and a HEMT amplifier at $4$~K.
}
\label{fig:exp_setup}
\end{figure*}

Following fabrication, the chip was wire-bonded into an SMA-8 sample holder and enclosed within a magnetic shield. The assembly was mounted on the mixing-chamber plate of a Bluefors XLD500 dilution refrigerator, which stabilized between $18$--$24$~mK during the measurements. An overview of the cryogenic wiring and measurement setup is shown in Supplementary Fig.~\ref{fig:exp_setup}.

We used four independent microwave drive lines for qubits Q1--Q4 and a single readout input line. Each qubit drive line included $62$~dB of cryogenic attenuation distributed across temperature stages and a low-pass filter (DC--$7.2$~GHz, Mini-Circuits VLF-7200+). The readout input line included $68$~dB total attenuation and a band-pass filter ($3.4$--$9.9$~GHz, Mini-Circuits VLF-3100+) together with a distributed-coaxial low-pass filter (DC--$12$~GHz, K\&L Microwave 6L250-12000). To suppress high-frequency radiation, Eccosorb infrared (IR) filters were installed on all inputs ($0.32$~dB/GHz) and on the readout output line ($0.17$~dB/GHz).

DC biases for the global coil and the traveling-wave parametric amplifier (TWPA) were delivered via a superconducting twisted pair with an RC low-pass filter at $4$~K. On the readout output line, a TWPA was mounted at the $20$~mK stage and operated in a three-wave-mixing configuration. The weak readout signal from the device passed through two cryogenic circulators (LNF CIC 4-12 A) and a directional coupler (Krytar, $4$--$12.4$~GHz). A strong TWPA pump tone was injected into the coupler’s $-20$~dB port and combined with the readout signal at the coupler output port before entering the TWPA. The amplified signal was then routed through low-pass (DC--$12$~GHz, K\&L Microwave 6L250-12000), band-pass ($3.4$--$9.9$~GHz, Mini-Circuits VLF-3100+), and IR ($0.17$~dB/GHz) filters. To suppress back-propagating thermal noise from higher-temperature stages, two isolators were placed at $20$~mK on the readout output line. Final cryogenic gain was provided by a high-electron-mobility transistor (HEMT) amplifier at $4$~K, followed by room-temperature amplification.

At room temperature, time-domain control and acquisition were provided by a Presto microwave platform (Intermodulation Products) with $16$ phase-coherent channels spanning DC to $10$~GHz (via DC-coupled ports), used to generate qubit-drive pulses, readout pulses with in-phase/quadrature demodulation, and fast-flux waveforms. Continuous-wave (CW) measurements used a vector network analyzer (Rohde \& Schwarz) to characterize the Purcell filter and readout resonators. For CW two-tone spectroscopy, the drive tone was generated with an SGMA SGS100A RF signal generator (Rohde \& Schwarz). The TWPA pump was supplied by an E8257D PSG microwave analog signal generator (Agilent/Keysight). Device DC biases were provided by a low-noise SIM900 source (Stanford Research Systems).

\section{\hspace{-9pt}: Device characterization and parameters}\label{supp_note:device_char}

The basic device parameters are summarized in Supplementary Table~\ref{tab:device_char}.

{\renewcommand{\arraystretch}{1}
\begin{table*}[ht]
\caption{Summary of measured device parameters.}
\label{tab:device_char}
\centering
\begin{tabular}{|p{4.2cm}|p{2.1cm}|p{1cm}|p{1cm}|p{1cm}|p{1cm}|}
\hline
\textbf{Parameters } & \textbf{Symbols} & \textbf{Q1} & \textbf{Q2} & \textbf{Q3} & \textbf{Q4}\\
\hline
Resonator frequency & \ensuremath{\mathrm{\omega}_\mathrm{r}/2\pi} (GHz) & 7.4920 & 7.6786 & 7.4551 & 7.1174\\
Resonator linewidth &  \ensuremath{\mathrm{\kappa}/2\pi} (MHz) & 1.863 & 1.158 & 2.521 & 5.672\\
Qubit frequency & \ensuremath{\omega_\mathrm{q}/2\pi} (GHz) & 5.4292 & 4.3231 & 3.9514 & 3.9003\\
Qubit-Resonator Coupling &  \ensuremath{g/2\pi} (MHz) & 100.7 & 109.1 & 107.2 & 109.8\\
Qubit anharmonicity &  \ensuremath{\alpha/2\pi} (MHz) & $-129.7$ & $-132.1$ & $-134.6$ & $-134.9$\\
Dispersive shift & \ensuremath{2\chi/2\pi} (MHz) & $-0.66$ & $-0.29$ & $-0.27$ & $-0.33$\\
\hline
\end{tabular}
\end{table*}
}

\subsection*{Filter characterization}\label{supp_note:filter_char}

\begin{figure*}[ht]
  \centering
  \includegraphics[width=\linewidth]{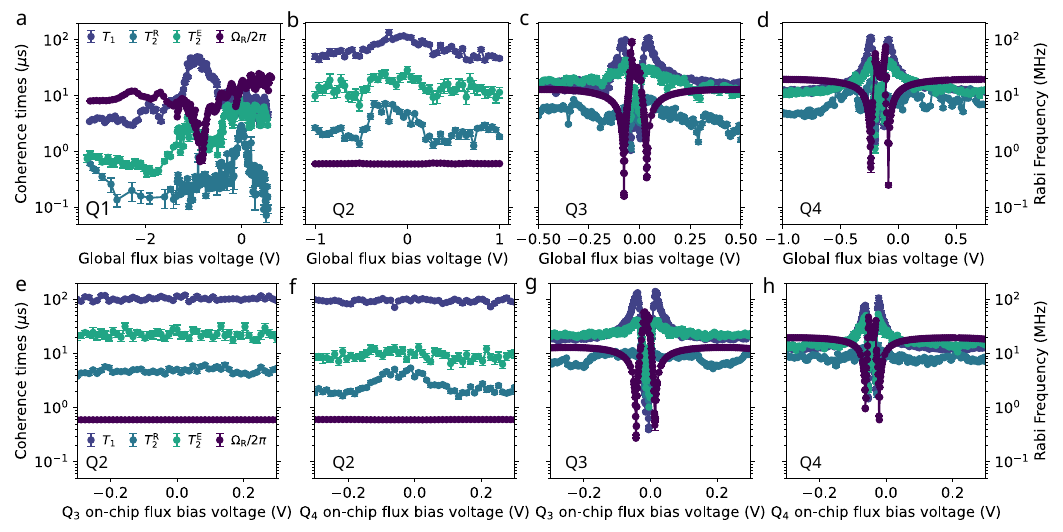}
  \caption{
  \textbf{Global- and local-flux characterization of the drive lines and filters.}
  \textbf{a--d}, Global-coil sweeps. For each qubit we extract the Rabi rate $\Omega/2\pi$ (right axis) and coherence times $T_1$, $T_2^{\mathrm{R}}$ (Ramsey), and $T_2^{\mathrm{E}}$ (Hahn echo) (left axis) as a function of global bias. Panel \textbf{a} shows Q1 with a fixed-frequency drive filter and a flux-tunable qubit; panels \textbf{b--d} show Q2--Q4, with a standard weakly coupled drive line for Q2 and tunable-frequency drive-line filters for Q3 and Q4. Qubit and readout-resonator frequencies are tracked during the global sweeps. \textbf{e--h}, Fast-flux sweeps with the global coil fixed at zero bias. Panels \textbf{e} and \textbf{f} show Q2 while sweeping the fast-flux line of Q3 and Q4, respectively; panels \textbf{g} and \textbf{h} show Q3 and Q4 under their respective fast-flux sweeps. Markers denote fit results and error bars indicate $1\sigma$ fit uncertainty.}
  \label{fig:filter_char_supp}
\end{figure*}

We characterize each drive line and filter by measuring the Rabi rate and the coherence times $T_1$, $T_2^{\mathrm{R}}$, and $T_2^{\mathrm{E}}$ as the applied flux bias is swept. Supplementary Fig.~\ref{fig:filter_char_supp}a--d summarizes measurements as a function of global-coil bias. Q1 is paired with a fixed-frequency drive filter and is therefore implemented as a flux-tunable qubit to match the fixed stopband. Q2--Q4 are fixed-frequency qubits, with Q2 coupled to a standard weakly coupled drive line and Q3 and Q4 coupled to tunable-frequency drive-line filters.

Supplementary Fig.~\ref{fig:filter_char_supp}e--h shows the corresponding characterization as a function of on-chip fast-flux bias with the global coil held fixed. Panels e and f use Q2 as a reference while sweeping the fast-flux lines local to Q3 and Q4. Panels g and h show the response of Q3 and Q4 under their respective fast-flux sweeps. The agreement between global and local tuning confirms that the intended filter tuning can be achieved with modest local bias.

Two trends are most relevant for operation. First, the tunable-frequency drive-line filters allow the qubits to remain at their flux-insensitive bias while the line environment is tuned, so that maxima in $T_1$ and $T_2$ coincide at the idle point. Second, sweeping the filter bias produces order-of-magnitude changes in both $\Omega/2\pi$ and $T_1$, enabling a low-coupling idle configuration and a higher-coupling configuration used for gates and dissipative reset. We also observe a common $T_1$ ceiling near $0.2~\mathrm{ms}$ for the fixed-frequency qubits, consistent with an internal loss channel that dominates over the extrinsic limit in this band. Under global bias we see a weak influence of the Q4 filter tuning on Q2, which is absent when biasing Q3 locally and reappears when biasing Q4 locally, consistent with a coupling path specific to Q4.

\begin{figure*}[ht]
  \centering
  \includegraphics[width=\linewidth]{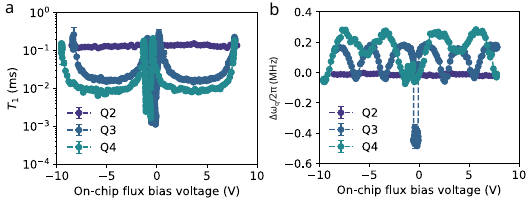}
  \caption{\textbf{Flux-tuned relaxation and change in qubit frequency.} \textbf{a}, Energy relaxation time $T_1$ as a function of on-chip flux bias for the three fixed-frequency qubits during the third cooldown. \textbf{b}, Change in qubit frequency $\Delta\omega_\mathrm{q}/2\pi$ as a function of the same bias. Q2 (no tunable element) is flat and serves as a reference, whereas Q3 and Q4 (coupled to the tunable environment) show small oscillatory shifts. The bias dependence is consistent with an environment-induced frequency shift: tuning the stopband changes the complex admittance seen by the qubit, linking $T_1$ and the change in qubit frequency.}
  \label{fig:t1_ramsey_freq_supp}
\end{figure*}

We also measure the change in qubit frequency while tuning the qubit--filter coupling with on-chip flux bias. In this cooldown we observed our best energy relaxation, with Q3 reaching $T_1 \approx 0.3~\mathrm{ms}$ as shown in Supplementary Fig.~\ref{fig:t1_ramsey_freq_supp}a. Q2 is coupled to a standard weakly coupled drive line and serves as a reference under the same applied on-chip bias, showing no measurable frequency shift.

In contrast, Q3 and Q4 show a small but reproducible bias-dependent shift of the Ramsey frequency across the tuning range, shown in Supplementary Fig.~\ref{fig:t1_ramsey_freq_supp}b. The shift is consistent with an environment-induced frequency shift from the flux-tuned stopband. Changing flux modifies the complex admittance $Y_{\mathrm{env}}(\omega_\text{q},\Phi)$ seen through the coupling capacitor, giving
\begin{equation*}
    \Gamma_1(\Phi)\propto \mathrm{Re}\{Y_{\mathrm{env}}(\omega_\text{q},\Phi)\},\qquad
    \delta f_\text{q}(\Phi)\propto -\,\mathrm{Im}\{Y_{\mathrm{env}}(\omega_\text{q},\Phi)\}.
\end{equation*}
Because $\mathrm{Re}\{Y\}$ and $\mathrm{Im}\{Y\}$ are linked by Kramers--Kronig relations, tuning $T_1$ necessarily produces a small shift in qubit frequency. Over the practical tuning window the frequency shift is at the sub-MHz level and is readily compensated with standard phase updates. Although no such phase updates is required as qubit control is operated at a fixed flux-bias, where the drive frequency is chosen to be in resonance with the qubit frequency at that bias.

\subsection*{Coherence measurements}\label{supp_note:long_coherence}

\begin{figure*}[ht]
  \centering
  \includegraphics[width=\linewidth]{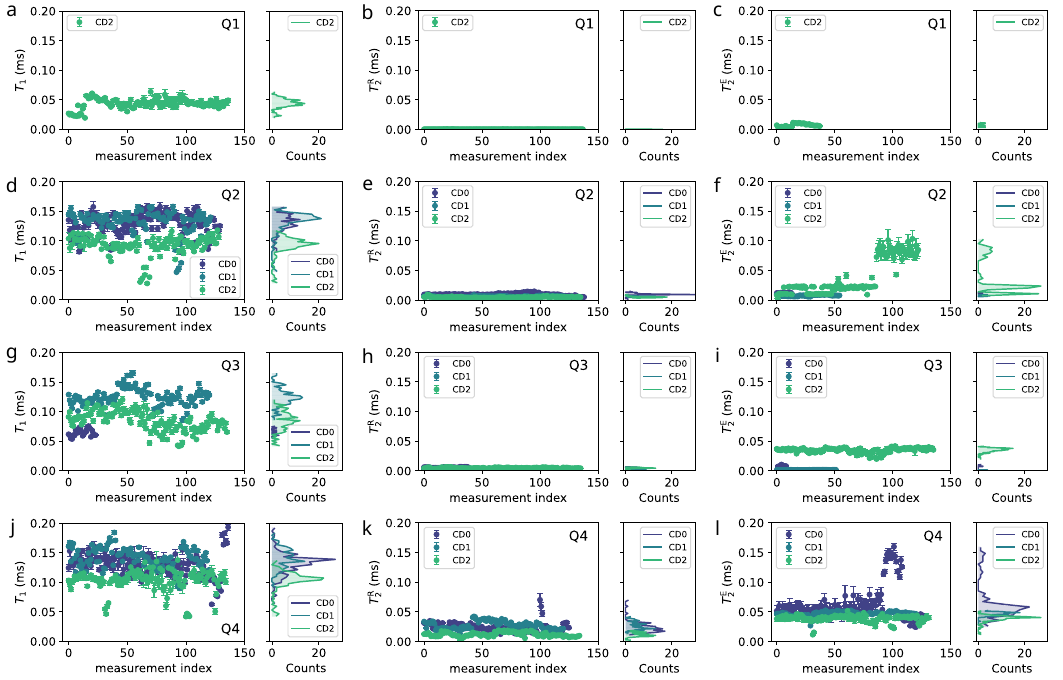}
  \caption{\textbf{Coherence across three cooldowns.}
  Measured $T_1$, $T_2^{\mathrm{R}}$ (Ramsey), and $T_2^{\mathrm{E}}$ (Hahn echo) for qubits Q1--Q4 across three cooldowns.
  Panels \textbf{a--c} show Q1, \textbf{d--f} Q2, \textbf{g--i} Q3, and \textbf{j--l} Q4.
  In each panel, the left subpanel shows individual coherence measurements as a function of acquisition index spanning approximately three months, and the right subpanel shows the corresponding histograms grouped by cooldown.
  Each data point is averaged over five repeats prior to fitting; error bars indicate $1\sigma$ fit uncertainty.
  }
  \label{fig:long_coherence_supp}
\end{figure*}

We track $T_1$, $T_2^{\mathrm{R}}$, and $T_2^{\mathrm{E}}$ for all four qubits over three cooldowns spanning approximately three months, as summarized in Supplementary Fig.~\ref{fig:long_coherence_supp}. The acquisition index labels consecutive measurements taken over this period. Q1, implemented as a flux-tunable qubit to match a fixed-frequency drive filter, exhibits slightly lower $T_1$ than the fixed-frequency qubits. All fixed-frequency qubits maintain $T_1>0.1~\mathrm{ms}$ across cooldowns. Ramsey coherence times are generally shorter for Q1--Q3, whereas Q4 exhibits consistently higher $T_2^{\mathrm{R}}$. Echo coherence times vary more strongly over time, indicating sensitivity to slow fluctuations in the dephasing environment.

\subsection*{Dynamical decoupling measurements}\label{supp_note:dynamical_decoupling}

\begin{figure*}[ht]
  \centering
  \includegraphics[width=\linewidth]{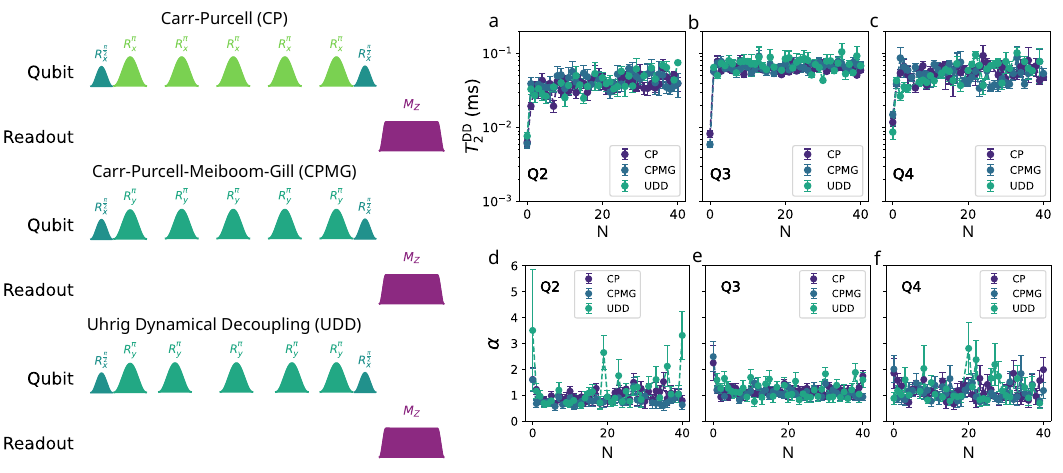}
  \caption{\textbf{Dynamical decoupling coherence and noise characterization.}
  Panels \textbf{a--c} show the extracted dynamical-decoupling coherence time $T_2^{\mathrm{DD}}$ as a function of the number of refocusing pulses $N$ for Carr--Purcell (CP), Carr--Purcell--Meiboom--Gill (CPMG), and Uhrig dynamical decoupling (UDD) sequences applied to qubits Q2, Q3, and Q4.
  Panels \textbf{d--f} show the corresponding stretch exponent $\alpha$ obtained from stretched-exponential fits.
  Error bars indicate $1\sigma$ fit uncertainty.
  }
  \label{fig:dynamical_decoup_supp}
\end{figure*}

To further probe the spectral content of dephasing noise, we perform dynamical decoupling (DD) measurements using CP, CPMG, and UDD pulse sequences. These sequences differ in the timing and axis of the refocusing $\pi$ pulses and therefore implement distinct noise filter functions that suppress low-frequency dephasing to varying degrees~\cite{Meiboom1958ModifiedTimes,Cywinski2008HowQubits,Biercuk2011DynamicalProblem,Uhrig2007KeepingSequences,Yang2008UniversalityRelaxation}.

We extract the coherence under DD by fitting the normalized coherence signal to a stretched-exponential envelope,
\begin{align}
W(t)=\frac{\rho_{01}(t)}{\rho_{01}(0)}
\simeq \exp\!\left[-\left(\frac{t}{T_2^{\mathrm{DD}}}\right)^{\alpha}\right],
\end{align}
where $T_2^{\mathrm{DD}}$ is an effective coherence time and $\alpha$ is a stretch exponent~\cite{Cywinski2008HowQubits,Biercuk2011DynamicalProblem,Bylander2011NoiseQubit,Medford2012ScalingQubits}. In this description, $\alpha$ provides qualitative information about the dominant noise after decoupling: $\alpha\approx1$ corresponds to nearly exponential decay associated with fast or Markovian noise, whereas $\alpha>1$ indicates increasing suppression of slow noise components.

Supplementary Fig.~\ref{fig:dynamical_decoup_supp}a--c show $T_2^{\mathrm{DD}}$ as a function of pulse number $N$ for qubits Q2, Q3, and Q4. In all cases, $T_2^{\mathrm{DD}}$ increases with $N$, consistent with the suppression of low-frequency dephasing noise. The corresponding stretch exponents, shown in panels d--f, are typically greater than unity, indicating that after dynamical decoupling the residual dephasing is dominated by faster noise components rather than quasi-static detuning noise.

\section{\hspace{-9pt}: Qubit control}\label{supp_note:qubit_control}

\begin{figure*}[ht]
  \centering
  \includegraphics[width=0.8\linewidth]{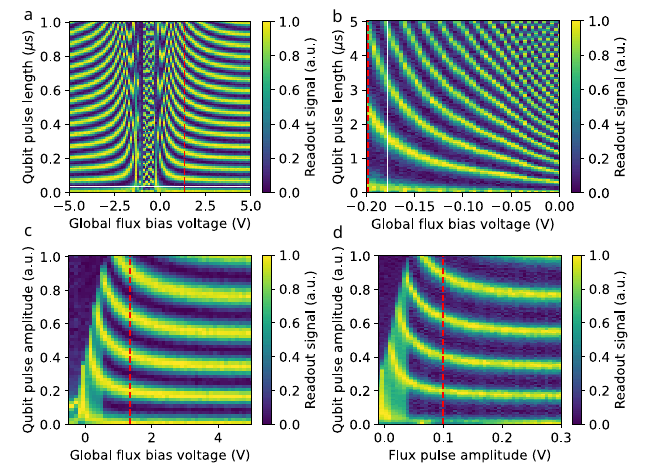}
  \caption{
  \textbf{Calibration of single-qubit drive conditions for randomized benchmarking (RB).} \textbf{a}, Two-dimensional (2D) Rabi oscillations of qubit Q4 as a function of global-coil bias voltage and pulse length. Each trace is taken at a fixed global bias and shows the excited-state probability as a function of drive duration; faster oscillations indicate stronger effective qubit--drive coupling. \textbf{b}, Fine sweep of Rabi oscillations near the filter-frequency bias of the tunable-frequency drive filter. As the global bias approaches the idle point (filter tuned to the qubit), the oscillations slow strongly, indicating suppressed coupling and a long $T_1$. We use these data to choose $1.35~\text{V}$ as the operating point for the \emph{static dialing} protocol, and $-0.2~\text{V}$ as the idle bias for the \emph{dynamic dialing} protocol. \textbf{c}, Amplitude-calibrated Rabi map at fixed pulse length using shaped $\sin^2$ envelopes with DRAG correction, as used in RB. Here we sweep the drive amplitude rather than the pulse length. Near half-flux quantum, where the SQUID-array junction critical current is reduced, the oscillations are suppressed at high drive, consistent with nonlinear response of the array. To avoid this regime, we select $1.35~\text{V}$ for static dialing. \textbf{d}, Rabi calibration for dynamic dialing. The global coil is held at $-0.2~\text{V}$ (idle point, long $T_1$), and we simultaneously apply a fast-flux pulse and a microwave drive pulse. We sweep both fast-flux amplitude and drive amplitude to extract the effective Rabi frequency. From this map we choose a fast-flux amplitude of $0.1~\text{V}$, which yields a Rabi frequency of $12.5~\text{MHz}$ for a $40~\text{ns}$ $\pi$-pulse at about $-110~\text{dBm}$ drive power at the sample. These calibrated parameters are used for the RB experiments in the main text.
  }
  \label{fig:RB_Rabi_supp}
\end{figure*}

We benchmark single-qubit gates using two drive configurations: \emph{static dialing} and \emph{dynamic dialing}. In static dialing, the tunable-frequency drive filter is held at a fixed global-coil bias that provides moderate qubit--drive coupling throughout the sequence. In dynamic dialing, the global-coil bias parks the filter at the idle point (suppressed coupling, long $T_1$), and a fast-flux pulse is applied only during each gate to momentarily increase the coupling.

To choose suitable bias points for static and dynamic dialing, we first measure two-dimensional (2D) Rabi oscillations as a function of global-coil bias voltage and qubit pulse length (Supplementary Fig.~\ref{fig:RB_Rabi_supp}a). We also carry out a fine sweep near the filter-frequency bias and observe that the Rabi oscillations slow down strongly there, indicating very weak coupling (Supplementary Fig.~\ref{fig:RB_Rabi_supp}b). From these data, we select a global-coil bias of $1.35~\text{V}$ for static dialing and $-0.2~\text{V}$ for dynamic dialing. In static dialing we remain at $1.35~\text{V}$ throughout the experiment. In dynamic dialing we remain at $-0.2~\text{V}$ (idle point) and use a fast-flux pulse to temporarily pull the filter away from the idle point only during the gate.

The measurements in Supplementary Fig.~\ref{fig:RB_Rabi_supp}a,b use rectangular pulses with finite rise and fall ramps. For randomized benchmarking (RB) we use shaped $\sin^2$ envelopes with DRAG correction, so we repeat the calibration by sweeping the pulse amplitude at fixed pulse length (Supplementary Fig.~\ref{fig:RB_Rabi_supp}c). Near half-flux quantum, where the SQUID-array junctions have reduced critical current, the oscillations are strongly suppressed at high drive, consistent with nonlinear response under large microwave current. To avoid this regime, we keep the static-dialing operating point at $1.35~\text{V}$.

For dynamic dialing, we hold the global-coil bias at $-0.2~\text{V}$ (idle point) and apply a fast-flux pulse together with the microwave drive pulse. We sweep both fast-flux amplitude and drive amplitude to map the effective Rabi frequency (Supplementary Fig.~\ref{fig:RB_Rabi_supp}d). From this map, we choose a fast-flux pulse amplitude of $0.1~\text{V}$, yielding a Rabi frequency of about $12.5~\text{MHz}$ for a $40~\text{ns}$ $\pi$-pulse while keeping the required drive power at the sample near $-110~\text{dBm}$. We use the same $40~\text{ns}$ gate length and similar drive power targets for both static and dynamic dialing in the RB experiments.

\begin{figure*}[ht]
  \centering
  \includegraphics[width=0.95\linewidth]{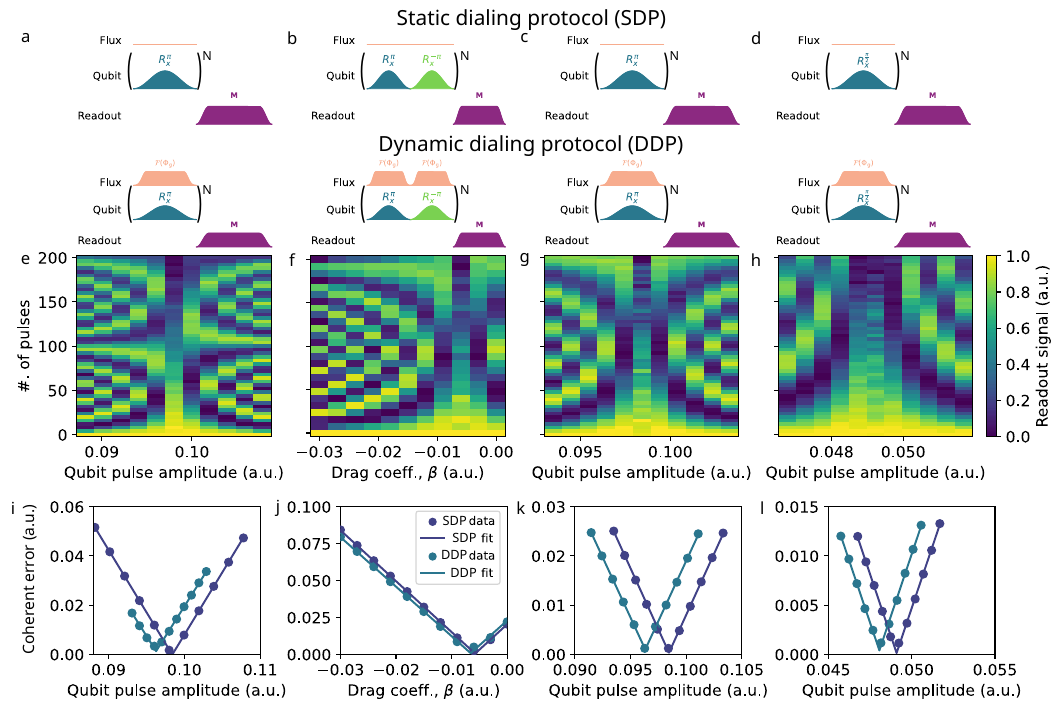}
  \caption{\textbf{Calibration of single-qubit pulse parameters using error amplification techniques.} \textbf{a}, Amplitude-calibration sequence. We apply an even number $N$ of nominal $\pi$-pulses about $X$, $(R_\mathrm{X}^{\pi})^N$ with $N=2,4,6,\ldots$, and sweep both $N$ and the pulse amplitude. In the ideal case, each pair of $\pi$-pulses returns the qubit to $\ket{\text{g}}$, so after any even $N$ the qubit should be in the ground state. A small angle error accumulates coherently with repetition and produces oscillations as a function of $N$. \textbf{b}, DRAG / quadrature calibration. We apply an even number of alternating $\pi$- and $-\pi$-rotations, $(R_\mathrm{X}^{\pi} R_\mathrm{X}^{-\pi})^{N/2}$, while sweeping both $N$ and the DRAG coefficient $\beta$ that sets the $Q$-quadrature correction. \textbf{c,d}, Final calibration sequences for $R_\mathrm{X}^{\pi}$ and $R_\mathrm{X}^{\pi/2}$ after DRAG is set. \textbf{e--h}, Example 2D calibration data for static dialing. \textbf{i--l}, Extraction of optimal parameters for static and dynamic dialing by fitting the oscillation frequency as a function of the swept pulse parameter; the minimum defines the calibrated pulse amplitudes and DRAG coefficient $\beta$. These calibrated parameters are used in the randomized benchmarking experiments.
  }
  \label{fig:RB_calib_supp}
\end{figure*}

After choosing the operating points of the tunable-frequency drive filter (global-coil bias, fast-flux pulse amplitude, gate length of $40~\text{ns}$, and drive amplitude), we run a pulse-parameter optimization routine to further calibrate the single-qubit gates. The goal is to minimize coherent control errors from (i) amplitude miscalibration, (ii) ac-Stark shifts and residual $Z$-phase accumulation during the pulse, (iii) leakage out of the computational subspace, and (iv) drive-line non-idealities such as quadrature imbalance. We use standard error amplification sequences in which a small systematic error per pulse is repeated many times so that it becomes measurable~\cite{Motzoi2009SimpleQubits,Magesan2012EfficientBenchmarking,Sheldon2016ProcedureGate,Ding2023High-FidelityCoupler}. Throughout these measurements we always use an even number of $\pi$-pulses so that, in the ideal case, the final state is $\ket{\text{g}}$.

We perform these calibrations for both single-qubit drive configurations: static dialing and dynamic dialing. The procedure is as follows.

\textbf{Amplitude calibration.} We apply an even number $N$ of nominal $\pi$-pulses about $X$,
\begin{equation*}
    (R_\mathrm{X}^{\pi})^N,\quad N = 2,4,6,\ldots,
\end{equation*}
and sweep both the pulse amplitude and $N$ (Supplementary Fig.~\ref{fig:RB_calib_supp}a). A small amplitude miscalibration by $\epsilon$ (each pulse is $R_\mathrm{X}^{\pi+\epsilon}$) accumulates coherently and produces oscillations in the measured population as a function of $N$. We fit the signal to a decaying cosine and take the fitted oscillation frequency as a coherent-error metric~\cite{Sheldon2016ProcedureGate}. An example 2D scan for static dialing is shown in Supplementary Fig.~\ref{fig:RB_calib_supp}e, and the corresponding extraction and optimum in Supplementary Fig.~\ref{fig:RB_calib_supp}i.

\textbf{DRAG / quadrature calibration.} We tune the DRAG correction to suppress ac-Stark shifts, leakage, and unwanted phase accumulation. We apply an even number of alternating $\pi$- and $-\pi$-pulses,
\begin{equation*}
    (R_\mathrm{X}^{\pi} R_\mathrm{X}^{-\pi})^{N/2},\quad N = 2,4,6,\ldots,
\end{equation*}
and sweep both $N$ and the DRAG coefficient $\beta$, which sets the $Q$-quadrature component $Q(t)=\beta\,\dot{I}(t)$ for a shaped $\sin^2$ envelope~\cite{Motzoi2009SimpleQubits,Gambetta2011AnalyticOscillator}. Example data (static dialing) appear in Supplementary Fig.~\ref{fig:RB_calib_supp}f, and the extracted optima for static and dynamic dialing in Supplementary Fig.~\ref{fig:RB_calib_supp}j.

\textbf{Final calibration of $R_\mathrm{X}^{\pi}$ and $R_\mathrm{X}^{\pi/2}$.} After setting DRAG, we repeat the amplitude calibration to update both the $\pi$- and $\pi/2$-pulse amplitudes. The pulse sequences are summarized in Supplementary Fig.~\ref{fig:RB_calib_supp}c,d. Example scans and fits are shown in Supplementary Fig.~\ref{fig:RB_calib_supp}g,h and the resulting optima in Supplementary Fig.~\ref{fig:RB_calib_supp}k,l.

In all steps, the analysis is two-stage: (i) for each trial parameter we fit the measured signal as a function of even $N$ to a decaying cosine and extract the oscillation frequency; (ii) we fit that extracted metric as a function of the swept parameter with a smooth curve with a single minimum. The minimum defines the calibrated setting. This procedure yields the final pulse amplitudes, DRAG coefficient, and $40~\text{ns}$ gate length used in the randomized benchmarking experiments for both static and dynamic dialing.

\begin{figure*}[ht]
  \centering
  \includegraphics[width=\linewidth]{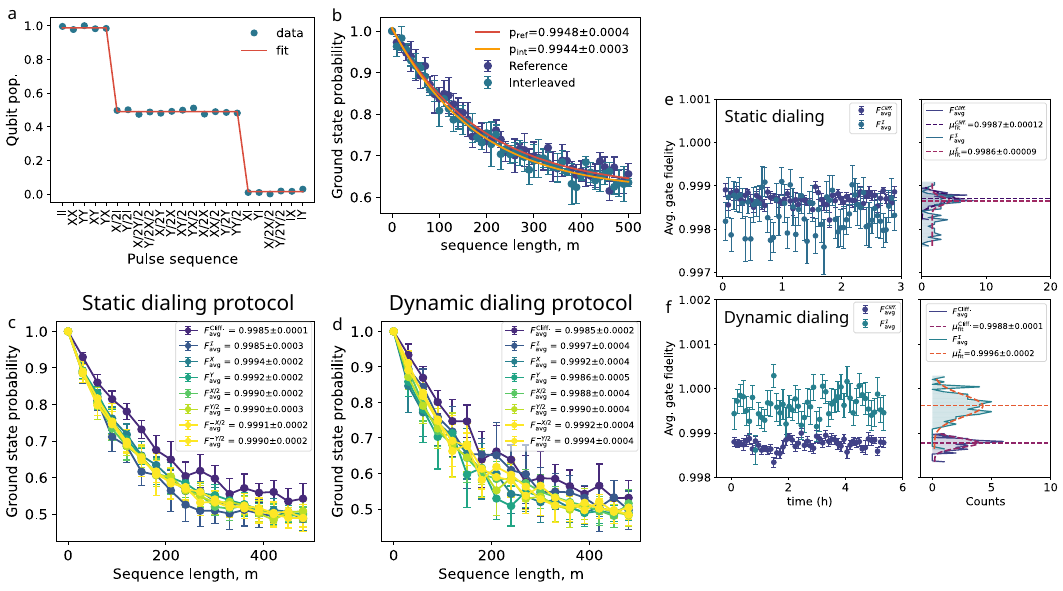}
  \caption{\textbf{Randomized benchmarking (RB) and interleaved RB results.}
  \textbf{a}, ALLXY-style validation of the calibrated single-qubit pulse set.
  \textbf{b}, Example reference RB and interleaved RB for dynamic dialing. We fit $P_\mathrm{g}(m)=A p^m + B$ to extract $p_\mathrm{ref}$ and $p_\mathrm{int}$ and compute average fidelity per Clifford operation $F_\mathrm{avg}^\mathrm{Cliff.}$ and the interleaved-gate fidelity $F_\mathrm{avg}^G$ using standard RB analysis. \textbf{c,d}, Summary of extracted fidelities for all calibrated single-qubit Clifford-equivalent operations under static dialing and dynamic dialing. \textbf{e,f}, Stability measurements over many hours: extracted fidelities as a function of time (left) and histograms with Gaussian fits (right).}
  \label{fig:RB_results_supp}
\end{figure*}

We assess single-qubit gate performance using standard randomized benchmarking (RB) and interleaved RB~\cite{Magesan2012EfficientBenchmarking,Magesan2012CharacterizingBenchmarking,Barends2014SuperconductingTolerance,Sheldon2016ProcedureGate}. Before running RB, we validate the pulse set using an ALLXY-style sequence.

We define a set of calibrated physical pulses
\begin{equation*}
    \{ I,\ R_\mathrm{X}^{\pi},\ R_\mathrm{Y}^{\pi},\ R_\mathrm{X}^{\pi/2},\ R_\mathrm{Y}^{\pi/2},\ R_\mathrm{X}^{-\pi/2},\ R_\mathrm{Y}^{-\pi/2} \},
\end{equation*}
which we also refer to informally as $\{ I, X, Y, X/2, Y/2, -X/2, -Y/2 \}$.

After completing the amplitude and DRAG calibrations, we run an ALLXY check and confirm the expected staircase pattern for both static and dynamic dialing (Supplementary Fig.~\ref{fig:RB_results_supp}a).

We then run RB. For each drive configuration we generate random sequences of Cliffords from the single-qubit Clifford group (24 elements)~\cite{Barends2014SuperconductingTolerance}, compile each Clifford into the physical pulse set above, and append a recovery Clifford that ideally returns the qubit to $\ket{\text{g}}$. The ground-state survival is fit to
\begin{equation*}
    P_\mathrm{g}(m) = A\, p_\mathrm{ref}^m + B,
\end{equation*}
where $p_\mathrm{ref}$ is the reference RB decay parameter and $A,B$ absorb state-preparation and measurement effects~\cite{Magesan2012CharacterizingBenchmarking,Magesan2012EfficientBenchmarking}.

For interleaved RB, we insert a target gate $G$ after every Clifford and fit the decay to
\begin{equation*}
    P_\mathrm{g}^\mathrm{(int)}(m) = A'\, p_\mathrm{int}^m + B'.
\end{equation*}
From $p_\mathrm{ref}$ we compute
\begin{equation*}
    F_\mathrm{avg}^\mathrm{Cliff.} = 1 - \frac{1}{2k}(1 - p_\mathrm{ref}),
\end{equation*}
where $k = 45/24 \approx 1.875$ is the average number of physical pulses per Clifford in our compilation~\cite{Barends2014SuperconductingTolerance}, and from $p_\mathrm{ref}$ and $p_\mathrm{int}$ we compute
\begin{equation*}
    F_\mathrm{avg}^G = 1 - \frac{1}{2}\left(1 - \frac{p_\mathrm{int}}{p_\mathrm{ref}}\right),
\end{equation*}
following standard RB analysis~\cite{Magesan2012EfficientBenchmarking,Magesan2012CharacterizingBenchmarking}.

An example reference and interleaved RB for dynamic dialing is shown in Supplementary Fig.~\ref{fig:RB_results_supp}b, and the summarized results for static and dynamic dialing are shown in Supplementary Fig.~\ref{fig:RB_results_supp}c,d. The interleaved idle fidelity is higher for dynamic dialing, consistent with the qubit spending most of its time at the decoupled (long-$T_1$) idle point and only pulsing to stronger coupling during gates. Stability data acquired overnight are shown in Supplementary Fig.~\ref{fig:RB_results_supp}e,f.

\begin{figure*}[ht]
  \centering
  \includegraphics[width=\linewidth]{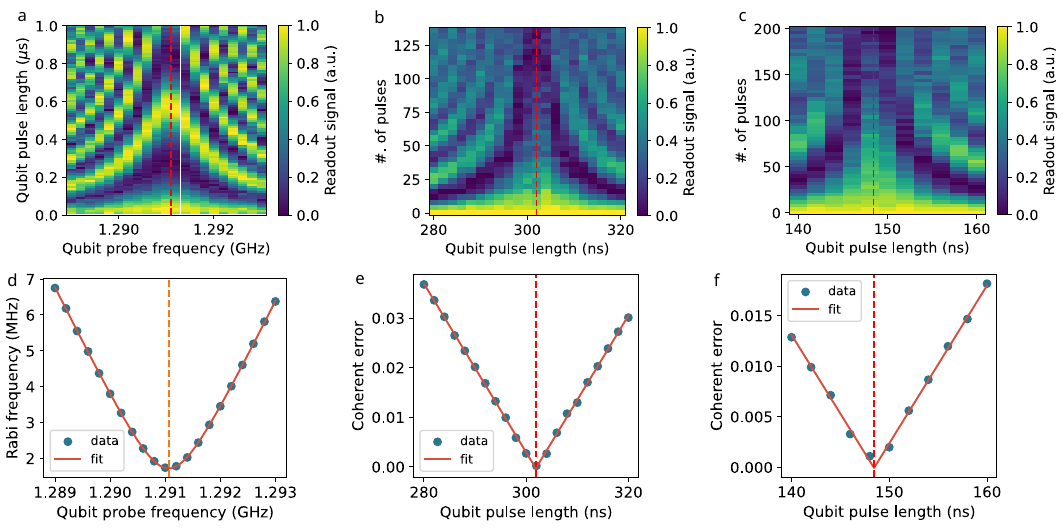}
  \caption{\textbf{Subharmonic driving of qubit Q4.} \textbf{a}, Two-dimensional Rabi oscillations as functions of qubit probe frequency and qubit-pulse length for Q4 with resonance frequency $\sim 3.9~\text{GHz}$. \textbf{b,c}, Pulse-length calibration sequences for the $\pi$-pulse in \textbf{b} and the $\pi/2$-pulse in \textbf{c}. \textbf{d--f}, Corresponding fits to the data in \textbf{a--c}, yielding the Rabi frequency as a function of qubit probe frequency in \textbf{d} and the coherent error as a function of pulse length for the $\pi$-pulse in \textbf{e} and the $\pi/2$-pulse in \textbf{f}.}
  \label{fig:Q4_subharmonic_supp}
\end{figure*}

We also demonstrate qubit control using subharmonic driving via a three-photon process at $f_\text{q}/3$. Supplementary Fig.~\ref{fig:Q4_subharmonic_supp}a shows a two-dimensional map of Rabi oscillations as functions of qubit probe frequency and qubit-pulse length for qubit Q4, which has a resonance frequency of $\sim 3.9~\text{GHz}$. We then perform pulse-length calibration sequences to determine the optimal $\pi$- and $\pi/2$-pulse durations, shown in Supplementary Fig.~\ref{fig:Q4_subharmonic_supp}b,c. The corresponding fits, displayed in Supplementary Fig.~\ref{fig:Q4_subharmonic_supp}d--f, give the Rabi frequency as a function of qubit probe frequency and the coherent error as a function of pulse length for the $\pi$- and $\pi/2$-pulses.

\section{\hspace{-9pt}: Multilevel qubit characterization}\label{supp_note:multilevel_char}

\begin{figure*}[ht]
  \centering
  \includegraphics[width=\linewidth]{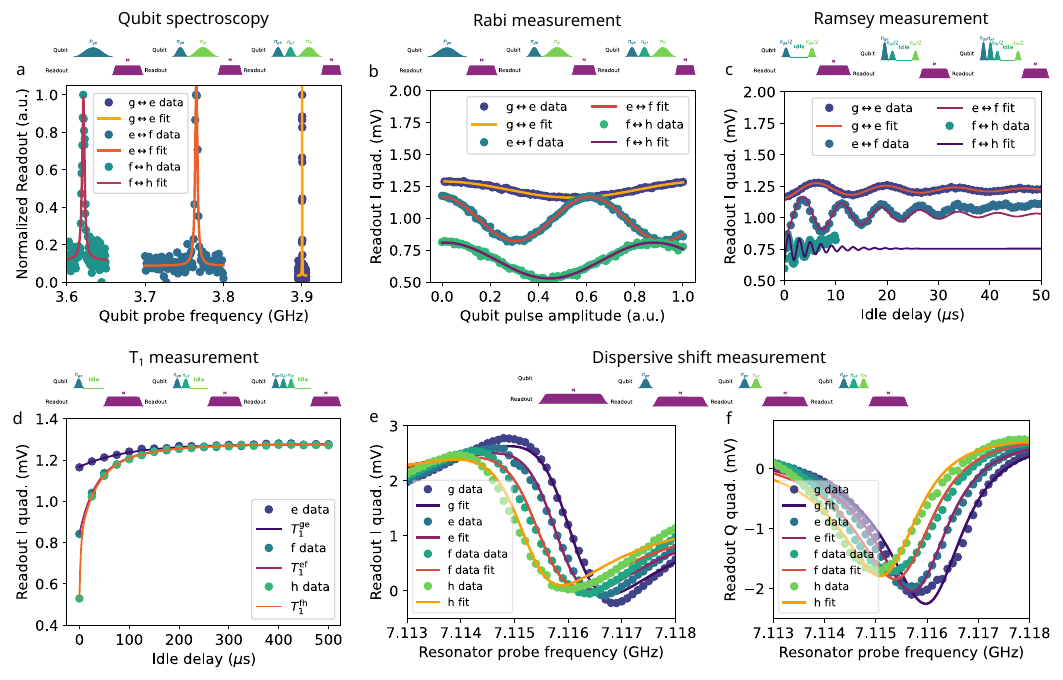}
  \caption{\textbf{Multilevel characterization of qubit Q4.} Each panel shows the pulse sequence (top) and representative data with fits (bottom). \textbf{a}, Qubit spectroscopy of adjacent transitions by sweeping a weak probe tone to locate $f_\mathrm{ge}$, $f_\mathrm{ef}$, and higher transitions. \textbf{b}, Rabi oscillations on the identified transition, used to calibrate $\pi$- and $\pi/2$-pulses for that transition. \textbf{c}, Ramsey measurement on the same transition, yielding $T_2^\mathrm{R}$ and the residual detuning. \textbf{d}, Energy relaxation measurement, yielding the transition lifetime. \textbf{e,f}, Readout calibration by preparing $\ket{\mathrm{g}}$, $\ket{\mathrm{e}}$, $\ket{\mathrm{f}}$, or $\ket{\mathrm{h}}$ and recording the corresponding IQ response to determine state-dependent dispersive shifts and single-shot reference clusters. The characterization proceeds up the ladder, using the lower transition to prepare the next excited state.}
  \label{fig:Q4_multilevel_char_supp}
\end{figure*}

Dissipative reset and thermometry require controlled preparation and readout of multiple transmon levels beyond $\ket{\mathrm{g}}$ and $\ket{\mathrm{e}}$. We therefore characterize adjacent transitions up the ladder, extracting (i) transition frequencies, (ii) Rabi calibrations for selective state preparation, (iii) relaxation times, and (iv) state-dependent dispersive responses needed for multi-level single-shot classification. Supplementary Fig.~\ref{fig:Q4_multilevel_char_supp} shows the procedure for qubit Q4.

We first identify each adjacent transition by spectroscopy (Supplementary Fig.~\ref{fig:Q4_multilevel_char_supp}a) and then calibrate resonant control pulses using Rabi oscillations (Supplementary Fig.~\ref{fig:Q4_multilevel_char_supp}b). We extract dephasing times from Ramsey measurements (Supplementary Fig.~\ref{fig:Q4_multilevel_char_supp}c) and relaxation times from population decay measurements (Supplementary Fig.~\ref{fig:Q4_multilevel_char_supp}d). Finally, we calibrate the dispersive response of the readout for $\ket{\mathrm{g}}$, $\ket{\mathrm{e}}$, $\ket{\mathrm{f}}$, and $\ket{\mathrm{h}}$ by preparing each state and recording its IQ response (Supplementary Fig.~\ref{fig:Q4_multilevel_char_supp}e,f), defining the reference clusters used for multi-state classification in the reset and thermometry experiments. This laddered sequence is repeated for each adjacent pair of levels, using the calibrated lower transition to prepare the next excited state.

We perform this multilevel characterization for all qubits. Supplementary Table~\ref{tab:multilevel_device_char} summarizes the extracted parameters for Q2--Q4 (transition frequencies, relaxation and dephasing times, and readout separation). For Q3 we report values at the idle configuration (filter tuned to the qubit), and for Q4 we report values near, but not exactly at, the idle configuration. These parameters are used to prepare higher states prior to reset, to model multilevel decay, and to extract populations across $\ket{\mathrm{g}}$, $\ket{\mathrm{e}}$, $\ket{\mathrm{f}}$, and $\ket{\mathrm{h}}$ in thermometry.

{\renewcommand{\arraystretch}{1}
\begin{table*}[ht]
\caption{Multilevel device parameters.}
\label{tab:multilevel_device_char}
\centering
\begin{tabular}{|p{4cm}|p{2.2cm}|p{1cm}|p{1cm}|p{1cm}|}
\hline
\textbf{Parameters } & \textbf{Symbols}  & \textbf{Q2} & \textbf{Q3} & \textbf{Q4}\\
\hline
g$\leftrightarrow$e transition frequency & \ensuremath{\mathrm{\omega}_\mathrm{ge}/2\pi} (GHz) & 4.3231 & 3.9514 & 3.9003\\
e$\leftrightarrow$f transition frequency & \ensuremath{\mathrm{\omega}_\mathrm{ef}/2\pi} (GHz) & 4.1910 & 3.8167 & 3.7654\\
f$\leftrightarrow$h transition frequency & \ensuremath{\mathrm{\omega}_\mathrm{fh}/2\pi} (GHz) & 4.0509 & 3.6730 & 3.6211\\
g$\leftrightarrow$e Relaxation time & \ensuremath{T_1^\mathrm{ge}} ($\mu\text{s}$) & 109.30 & 144.20 & 98.56\\
e$\leftrightarrow$f Relaxation time & \ensuremath{T_1^\mathrm{ef}} ($\mu\text{s}$) & 72.63 & 41.07 & 29.46\\
f$\leftrightarrow$h Relaxation time & \ensuremath{T_1^\mathrm{fh}} ($\mu\text{s}$) & 35.21 & 17.55 & 2.64\\
\hline
\end{tabular}
\end{table*}
}

Obtaining relaxation times for higher transmon levels requires a more elaborate procedure than the single-exponential fit applied after initializing $\ket{\mathrm{e}}$. In the following, we outline the multilevel rate-equation model used to extract state-dependent relaxation times from the measured dynamics after preparation in $\ket{\mathrm{e}}$, $\ket{\mathrm{f}}$, and $\ket{\mathrm{h}}$.

\subsection*{Rate equations}
We write down the rate equations for downward transition rates $\Gamma_{ij}$ from level $j$ to level $i<j$ for qubit populations $P_i(t)$ with $\sum_i P_i=1$, where $i \in \{\mathrm{g}, \mathrm{e}, \mathrm{f}, \mathrm{h}\}$

\begin{align}
\dot P_\mathrm{g} &= \Gamma_\mathrm{ge}\,P_\mathrm{e} + \Gamma_\mathrm{gf}\,P_\mathrm{f} + \Gamma_\mathrm{gh}\,P_\mathrm{h}, \nonumber\\
\dot P_\mathrm{e} &= -\Gamma_\mathrm{ge}\,P_\mathrm{e} + \Gamma_\mathrm{ef}\,P_\mathrm{f} + \Gamma_\mathrm{eh}\,P_\mathrm{h}, \nonumber\\
\dot P_\mathrm{f} &= -(\Gamma_\mathrm{gf}+\Gamma_\mathrm{ef})\,P_\mathrm{f} + \Gamma_\mathrm{fh}\,P_\mathrm{h}, \label{eq:master}\\
\dot P_\mathrm{h} &= -(\Gamma_\mathrm{gh}+\Gamma_\mathrm{eh}+\Gamma_\mathrm{fh})\,P_\mathrm{h}. \nonumber
\end{align}

We use the shorthand
\[
A_\mathrm{f} = \Gamma_\mathrm{gf}+\Gamma_\mathrm{ef},\qquad
A_\mathrm{h} = \Gamma_\mathrm{gh}+\Gamma_\mathrm{eh}+\Gamma_\mathrm{fh}.
\]

The closed-form solution $P^{j}_\mathrm{g}(t)$ for qubit prepared in states $j \in \{\mathrm{e}, \mathrm{f}, \mathrm{h}\}$ 
\begin{align}
    P^\mathrm{(e)}_\mathrm{g}(t) &= 1 - e^{-\Gamma_\mathrm{ge}t},\nonumber\\
    P^\mathrm{(f)}_\mathrm{g}(t) &= 1 - e^{-A_\mathrm{f}t} - \frac{\Gamma_\mathrm{ef}}{A_\mathrm{f}-\Gamma_\mathrm{ge}}e^\mathrm{-\Gamma_\mathrm{ge}t}\big[1-e^{-(A_\mathrm{f}-\Gamma_\mathrm{ge})t}\big],\nonumber\\
    P^\mathrm{(h)}_\mathrm{g}(t) &= 1 - e^{-A_\mathrm{h}t} - \frac{\Gamma_\mathrm{fh}}{A_\mathrm{h}-A_\mathrm{f}}e^\mathrm{-A_\mathrm{f}t}\big[1-e^{-(A_\mathrm{h}-A_\mathrm{f})t}\big]\label{eq:rate_eqn_soln}\\
    &-\frac{\Gamma_\mathrm{eh}}{A_\mathrm{h}-\Gamma_\mathrm{ge}}e^\mathrm{-\Gamma_\mathrm{ge}t}\big[1-e^{-(A_\mathrm{h}-\Gamma_\mathrm{ge})t}\big]\nonumber\\
    &-\frac{\Gamma_\mathrm{ef}\Gamma_\mathrm{fh}}{A_\mathrm{h}-A_\mathrm{f}}e^\mathrm{-\Gamma_\mathrm{ge}t}\bigg[\frac{1-e^{-(A_\mathrm{f}-\Gamma_\mathrm{ge})t}}{A_\mathrm{f}-\Gamma_\mathrm{ge}}-\frac{1-e^{-(A_\mathrm{h}-\Gamma_\mathrm{ge})t}}{A_\mathrm{h}-\Gamma_\mathrm{ge}}\bigg]\nonumber.
\end{align}

Due to highly suppressed non-sequential decay rates, we set $\Gamma_\mathrm{gf}=\Gamma_\mathrm{gh}=\Gamma_\mathrm{eh}=0$, resulting in $A_\mathrm{f} = \Gamma_\mathrm{ef},\, \text{and}\,A_\mathrm{h} = \Gamma_\mathrm{fh}$. We obtain the closed-form solution for only sequential rates
\begin{align}
    P^\mathrm{(e)}_\mathrm{g}(t) &= 1 - e^{-\Gamma_\mathrm{ge}t},\nonumber\\
    P^\mathrm{(f)}_\mathrm{g}(t) &= 1 - e^{-\Gamma_\mathrm{ef}t} - \frac{\Gamma_\mathrm{ef}}{\Gamma_\mathrm{ef}-\Gamma_\mathrm{ge}}e^\mathrm{-\Gamma_\mathrm{ge}t}\big[1-e^{-(\Gamma_\mathrm{ef}-\Gamma_\mathrm{ge})t}\big],\nonumber\\
    P^\mathrm{(h)}_\mathrm{g}(t) &= 1 - e^{-\Gamma_\mathrm{fh}t} - \frac{\Gamma_\mathrm{fh}}{\Gamma_\mathrm{fh}-\Gamma_\mathrm{ef}}e^\mathrm{-\Gamma_\mathrm{ef}t}\big[1-e^{-(\Gamma_\mathrm{fh}-\Gamma_\mathrm{ef})t}\big]\label{eq:rate_eqn_seq}\\
    &-\frac{\Gamma_\mathrm{ef}\Gamma_\mathrm{fh}}{\Gamma_\mathrm{fh}-\Gamma_\mathrm{ef}}e^\mathrm{-\Gamma_\mathrm{ge}t}\bigg[\frac{1-e^{-(\Gamma_\mathrm{ef}-\Gamma_\mathrm{ge})t}}{\Gamma_\mathrm{ef}-\Gamma_\mathrm{ge}}-\frac{1-e^{-(\Gamma_\mathrm{fh}-\Gamma_\mathrm{ge})t}}{\Gamma_\mathrm{fh}-\Gamma_\mathrm{ge}}\bigg]\nonumber.
\end{align}

We fit the experimental data to only-sequential decay rate model in Eq.~\eqref{eq:rate_eqn_seq} and obtain measured relaxation times $T_1^{ij}=1/\Gamma_{ij}$ shown in Supplementary Table~\ref{tab:multilevel_device_char}.

In our analysis, we treat both averaged readout data and single-shot readout data within the same multilevel rate-equation framework. For the averaged measurements, we record the dispersive readout response of the resonator after preparing the transmon in $\ket{\mathrm{e}}$, $\ket{\mathrm{f}}$, or $\ket{\mathrm{h}}$ and waiting a variable delay time $t$. Each transmon level $\ket{\mathrm{g}}$, $\ket{\mathrm{e}}$, $\ket{\mathrm{f}}$, and $\ket{\mathrm{h}}$ corresponds to a characteristic complex pointer value $s_{\mathrm{g}}, s_{\mathrm{e}}, s_{\mathrm{f}}, s_{\mathrm{h}}$ in the IQ plane, and the averaged signal is modeled as
\begin{equation}
    S(t)
    =
    \sum_{i \in \{\mathrm{g},\mathrm{e},\mathrm{f},\mathrm{h}\}}
    P_i(t)\, s_i
    =
    s_{\mathrm{g}}
    + (s_{\mathrm{e}}-s_{\mathrm{g}}) P_{\mathrm{e}}(t)
    + (s_{\mathrm{f}}-s_{\mathrm{g}}) P_{\mathrm{f}}(t)
    + (s_{\mathrm{h}}-s_{\mathrm{g}}) P_{\mathrm{h}}(t),
\end{equation}
where $P_{\mathrm{g}}(t)$, $P_{\mathrm{e}}(t)$, $P_{\mathrm{f}}(t)$, and $P_{\mathrm{h}}(t)$ are the level populations at time $t$. We fit $S(t)$ (or equivalently its $I$ and $Q$ quadratures separately) using the analytic solutions of the coupled rate equations for these populations. For the single-shot measurements, we instead classify each individual readout shot as $\ket{\mathrm{g}}$, $\ket{\mathrm{e}}$, $\ket{\mathrm{f}}$, or $\ket{\mathrm{h}}$ and directly construct the populations as empirical frequencies,
\begin{equation}
    P_{\mathrm{g}}(t) = \frac{N_{\mathrm{g}}(t)}{N_{\mathrm{tot}}(t)}, \qquad
    P_{\mathrm{e}}(t) = \frac{N_{\mathrm{e}}(t)}{N_{\mathrm{tot}}(t)}, \qquad
    P_{\mathrm{f}}(t) = \frac{N_{\mathrm{f}}(t)}{N_{\mathrm{tot}}(t)}, \qquad
    P_{\mathrm{h}}(t) = \frac{N_{\mathrm{h}}(t)}{N_{\mathrm{tot}}(t)},
\end{equation}
where $N_{\mathrm{g}}(t)$, $N_{\mathrm{e}}(t)$, $N_{\mathrm{f}}(t)$, and $N_{\mathrm{h}}(t)$ are the number of shots classified into each state after a delay time $t$, and $N_{\mathrm{tot}}(t)$ is the total number of shots. These extracted $P_i(t)$ are then fit to the same multilevel rate-equation model. This unified treatment allows us to extract state-dependent relaxation rates $\Gamma_{ij}$ and the corresponding lifetimes $T_1^{ij}$ for all observed transitions.

\section{\hspace{-9pt}: Qubit reset}\label{supp_note:qubit_reset}

\subsection*{Optimization of readout frequency}
\begin{figure*}[ht]
  \centering
  \includegraphics[width=0.8\linewidth]{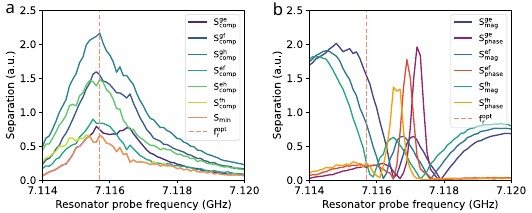}
  \caption{
  \textbf{Optimization of the readout probe frequency for four-level discrimination (Q4).} \textbf{a}, Pairwise state-separation analysis used to choose the probe frequency $f_\mathrm{r}^\mathrm{opt}$. For each probe frequency, we measure the \emph{averaged} complex readout response in the IQ plane for the transmon prepared in $\ket{\mathrm{g}}$, $\ket{\mathrm{e}}$, $\ket{\mathrm{f}}$, and $\ket{\mathrm{h}}$. We compute the complex distance for every state pair and define $S_\mathrm{min}$ as the minimum pairwise separation (worst-case distinguishability) at that frequency. The chosen probe frequency $f_\mathrm{r}^\mathrm{opt}$ (dashed line) maximizes $S_\mathrm{min}$, ensuring that all four states remain simultaneously distinguishable. \textbf{b}, The same analysis performed using only the magnitude or only the phase of the readout response. Using the full complex IQ distance yields a more uniform separation across state pairs than magnitude- or phase-only metrics.
  }
  \label{fig:Q4_singleshot_ro_freq_opt_supp}
\end{figure*}

To optimize the resonator probe frequency for multi-state discrimination, we quantify state separation as a function of probe frequency using averaged IQ responses for preparations in $\ket{\mathrm{g}}$, $\ket{\mathrm{e}}$, $\ket{\mathrm{f}}$, and $\ket{\mathrm{h}}$. For each probe frequency, we compute the complex-distance separation between all state pairs and take the minimum separation $S_\mathrm{min}$ as a conservative proxy for the worst-case distinguishability. We select $f_\mathrm{r}^\mathrm{opt}$ by maximizing $S_\mathrm{min}$ (Supplementary Fig.~\ref{fig:Q4_singleshot_ro_freq_opt_supp}a). For comparison, Supplementary Fig.~\ref{fig:Q4_singleshot_ro_freq_opt_supp}b shows the same procedure using magnitude-only and phase-only distances; in the reset experiments we use the complex-distance criterion. We also optimize the readout amplitude by maximizing the multi-state assignment fidelity across $\{\ket{\mathrm{g}},\ket{\mathrm{e}},\ket{\mathrm{f}},\ket{\mathrm{h}}\}$.

\subsection*{Four-level single-shot readout and initialization strategies}
\begin{figure*}[ht]
  \centering
  \includegraphics[width=\linewidth]{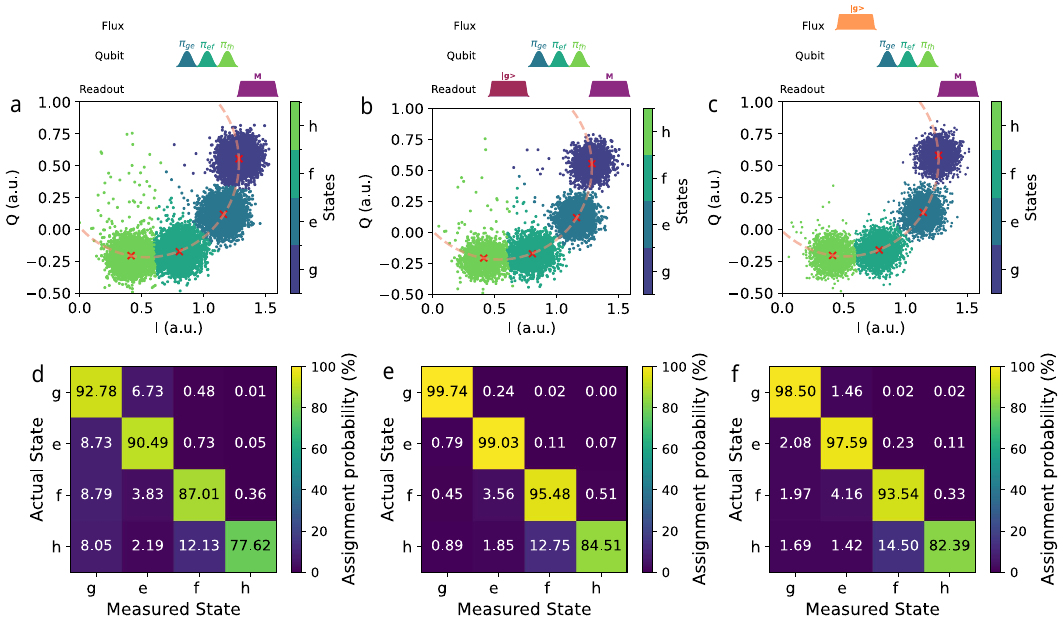}
  \caption{ \textbf{Four-level single-shot readout and ground-state initialization strategies (Q4).}
  \textbf{a}, Single-shot IQ outcomes for preparations in $\ket{\mathrm{g}}$, $\ket{\mathrm{e}}$, $\ket{\mathrm{f}}$, and $\ket{\mathrm{h}}$ using calibrated $\pi$-pulses on successive transitions ($\pi_\mathrm{ge}$, $\pi_\mathrm{ef}$, $\pi_\mathrm{fh}$). Red crosses mark the Gaussian-mixture model (GMM) means; the dashed curve highlights the approximate readout trajectory of the means in IQ space. \textbf{b}, Heralded initialization by post-selection: a weak pre-readout is used to identify shots starting in $\ket{\mathrm{g}}$, and only shots with $P(\ket{\mathrm{g}})>0.995$ are retained.
  \textbf{c}, Dissipative initialization: a short flux pulse biases the tunable-frequency drive filter to a strong-coupling (short-$T_1$) point to relax into $\ket{\mathrm{g}}$ before state preparation, eliminating post-selection.
  Pulse sequences are shown above the corresponding panels. \textbf{d}, Assignment matrix without initialization (panel \textbf{a}). \textbf{e}, Assignment matrix after heralded initialization (panel \textbf{b}). \textbf{f}, Assignment matrix after dissipative initialization (panel \textbf{c}). Off-diagonal weight reflects residual thermal population and relaxation during the readout window.}
  \label{fig:Q4_singleshot_supp}
\end{figure*}

Reset characterization requires reliable single-shot discrimination of multiple transmon levels.
We prepare $\ket{\mathrm{g}}$, $\ket{\mathrm{e}}$, $\ket{\mathrm{f}}$, and $\ket{\mathrm{h}}$ using calibrated pulses on successive transitions and record one complex IQ point per shot (Supplementary Fig.~\ref{fig:Q4_singleshot_supp}a).
We fit the joint IQ distribution with a four-component GMM to obtain, for each level, a mean IQ point and a $2\times2$ covariance matrix, and we use maximum-likelihood classification to assign each shot. From these assignments we construct the $4\times4$ assignment matrix (Supplementary Fig.~\ref{fig:Q4_singleshot_supp}d).

Residual thermal excitation produces noticeable off-diagonal weight in the raw assignment matrix. To mitigate this, we use two ground-state initialization strategies prior to state preparation. In heralded initialization, a weak pre-readout is used to post-select shots that begin in $\ket{\mathrm{g}}$ (Supplementary Fig.~\ref{fig:Q4_singleshot_supp}b,e), improving the assignment fidelities at the cost of discarding a fraction of shots. In dissipative initialization, we apply a short flux pulse that biases the tunable-frequency drive filter to a strong-coupling point (short $T_1$), allowing rapid relaxation into $\ket{\mathrm{g}}$ before state preparation (Supplementary Fig.~\ref{fig:Q4_singleshot_supp}c,f). This removes post-selection overhead and is the default initialization method used when we report reset performance without heralding.

\subsection*{Readout-limited assignment floor from signal-to-noise ratio (SNR)}
\begin{figure*}[ht]
  \centering
  \includegraphics[width=\linewidth]{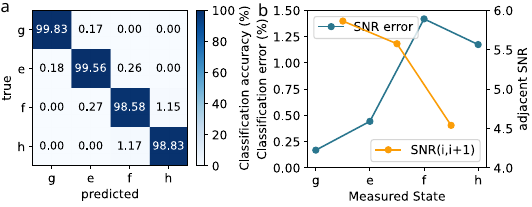}
  \caption{
  \textbf{Assignment floor expected from readout discrimination alone.} \textbf{a}, Using the fitted GMM parameters (means $\boldsymbol{\mu}_i$ and covariances $\Sigma_i$), we generate synthetic IQ shots for each state and reclassify them by maximum-likelihood to obtain the confusion matrix expected from stationary Gaussian readout noise. \textbf{b}, We summarize the resulting discrimination performance as an effective two-class separation by mapping the average misclassification probability to an equivalent binary Gaussian separation (Methods). This defines a readout-limited floor on inferred populations, independent of thermal occupation and relaxation during the readout window.}
  \label{fig:Q4_singleshot_classification_acc_supp}
\end{figure*}

To separate readout-discrimination limits from physical effects (thermal population and in-flight relaxation), we estimate the confusion matrix expected from readout noise alone. We sample synthetic IQ points from the GMM components fitted to the calibration data and classify them by maximum-likelihood, yielding the confusion matrix in Supplementary Fig.~\ref{fig:Q4_singleshot_classification_acc_supp}a. We also report an effective binary separation by mapping the average misclassification probability to an equivalent two-Gaussian separation (Supplementary Fig.~\ref{fig:Q4_singleshot_classification_acc_supp}b; Methods). In the experiment, the measured assignment matrices (Supplementary Fig.~\ref{fig:Q4_singleshot_supp}d--f) show that residual thermal population and relaxation during the readout window dominate over this readout-only floor. Accordingly, the reported reset floor is ultimately set by the residual bath temperature rather than by discrimination error.

\subsection*{Reset dial calibration: tuning $T_1$ with fast flux}
\begin{figure*}[ht]
  \centering
  \includegraphics[width=0.6\linewidth]{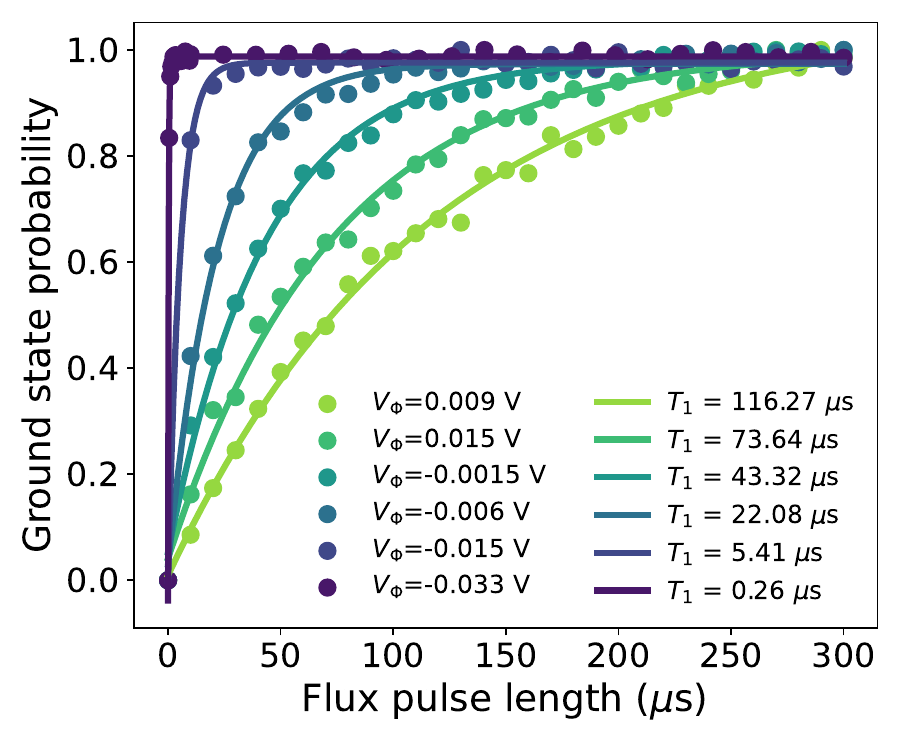}
  \caption{ \textbf{Calibrating the reset configuration through $T_1$ as a function of fast-flux pulse parameters (Q4).} The qubit is prepared in $\ket{\mathrm{e}}$, held at a chosen reset bias $V_\Phi$ for a chosen flux-pulse length, and then measured. Repeating this sequence for multiple pulse amplitudes yields decays that are fit to extract an effective $T_1(\Phi)$ at each bias point.}
  \label{fig:Q4_t1_fpl_supp}
\end{figure*}

We calibrate the reset configuration by measuring the relaxation time as functions of flux-pulse amplitude $V_\Phi$ and flux-pulse length. For a given $V_\Phi$, we initialize the qubit in $\ket{\mathrm{e}}$, apply a flux pulse of varying length, and measure the remaining excited-state population after returning to the idle point. Fitting the decay as a function of flux-pulse length yields an effective $T_1$ at a given $V_\Phi$ as shown in Supplementary Fig.~\ref{fig:Q4_t1_fpl_supp}. We select the reset operating point at the flux amplitude that minimizes $T_1$, in this case $V_\Phi = -0.033$~V, and use this bias in the reset experiments reported in the main text.

\subsection*{Extended single-shot reset data and global decay fits}
\begin{figure*}[ht]
  \centering
  \includegraphics[width=\linewidth]{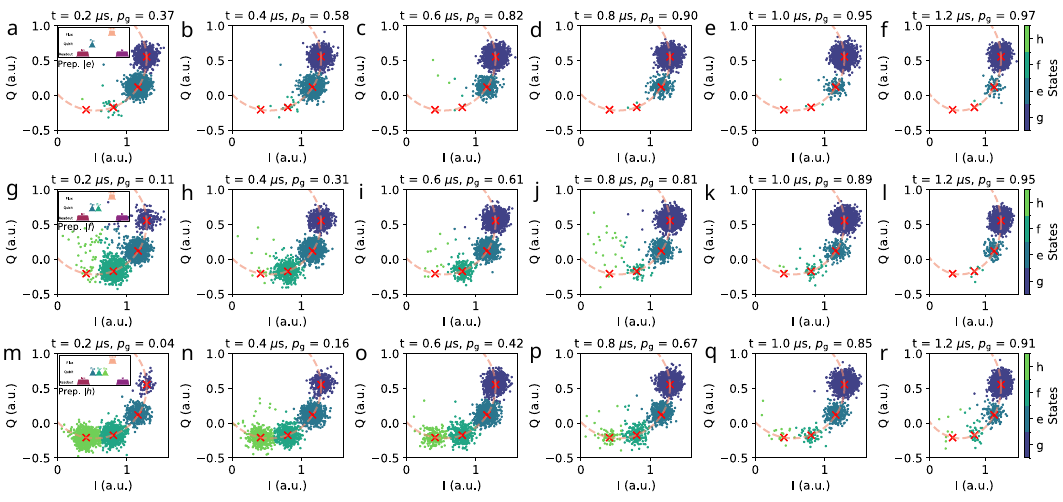}
  \caption{
  \textbf{Single-shot reset outcomes as a function of reset-pulse duration (Q4).}
  Each panel shows the pulse sequence, post-reset single-shot IQ points, and the extracted ground-state population $P_\mathrm{g}$. \textbf{a--f}, initial preparation in $\ket{\mathrm{e}}$; \textbf{g--l}, initial preparation in $\ket{\mathrm{f}}$; \textbf{m--r}, initial preparation in $\ket{\mathrm{h}}$. Increasing the reset-pulse duration increases $P_\mathrm{g}$, demonstrating fast relaxation toward $\ket{\mathrm{g}}$ from multiple initial states.}
  \label{fig:Q4_reset_IQ_fpl_supp}
\end{figure*}

Supplementary Fig.~\ref{fig:Q4_reset_IQ_fpl_supp} shows extended single-shot reset data complementing Fig.~\ref{fig:qubit_reset}c. For each initial state, we apply the reset pulse at fixed amplitude and vary its duration, then infer post-reset populations from GMM classification. Across $\ket{\mathrm{e}}$, $\ket{\mathrm{f}}$, and $\ket{\mathrm{h}}$ preparations, the ground-state population increases monotonically with reset duration, consistent with sequential relaxation through the multilevel ladder.

\begin{figure*}[ht]
  \centering
  \includegraphics[width=\linewidth]{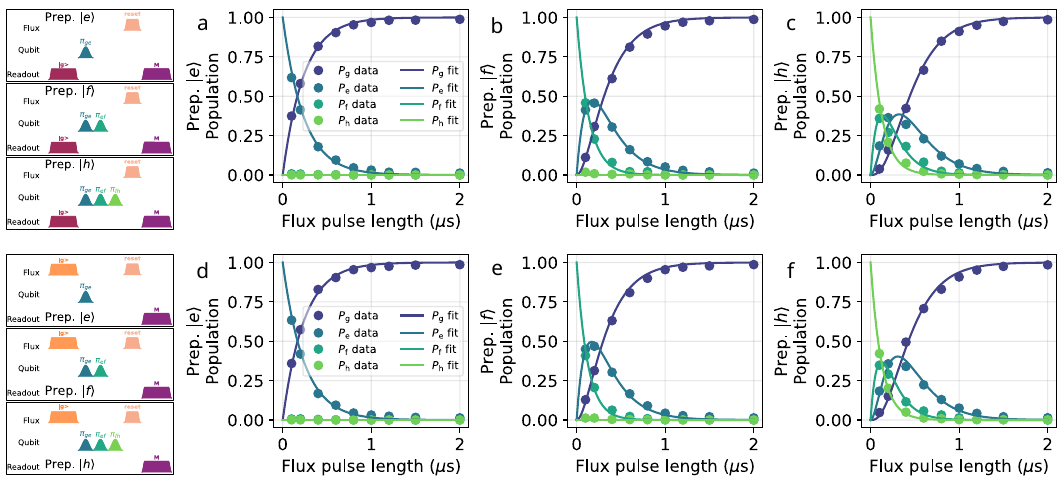}
  \caption{\textbf{Reset dynamics and global multilevel decay fits (Q4).} \textbf{a--c}, Heralded initialization: extracted populations $P_\mathrm{g}$, $P_\mathrm{e}$, $P_\mathrm{f}$, $P_\mathrm{h}$ as a function of reset-pulse duration for initial preparations in $\ket{\mathrm{e}}$, $\ket{\mathrm{f}}$, and $\ket{\mathrm{h}}$. \textbf{d--f}, Dissipative initialization: the same analysis without post-selection. Solid curves are global fits to a four-level rate-equation model with sequential downward transitions $\ket{\mathrm{h}}\rightarrow\ket{\mathrm{f}}\rightarrow\ket{\mathrm{e}}\rightarrow\ket{\mathrm{g}}$, yielding effective decay rates $\Gamma_{ij}$ and $T_1^{ij}=1/\Gamma_{ij}$. Pulse sequences for the two initialization procedures are shown at left.
  }
  \label{fig:Q4_reset_decayfit_supp}
\end{figure*}

We fit the extracted populations to a four-level rate-equation model with sequential downward transitions
$\ket{\mathrm{h}}\rightarrow\ket{\mathrm{f}}\rightarrow\ket{\mathrm{e}}\rightarrow\ket{\mathrm{g}}$ (\ref{supp_note:multilevel_char}). We perform global fits for both heralded and dissipative initialization datasets (Supplementary Fig.~\ref{fig:Q4_reset_decayfit_supp}), yielding consistent decay rates and confirming that dissipative initialization reproduces the reset dynamics without post-selection overhead. Fitted relaxation times are summarized in Supplementary Table~\ref{tab:qubit_reset_ps_fit}.

{\renewcommand{\arraystretch}{1}
\begin{table*}[ht]
\caption{Summary of state-dependent relaxation times extracted from reset dynamics for heralded and dissipative initialization.}
\label{tab:qubit_reset_ps_fit}
\centering
\begin{tabular}{|p{2cm}|p{4cm}|p{4.5cm}|}
\hline
\textbf{Parameter} & \textbf{Heralded initialization} & \textbf{Dissipative initialization}\\
\hline
\ensuremath{T_1^\mathrm{ge}} & $238.22\pm3.28~\text{ns}$ & $241.40\pm2.78~\text{ns}$\\
\ensuremath{T_1^\mathrm{ef}} & $136.80\pm2.80~\text{ns}$ & $123.07\pm2.23~\text{ns}$\\
\ensuremath{T_1^\mathrm{fh}} & $128.84\pm3.38~\text{ns}$ & $123.18\pm2.80~\text{ns}$\\
\hline
\end{tabular}
\end{table*}
}

\section{\hspace{-9pt}: Qubit thermometry}\label{supp_note:qubit_thermometry}

\subsection*{Single-shot readout and SNR}

Population-based thermometry is limited primarily by (i) the separation of multilevel single-shot IQ clusters and (ii) relaxation during the readout window. Our protocol estimates temperature from populations in the first four transmon levels. For qubit Q3, the minimum nearest-neighbour separation extracted from the Gaussian-mixture model (GMM; see Methods) corresponds to an \emph{adjacent} SNR $\ge 6$, which for approximately Gaussian clusters implies an ideal nearest-boundary misclassification probability $\lesssim 0.15\%$. This is small compared with the sampling noise of a $5000$-shot window used throughout this work and therefore does not limit the reported thermometry precision.

The $\ket{\mathrm{e}}$ and $\ket{\mathrm{f}}$ levels have sufficiently long lifetimes that their contribution to the thermometry error budget is negligible. The $\ket{\mathrm{h}}$ level relaxes more rapidly, but over the relevant temperature range ($50$--$200~\mathrm{mK}$) its equilibrium population is small and therefore weakly weighted in the estimator.

\begin{figure*}[ht]
  \centering
  \includegraphics[width=\linewidth]{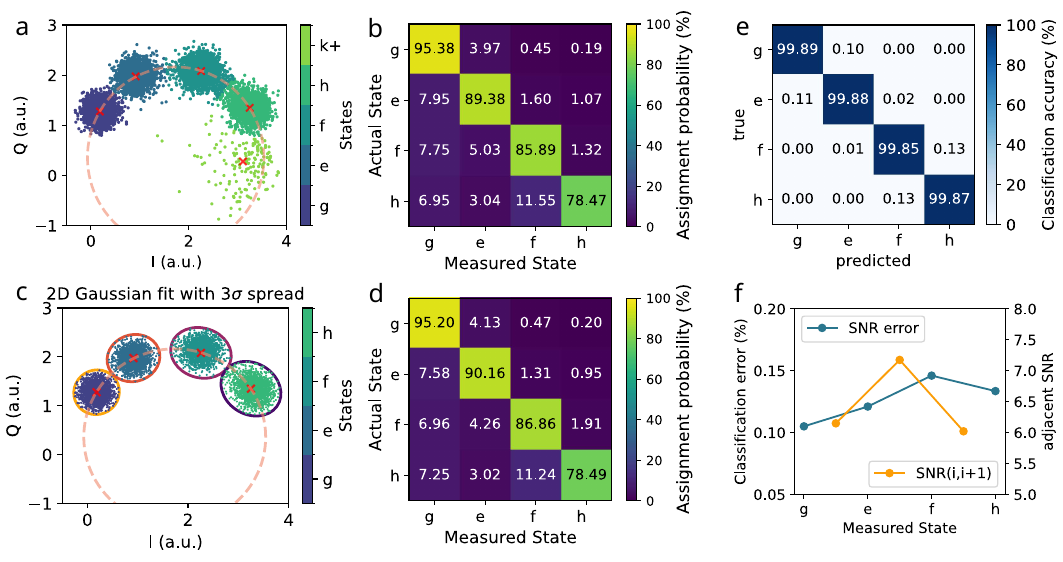}
  \caption{\textbf{Single-shot readout of qubit Q3.} \textbf{a}, Single-shot IQ outcomes for preparations in $\ket{\mathrm{g}}$, $\ket{\mathrm{e}}$, $\ket{\mathrm{f}}$, and $\ket{\mathrm{h}}$. \textbf{b}, Assignment matrix from a five-component GMM fit to the full dataset. \textbf{c}, IQ points retained after truncation to within $3\sigma$ of each Gaussian component. \textbf{d}, Assignment matrix after re-fitting the truncated dataset in \textbf{c}. \textbf{e}, Confusion matrix expected from readout SNR alone. \textbf{f}, Per-state classification error (left axis) and adjacent SNR (right axis).}
  \label{fig:Q3_thermometry_singleshot_supp}
\end{figure*}

Supplementary Fig.~\ref{fig:Q3_thermometry_singleshot_supp}\textbf{a} shows representative single-shot IQ clouds for preparations in $\ket{\mathrm{g}}$, $\ket{\mathrm{e}}$, $\ket{\mathrm{f}}$, and $\ket{\mathrm{h}}$, and the corresponding assignment matrix from a five-component GMM fit is shown in panel~\textbf{b}. To distinguish discrimination limits from physical effects (residual thermal population and in-flight relaxation), we apply a flux-pulse ground-state initialization and truncate the dataset to shots within $3\sigma$ of each fitted Gaussian (panel~\textbf{c}). Re-fitting and reclassifying this truncated dataset yields the assignment matrix in panel~\textbf{d}, which closely matches panel~\textbf{b}. This indicates that the observed off-diagonal weight is not set by readout SNR but by residual thermal occupation (consistent with a qubit temperature of $\approx 60~\mathrm{mK}$) together with $T_1$ processes in higher levels.

For comparison, panel~\textbf{e} shows the confusion matrix predicted from SNR alone, and panel~\textbf{f} plots the per-state classification error alongside the adjacent SNR. Together, these data confirm that readout discrimination contributes only a minor fraction of the total thermometry error; the dominant floor is set by residual thermal excitation and relaxation dynamics during readout.

\subsection*{Effective qubit temperature}

Across multiple cooldowns we observe elevated effective qubit temperatures, consistent with coupling to a comparatively hot electromagnetic environment. When operating in the reset configuration, the quantum dial provides a strongly coupled channel that can act as an effective cold reservoir.

\begin{figure*}[ht]
  \centering
  \includegraphics[width=\linewidth]{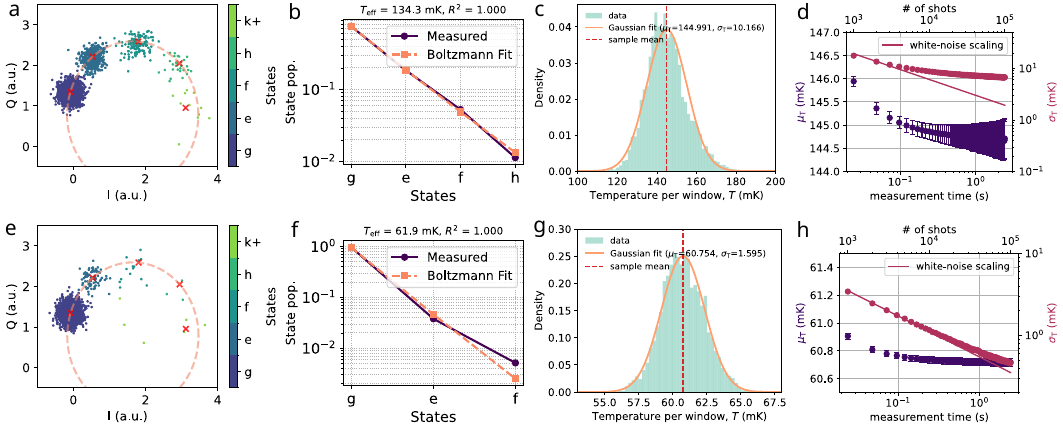}
  \caption{\textbf{Qubit temperature with and without pre-readout thermalization using the tunable-frequency drive filter (Q3).} \textbf{a}, Single-shot IQ outcomes at the lowest coupling point (idle configuration). \textbf{b}, Populations of the first four levels from $5000$ shots with a Boltzmann fit giving $T_{\mathrm{eff}}\approx 134~\mathrm{mK}$. \textbf{c}, Histogram of $T_{\mathrm{eff}}$ from non-overlapping $5000$-shot windows over $25\times10^{6}$ shots; Gaussian fit yields $\mu_\mathrm{T}$ and $\sigma_\mathrm{T}$ (red dashed line: sample mean). \textbf{d}, $\mu_\mathrm{T}$ (left axis) and $\sigma_\mathrm{T}$ (right axis) as functions of measurement time (shot count); solid line indicates $1/\sqrt{t}$ scaling. \textbf{e}, Single-shot IQ outcomes after a $10~\mu\mathrm{s}$ flux pulse that biases the filter to the strongest coupling immediately before readout. \textbf{f}, Populations (first three levels) with a Boltzmann fit giving $T_{\mathrm{eff}}\approx 60~\mathrm{mK}$. \textbf{g}, Histogram of $T_{\mathrm{eff}}$ as in \textbf{c}. \textbf{h}, $\mu_\mathrm{T}$ and $\sigma_\mathrm{T}$ as functions of measurement time as in \textbf{d}.}
  \label{fig:Q3_temp_w_wo_therm}
\end{figure*}

At the weakest-coupling point of the dial (idle configuration), the qubit equilibrates to an effective temperature of $\sim 140~\mathrm{mK}$, extracted from Boltzmann fits to the populations of the first four transmon levels (Supplementary Fig.~\ref{fig:Q3_temp_w_wo_therm}\textbf{a--d}). Panel~\textbf{a} shows representative single-shot IQ data and panel~\textbf{b} shows the extracted populations and Boltzmann fit. Panels~\textbf{c} and \textbf{d} summarize windowed temperature statistics computed from \emph{non-overlapping} windows, yielding a Gaussian distribution of $T_\mathrm{eff}$ with mean $\mu_\mathrm{T}$ and standard deviation $\sigma_\mathrm{T}$ that decreases approximately as $1/\sqrt{N}$.

When the dial is biased to its strongest-coupling point for $10~\mu\mathrm{s}$ immediately before readout, the qubit thermalizes to a colder bath and the effective temperature drops to $\sim 60~\mathrm{mK}$, as shown in Supplementary Fig.~\ref{fig:Q3_temp_w_wo_therm}\textbf{e--h}. The corresponding single-shot IQ clouds and population fits appear in panels~\textbf{e} and \textbf{f}, with windowed statistics shown in panels~\textbf{g} and \textbf{h}. These results demonstrate that the dial can transiently cool the qubit when engaged prior to readout.

\subsection*{Thermometry setup and heating protocol}

In the thermometry experiments, we use the dissipative elements of the fast-flux line as a controllable hot environment. We do not use the tunable-frequency drive-filter line for heating because it is heavily attenuated for thermalization and, away from the filter frequency, it can be strongly coupled to the qubit such that a large off-resonant tone would induce unwanted driving and dephasing.

As shown in Supplementary Fig.~\ref{fig:Q3_thermometry_setup_supp}\textbf{a}, a continuous-wave heating tone at $5~\mathrm{GHz}$ is injected through the fast-flux line and dissipated in its attenuators. The tone is generated at room temperature and amplified by a cascade of microwave amplifiers with a saturated output of $18$--$20~\mathrm{dBm}$. We parameterize the heating amplitude by a dimensionless scale between $0$ and $1$. Temperature traces of the dilution refrigerator during extended runs are shown in panel~\textbf{b}, where increases in the mixing-chamber and $4~\mathrm{K}$ stages are consistent with the attenuation distribution along the line.

We perform measurements in four configurations, summarized in Supplementary Fig.~\ref{fig:Q3_thermometry_setup_supp}\textbf{e}: $C0$ (no flux pulse, no heating), $C1$ (flux pulse only), $C2$ (flux pulse and heating tone), and $C3$ (heating tone only). Panels~\textbf{c} and \textbf{d} show the evolution of the ground-state population and effective temperature as functions of the combined pulse duration. In $C2$, the qubit thermalizes rapidly to an effective temperature of $\sim 700~\mathrm{mK}$ on a timescale of $\sim 8~\mu\mathrm{s}$, whereas in $C3$ thermalization is slower and saturates at a lower temperature. The contrast between these cases shows that strong coupling mediated by the quantum dial both accelerates thermalization and enhances the influence of the heated environment.

\begin{figure*}[ht]
  \centering
  \includegraphics[width=\linewidth]{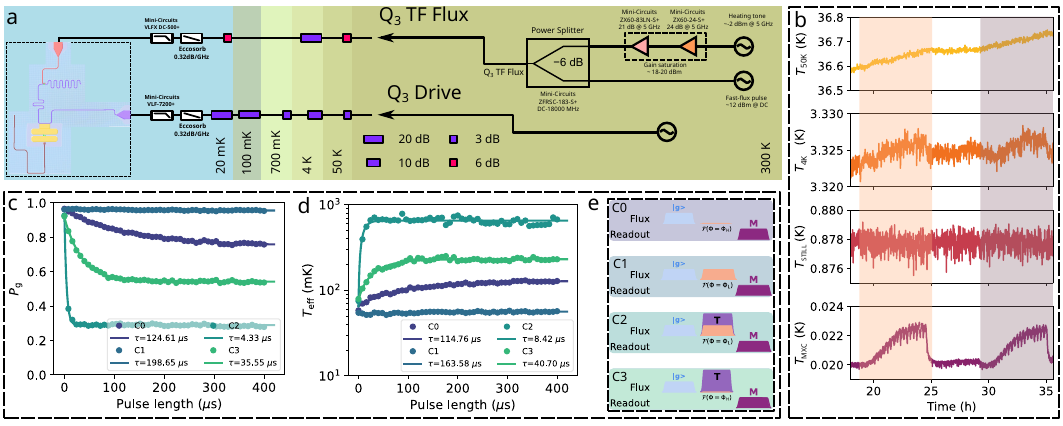}
  \caption{\textbf{Thermometry setup and heating protocol.}
  \textbf{a}, Cryogenic wiring diagram for the thermometry experiment with Q3. A continuous-wave $5~\mathrm{GHz}$ heating tone is injected into the fast-flux line and dissipated in its attenuators, increasing the microwave radiation temperature seen by the qubit. The same line applies flux pulses that tune the coupling between Q3 and the tunable-frequency drive filter. Multilevel single-shot readout yields the effective qubit temperature. \textbf{b}, Temperature traces of several dilution-refrigerator stages during extended thermometry runs; shaded regions indicate the acquisition windows used in Fig.~\ref{fig:qubit_thermometry}. \textbf{c}, Ground-state population $P_\mathrm{g}$ as a function of the combined flux/heating pulse length for configurations $C0$--$C3$; solid curves are single-exponential fits. \textbf{d}, Effective temperature $T_\mathrm{eff}$ extracted from Boltzmann fits to the four-level populations for the same pulse lengths as in \textbf{c}; solid curves are single-exponential fits. \textbf{e}, Pulse sequences defining $C0$--$C3$: $C0$ (no flux pulse, no heating tone), $C1$ (flux pulse only), $C2$ (flux pulse and heating tone), and $C3$ (heating tone only).}
  \label{fig:Q3_thermometry_setup_supp}
\end{figure*}

\subsection*{Extended thermometry data}

\begin{figure*}[ht]
  \centering
  \includegraphics[width=\linewidth]{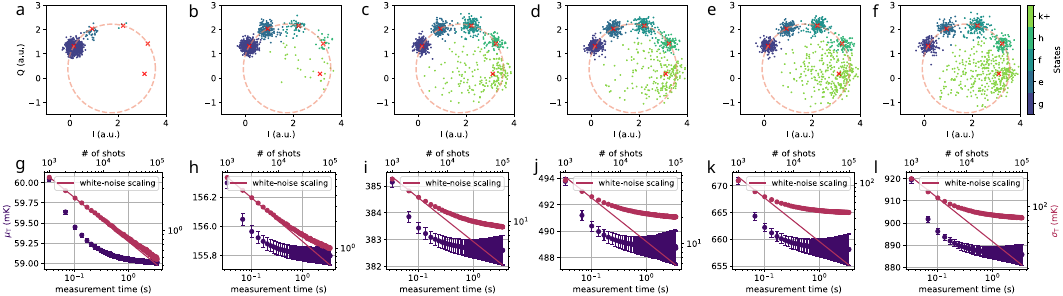}
  \caption{\textbf{Full thermometry analysis for varying heating-tone amplitudes.} \textbf{a--f}, Single-shot IQ clouds for Q3 for six heating-tone amplitudes in configuration $C2$ (increasing left to right). Points are coloured by the assigned state; the dashed orange circle indicates the ideal readout trajectory. Increasing heating populates higher levels, consistent with an increasing effective temperature. \textbf{g--l}, Extracted mean effective temperature $\mu_\mathrm{T}$ (left axis) and standard deviation $\sigma_\mathrm{T}$ (right axis) as functions of total measurement time (equivalently, number of single shots) for the same amplitudes as in \textbf{a--f}. Solid lines indicate the expected $1/\sqrt{t}$ white-noise scaling for $\sigma_\mathrm{T}$.}
  \label{fig:Q3_qt_full_supp}
\end{figure*}

Supplementary Fig.~\ref{fig:Q3_qt_full_supp} presents extended thermometry results for different heating-tone amplitudes using configuration $C2$. As the heating power increases, higher excited states become progressively populated, consistent with an increasing effective qubit temperature. Panels~\textbf{g--l} show the extracted mean temperature $\mu_\mathrm{T}$ and its standard deviation $\sigma_\mathrm{T}$ as functions of total measurement time. At low heating power, $\sigma_\mathrm{T}$ follows approximately white-noise scaling, whereas at higher heating power, the fluctuations saturate at long averaging times.

\subsection*{Readout in the strong-coupling regime}

Fast thermalization requires biasing the qubit--filter system into a strongly coupled configuration. In this regime, the Purcell-enhanced relaxation time is reduced to a few hundred nanoseconds, and the qubit decays significantly during the readout window. As a result, high-fidelity single-shot discrimination of the multilevel manifold is no longer possible, as illustrated in Supplementary Fig.~\ref{fig:Q3_singleshot_lowT1_supp}. This motivates the use of a separate, weakly coupled configuration for high-fidelity readout in thermometry measurements.

\begin{figure*}[ht]
  \centering
  \includegraphics[width=\linewidth]{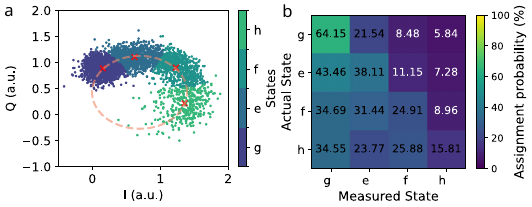}
  \caption{\textbf{Single-shot readout in the strong-coupling, low-$T_1$ regime.} \textbf{a}, Single-shot IQ cloud of Q3 in the strongly coupled configuration used for fast thermalization. Points are coloured by the prepared state $\ket{\text{g}},\ket{\text{e}},\ket{\text{f}},\ket{\text{h}}$; the dashed orange curve indicates the readout trajectory. The short $T_1$ during the measurement window causes substantial overlap of the IQ clusters. \textbf{b}, Corresponding assignment matrix, showing reduced state discrimination in this configuration and motivating the alternating strong-coupling (thermalization) and weak-coupling (readout) protocol used for thermometry.}
  \label{fig:Q3_singleshot_lowT1_supp}
\end{figure*}


\end{document}